\documentstyle[preprint,tighten,aps]{revtex}

\newcommand{\lb}{\left(}
\newcommand{\rb}{\right)}
\newcommand{\al}{\alpha}
\newcommand{\be}{\beta}
\newcommand{\ga}{\gamma}
\newcommand{\de}{\delta}
\newcommand{\rh}{\rho}
\newcommand{\si}{\sigma}
\newcommand{\ta}{\tau}
\newcommand{\et}{\eta}
\newcommand{\tchi}{\tilde{\chi}}

\newcommand{\beq}{\begin{equation}}
\newcommand{\eeq}[1]{\label{#1} \end{equation}}

\newcommand{\cpm}[1]{[#1]_\pm}
\newcommand{\cp}[1]{[#1]_+}
\newcommand{\cm}[1]{[#1]_-}
\newcommand{\Tr}[1]{Tr \lb #1 \rb}

\begin{document}

\preprint{TRI-PP-94-68}
\draft
\title{Extension of the Chiral Perturbation Theory
Meson Lagrangian to Order $p^6$}

\author{H.\ W.\ Fearing and S.\ Scherer\thanks{Address after Sept.\ 1, 1994:
Institut f\"ur Kernphysik, Johannes Gutenberg-Universit\"at, J.~J.~Becher-Weg
45, D-55099 Mainz, Germany} }

\address{TRIUMF, 4004 Wesbrook Mall, Vancouver, B.\ C.,\\ Canada V6T 2A3}

\date{August 22, 1994}
\maketitle
\begin{abstract}
We have derived the most general chirally invariant Lagrangian ${\cal L}_6$ for
the meson sector at order $p^6$. The result provides an extension of the
standard Gasser-Leutwyler Lagrangian ${\cal L}_4$ to one higher order,
including as well all the odd intrinsic parity terms in the Lagrangian. The
most difficult part of the derivation was developing a systematic strategy so
as to get all of the independent terms and eliminate the redundant ones in an
efficient way. The 'equation of motion' terms, which are redundant in the sense
that they can be transformed away via field transformations, are separated out
explicitly. The resulting Lagrangian has been separated into groupings of terms
contributing to increasingly more complicated processes, so that one does not
have to deal with the full result when calculating $p^6$ contributions to
simple processes.
\end{abstract}
\pacs{11.30.Rd, 12.39.Fe }

\narrowtext

\section{Introduction }
\label{intro}

Ever since the early days of current algebra and the $PCAC$ hypothesis (for an
overview see, e.g., Refs.\ \cite{Adler,Treiman,Cheng}), approximate chiral
symmetry and its application in terms of effective Lagrangians
\cite{Weinberg1,Weinberg2,Gasiorowicz,Dashen} have been cornerstones of the
description of the low--energy interactions of hadrons.  In our present
understanding, chiral symmetry is interpreted in terms of $QCD$, the $SU(3)$
gauge theory of the strong interaction involving quarks and gluons (see,
e.g., Refs.\ \cite{Marciano,Altarelli}).  In the limit of massless $u$, $d$ and
$s$ quarks, the $QCD$ Hamiltonian exhibits a global $SU(3)_L\times SU(3)_R$
symmetry which is assumed to be spontaneously broken to $SU(3)_V$, giving rise
to 8 massless Goldstone bosons.  The experimentally observed small masses of
the pseudoscalar octet ($\pi,K,\eta$) then originate from an explicit symmetry
breaking due to the finite quark masses (see Refs.\ \cite{Pagels,Gasser1} and
references therein).

Extending a method originally proposed by Weinberg for the analysis of
S--matrix elements \cite{Weinberg3}, Gasser and Leutwyler
\cite{Gasser2,Gasser3} developed a technique which allows an expansion of $QCD$
Green's functions of quark currents in terms of momenta and quark masses.
Their procedure, now known as chiral perturbation theory (ChPT),
makes use of an
effective Lagrangian approach for the interaction between the Goldstone bosons
$\pi,K,\eta$, and has been applied to a wide variety of processes, including
interactions with external electromagnetic and electroweak probes (for
pedagogical introductions and recent reviews, see Refs.\
\cite{Gasser4,Leutwyler1,Donoghue2,Meissner,Pich1,Ecker2,Leutwyler2}).  The
lowest--order Lagrangian corresponds to a nonlinear $\sigma$ model containing
two free parameters, namely, the pion decay constant and the scalar quark
condensate.  At $O(p^4)$ the most general Lagrangian which is consistent with
chiral symmetry, parity, and charge conjugation invariance contains 10
structures.  The renormalized coefficients $L_1 \dots L_{10}$ have
been empirically determined by
fitting experimental data \cite{Gasser2,Gasser3} (for a recent analysis see
Ref.\ \cite{Bijnens1} and references therein).  Theoretical predictions for the
chiral coefficients have also been obtained by several authors using various
techniques (see, e.g., Refs.\
\cite{Gasser2,Balog,Ebert,Ecker1,Donoghue1,Espriu,Belkov1}).

Without external fields (i.e., pure $QCD$) or including electromagnetic
processes only, the effective Lagrangian of Gasser and Leutwyler has an
additional symmetry resulting in the property that it contains interaction
terms involving exclusively an even number of Goldstone bosons
\cite{Witten,Bijnens2}.  Such interaction terms are sometimes also referred to
as being of normal or even intrinsic parity.  In Ref.\ \cite{Witten} Witten
discussed how to remove this symmetry which is not a symmetry of nature (for
example, $\pi^0\rightarrow \gamma\gamma, K^+K^-\rightarrow \pi^+\pi^-\pi^0,$
etc.).  He essentially rederived the Wess--Zumino anomalous effective action
describing the chiral anomaly \cite{Wess}.  The corresponding Lagrangian, which
is of $O(p^4)$, cannot be written as a standard local effective Lagrangian in
terms of the chiral matrix $U$ but can be expressed directly in terms of the
boson fields.  In particular, by construction it contains interaction terms
with an odd number of Goldstone bosons (odd intrinsic parity).  It was
subsequently shown by several authors that quantum corrections to the
Wess--Zumino classical action do not renormalize the coefficient of the
$O(p^4)$ Wess--Zumino term \cite{Donoghue3,Donoghue4,Issler,Bijnens3,Akhoury}.
Furthermore, the one--loop counter terms lead to conventional chirally
invariant structures at $O(p^6)$ \cite{Donoghue4,Issler,Bijnens3,Akhoury}.  For
a review of chiral perturbation theory involving the odd intrinsic parity
sector we refer the reader to Ref.\ \cite{Bijnens2}.

Chiral perturbation theory to $O(p^4)$ has become a well developed effective
representation of $QCD$ at low energies in terms of Goldstone bosons, with many
applications.  However, there exist cases where one--loop calculations to
$O(p^4)$ do not appear to lead to satisfactory agreement with experimental
data (see for example Refs.\ \cite{Ametller,Donoghue5}).  As
consistent two--loop calculations in the
even intrinsic parity sector for particular processes have started to become
available \cite{Bellucci} it is clearly important and timely to extend the most
general chiral effective Lagrangian to $O(p^6)$.  Moreover, it is now also
possible to calculate the $O(p^6)$ coefficients from phenomenological models
\cite{Belkov2}.

The purpose of this work is to derive the most general structures of the chiral
Lagrangian at $O(p^6)$ including the even, as well as the odd, intrinsic parity
sector.  The result provides an extension in analogous form of the standard
Gasser--Leutwyler $O(p^4)$ Lagrangian to $O(p^6)$.

In the even intrinsic parity sector we find 111 independent terms.  This part
of the $O(p^6)$ Lagrangian will be of relevance for processes which vanish at
$O(p^2)$, so that $O(p^4)$ is the leading order and the $O(p^6)$
terms provide the
leading--order correction.  A two--loop calculation of such a process, namely
$\gamma\gamma\rightarrow\pi^0\pi^0$, has recently been reported in Ref.\
\cite{Bellucci} (see also \cite{Knecht}).  It is a feature of this process, and
similarly of the decay $\eta\rightarrow\pi^0\gamma\gamma$ \cite{Ametller}, that
the $O(p^4)$ contribution is exclusively generated by one--loop diagrams.
According to the power counting scheme \cite{Weinberg3}, at $O(p^6)$
contributions result from two--loop diagrams with vertices from the $O(p^2)$
Lagrangian, one--loop diagrams with one vertex being $O(p^2)$ and the other
$O(p^4)$ and, finally, from tree--level diagrams from the $O(p^6)$ effective
Lagrangian.

In the odd intrinsic parity sector we find 32 independent structures of
$O(p^6)$.  Our result differs from previous determinations of these structures
\cite{Issler,Akhoury} which in turn are not in agreement with each other.  We
will provide a detailed comment on these differences.  This
sector has already been discussed in
the literature in quite some detail \cite{Bijnens2}, as it provides
leading--order corrections to processes originating in the Wess--Zumino term.

At present it appears extremely unlikely that all renormalized coefficients at
$O(p^6)$ may be determined empirically.  However, there is the hope that for
simple processes the much smaller subgroups of relevant coefficients can be
determined by one experiment and then provide predictive power for a related
process.  Furthermore, theoretical techniques which resulted in predictions of
the $O(p^4)$ coefficients might be extended to predict the coefficients at
$O(p^6)$.  For the corresponding discussion of techniques applying to the odd
intrinsic parity sector see Ref.\ \cite{Bijnens2}.  A derivation of the
$O(p^6)$  even intrinsic parity sector using the
Nambu--Jona-Lasinio model will be provided in Ref.\ \cite{Belkov2}.

Our main emphasis in this paper will be on developing a systematic procedure
for the derivation of the $O(p^6)$ Lagrangian so that the reader can be
confident that the large number of terms we have found are indeed independent
as well sufficient to describe the most general effective chiral Lagrangian at
$O(p^6)$.  The derivation involves various technical points, such as
total--derivative arguments, trace--relations, use of the classical equation of
motion in terms of field transformations, and special relations involving the
completely antisymmetric tensor in four dimensions.  We group the final
structures according to the minimal number of Goldstone boson fields, assuming
for the purpose of organizing the terms, a coupling to an external
electromagnetic field.  Our list will allow the reader to quickly identify the
relevant terms at $O(p^6)$ needed for particular physical processes.  The
structures will, of course, be given in their full generality without any
assumption concerning the nature of the external fields.

Our work is organized as follows. In Sec. \ref{basic} we discuss some of the
general foundations of chiral perturbation theory, define the basic building
blocks for the Lagrangian, discuss the strategy for obtaining a complete and
independent set of terms, and describe a number of simplifications, symmetries
and relations which will be used to reduce the result. In Sec. \ref{calc} we
describe the details of the explicit construction of the Lagrangian. Sec.
\ref{results} contains a discussion of the equation of motion terms, of some
simplifications, and of a scheme of organization which groups the most useful
terms together and a comparison with some previous works. The final results are
given in Tables \ref{2phi} - \ref{eps5phi}. Sec. \ref{concl} gives a brief
summary. Details of the derivation of trace relations used in the
simplifications, a table of the equation of motion terms, and some tables
showing relations among terms which are not independent are
relegated to the appendices.

The reader primarily interested in using the results should read Sec.
\ref{basic} to understand the notation and the general strategy and Secs.
\ref{results} and \ref{concl} for a listing and discussion of the results.

\section{Construction of the most general Lagrangian ${\cal L}_6$ -
Introduction and Basic Approach}
\label{basic}
\subsection{Introduction}
\label{ssintro}

Our aim in this section is to lay the groundwork for the construction of
${\cal L}_6$, the
most general chirally invariant Lagrangian at order $p^6$. This
Lagrangian will provide an extension to the next higher order of the standard
Gasser-Leutwyler ${\cal L}_4$, as well as the odd intrinsic parity
Lagrangian. Like ${\cal L}_4$ it will be expressed in terms of a
set of basic building blocks, arranged in traces or products of traces such
that the result is chirally invariant.

The main problem with such a construction is that it is far too easy to think
of terms satisfying the necessary criteria. One quickly obtains many more terms
than necessary, many that are not independent, though often not obviously so,
and many expressed in a more complicated fashion than necessary.

Thus the most difficult task in such a construction is to develop a systematic
strategy which allows one to get in an efficient manner all of the independent
terms while at the same time eliminating, or preferably never generating, those
terms which are redundant. Most of this section will be devoted to developing
the various ingredients of this strategy.  Then in the following major section
the strategy will be applied to derive the most general ${\cal L}_6$. Some of
this will be pedagogical, as it is important for the reader to be able to
follow and understand the steps of the derivation and thus be convinced in the
end that the result obtained is, in fact, complete and correct, something which
is not necessarily possible when only a table of results is given.

To our knowledge there have been no previous attempts to generate the complete
and most general ${\cal L}_6$, though there have been several attempts
\cite{Issler,Akhoury} to derive the set of odd intrinsic parity terms, i.e.,
those involving the completely antisymmetric tensor
$\epsilon_{\alpha\beta\gamma\delta}$.  In all
of these attempts it appears that terms were missed, redundant terms were
included, or both.

\subsection{Chiral symmetry of the $QCD$ Lagrangian}
\label{sscs}
Before discussing the construction of the effective chiral Lagrangian we
will shortly review the chiral symmetry of the $QCD$ Lagrangian coupled
to external fields.
This discussion is relevant for the identification of the transformation
properties of the external sources under the group, parity, and charge
conjugation.

Let us consider the $QCD$ Lagrangian\footnote{We omit the gauge fixing term
and the Faddeev--Popov ghost term, since they are not relevant for the
discussion of chiral symmetry.}
\cite{Marciano,Altarelli},
\begin{equation}
\label{lqcd}
{\cal L}_{QCD}=\sum_f \bar{q}_f\left(i\gamma^{\mu}D_\mu -m_f\right)q_f
-\frac{1}{4}G^a_{\mu\nu}G^{\mu\nu}_a.
\end{equation}
In Eq.\ (\ref{lqcd}) $q_f$ denotes a quark field of flavor $f$ with
current quark mass $m_f$.
The covariant derivative is defined as
\begin{equation}
\label{cdq}
D_\mu q_f=\left(\partial_\mu +ig\frac{\lambda^a}{2}A^a_\mu\right)q_f,
\end{equation}
where, for simplicity, we do not exhibit the fact that the quark fields
have three color indices.
In Eq.\ (\ref{cdq}) $g$ denotes the strong coupling constant,
$\lambda^a$ are the usual Gell--Mann matrices, and $A^a_\mu$ represents
a gluon field with color index $a$ which may take the values 1 to 8.
Summation over repeated color indices is assumed.
The gluon field strength tensor is defined as
\begin{equation}
\label{gfst}
G_{\mu\nu}^a=\partial_\mu A^a_\nu -\partial_\nu A^a_\mu-g
f_{abc} A^b_\mu A^c_\nu,
\end{equation}
where $f_{abc}$ are the $SU(3)$ structure constants.
We do not discuss the $\theta$--term which induces $P$, $T$ and $CP$ violations
in the strong interactions \cite{Altarelli}.

In the following we will restrict ourselves to three flavors, $u,d,s$, and
consider the limit $m_u,m_d,m_s\rightarrow 0$.
In this limit Eq.\ (\ref{lqcd}) reduces to
\begin{equation}
\label{lqcd0}
{\cal L}_{QCD}^0=\sum_{f=u,d,s}\left( i\bar{q}_{L,f}\gamma^{\mu}D_\mu q_{L,f}
+i\bar{q}_{R,f}\gamma^{\mu}D_\mu q_{R,f}\right)
-\frac{1}{4}G^a_{\mu\nu}G^{\mu\nu}_a,
\end{equation}
where the left--handed and right--handed quark fields are defined as
\begin{displaymath}
q_L=\frac{1}{2}(1-\gamma_5)q,\quad q_R=\frac{1}{2}(1+\gamma_5)q.
\end{displaymath}
The Lagrangian of Eq.\ (\ref{lqcd0}) is invariant under the following global
transformations of $q_L$ and $q_R$, respectively,
\begin{equation}
\label{trafoq}
\left( \begin{array}{c}u'_L\\d'_L\\s'_L\end{array}\right)
= U_L \left( \begin{array}{c}u_L\\d_L\\s_L\end{array}\right),
\quad
\left( \begin{array}{c}u'_R\\d'_R\\s'_R\end{array}\right)
= U_R \left( \begin{array}{c}u_R\\d_R\\s_R\end{array}\right),
\end{equation}
where $U_L$ and $U_R$ are independent $U(3)$ matrices.
This invariance, in principle, gives rise to 18 conserved currents.
However, due to quantum effects, the axial $U(1)_A$ current is not
conserved.
In the following we will only be concerned with the analysis of the
$G=SU(3)_L\times SU(3)_R$ symmetry of Eq.\ (\ref{lqcd0}).
It is generally accepted that this symmetry is spontaneously broken to
$SU(3)_V$ giving rise to 8 Goldstone bosons.
The finite quark masses in Eq.\ (\ref{lqcd}) then give rise to finite
-- but in comparison with other hadrons small -- masses of the Goldstone
bosons.

In order to systematically study the consequences of chiral symmetry and its
breaking through the quark masses, we follow the technique of Gasser and
Leutwyler \cite{Gasser2,Gasser3} and introduce external c--number fields
$v_\mu (x)$, $a_\mu (x)$, $s(x)$ and $p(x)$ into the Lagrangian,
\begin{equation}
\label{lqcds}
{\cal L}={\cal L}^0_{QCD}+{\cal L}_{ext}
={\cal L}^0_{QCD}+\bar{q}\gamma^\mu (v_\mu +\gamma_5 a_\mu )q
-\bar{q}(s-i\gamma_5 p)q.
\end{equation}
The external fields are color neutral, hermitian $3\times 3$ matrices,
where the matrix character, with respect to the (suppressed) flavor indices
$u,d,s$ of the quark fields, is
\begin{equation}
\label{mch}
v_\mu=\frac{\lambda^a}{2}v^a_\mu,\quad
a_\mu=\frac{\lambda^a}{2}a^a_\mu,\quad
s=\frac{\lambda^a}{2}s^a,\quad
p=\frac{\lambda^a}{2}p^a.
\end{equation}
Of course, the three flavor $QCD$ Lagrangian is recovered by setting
$v_\mu=a_\mu=p=0$ and $s=diag(m_u,m_d,m_s)$.

Requiring the total Lagrangian of Eq.\ (\ref{lqcds}) to be invariant
under $P$, $C$ and $T$ leads to constraints on the transformation
behavior of the external fields.
Under a parity transformation the quark fields transform
as\footnote{We suppress color indices since we are considering color
neutral external sources.}
\begin{equation}
\label{qtrafop}
q_f(\vec{x},t)\stackrel{P}{\rightarrow}\gamma^0 q_f(-\vec{x},t).
\end{equation}
The requirement of invariance under a parity transformation,
\begin{equation}
\label{parinv}
{\cal L}(\vec{x},t) \stackrel{P}{\rightarrow} {\cal L}(-\vec{x},t),
\end{equation}
results in the following transformation properties of the external fields
\begin{equation}
\label{eftrafop}
v^\mu\stackrel{P}{\rightarrow}v_\mu,\quad
a^\mu\stackrel{P}{\rightarrow}-a_\mu,\quad
s\stackrel{P}{\rightarrow}s,\quad
p\stackrel{P}{\rightarrow}-p,
\end{equation}
where it is understood that the arguments change from $(\vec{x},t)$
to $(-\vec{x},t)$.
Under charge conjugation the quark fields transform as
\begin{equation}
\label{qtrafc}
q_{\alpha,f}\stackrel{C}{\rightarrow}C_{\alpha\beta}\bar{q}_{\beta,f},
\quad
\bar{q}_{\alpha,f}\stackrel{C}{\rightarrow}-q_{\beta,f}C^{-1}_{\beta\alpha}.
\end{equation}
In Eq.\ (\ref{qtrafc}) the subscripts $\alpha$ and $\beta$ are Dirac
spinor indices, $C=i\gamma^2\gamma^0$ is the usual charge conjugation
matrix in the convention of Ref.\ \cite{Bjorken} and $f$ refers to flavor.
Using Eq.\ (\ref{qtrafc}) it is straightforward to show that invariance
of ${\cal L}_{ext}$ under charge conjugation requires the transformation
properties
\begin{equation}
\label{eftrafoc}
v_\mu\stackrel{C}{\rightarrow}-v_\mu^T,\quad
a_\mu\stackrel{C}{\rightarrow}a_\mu^T,\quad
s,p\stackrel{C}{\rightarrow}s^T,p^T,
\end{equation}
where the transposition refers to the flavor space.

Finally, we discuss the properties of ${\cal L}$  under
the group $G$. To that end we rewrite Eq.\ (\ref{lqcds}) in terms of
$q_L$ and $q_R$,
\begin{equation}
\label{lqcdsn}
{\cal L}={\cal L}_{QCD}^0
+\bar{q}_L\gamma^\mu L_\mu q_L
+\bar{q}_R\gamma^\mu R_\mu q_R
+\bar{q}_L(s-ip)q_R + \bar{q}_R(s+ip)q_L,
\end{equation}
with $R_\mu=v_\mu +a_\mu$ and $L_\mu =v_\mu -a_\mu$.
We then promote the global symmetry to a local symmetry.
Eq.\ (\ref{lqcdsn}) remains invariant under $q_R\rightarrow V_R(x) q_R$
and $q_L\rightarrow V_L(x) q_L$, provided the external fields transform as
\begin{eqnarray}
\label{sg}
&&R_\mu\stackrel{G}{\rightarrow} V_R R_\mu V_R^{\dagger}
+iV_R\partial_\mu V_R^{\dagger},\quad
L_\mu\stackrel{G}{\rightarrow} V_L L_\mu V_L^{\dagger}
+iV_L\partial_\mu V_L^{\dagger},
\nonumber\\
&&s-ip\stackrel{G}{\rightarrow} V_L(s-ip)V_R^{\dagger},\quad
s+ip\stackrel{G}{\rightarrow} V_R(s+ip)V_L^{\dagger}.
\end{eqnarray}

\subsection{Basic Building Blocks}
\label{ssbldblks}
The Lagrangian ${\cal L}_6$ is constructed from the same basic ingredients as
${\cal L}_4$, namely the Goldstone boson fields, external gauge fields,
scalar and pseudoscalar external sources, and their derivatives.
The 8 Goldstone bosons arising from the spontaneous symmetry breaking
are collected in a $SU(3)$ matrix
\beq
U(x)=\exp\lb  i\frac{\phi(x)}{F_0} \right ),
\eeq{u}
with
\beq
\phi(x)= \left (
\begin{array}{ccc}
\pi^0+\frac{1}{\sqrt{3}}\eta & \sqrt{2} \pi^+ & \sqrt{2} K^+ \\
\sqrt{2} \pi^- & -\pi^0+\frac{1}{\sqrt{3}}\eta & \sqrt{2} K^0 \\
\sqrt{2} K^- & \sqrt{2} \bar{K}^0 & -\frac{2}{\sqrt{3}}\eta
\end{array}
\right ),
\eeq{Phi}
and $F_0$ the pseudoscalar meson decay constant in the chiral limit
\cite{Gasser3}.
The matrix $U$ transforms linearly under the group $G=SU(3)_L\times
SU(3)_R: U\rightarrow U'=V_R U V_L^\dagger$ \cite{Gasser3}.
Furthermore, under charge conjugation and parity the Goldstone bosons
transform as $C:\phi\rightarrow \phi^T$ and $P:\phi(\vec{x},t)
\rightarrow -\phi(-\vec{x},t)$, or equivalently $C:U\rightarrow U^T$
and $P:U(\vec{x},t)\rightarrow U^\dagger(-\vec{x},t)$.

We define the field strength tensors associated with the external
gauge fields $R_\mu$ and $L_\mu$  as
\begin{eqnarray}
\label{Fdef}
F^R_{\mu\nu}&\equiv&\partial_\mu R_\nu-\partial_\nu R_\mu-i[R_\mu,R_\nu],
\nonumber\\
F^L_{\mu\nu}&\equiv&\partial_\mu L_\nu-\partial_\nu L_\mu-i[L_\mu,L_\nu].
\end{eqnarray}
Note that $F^R_{\mu\nu}$ and $F^L_{\mu\nu}$ are hermitian,
traceless, and gauge invariant, i.e., under the group they transform as
$F^R_{\mu\nu}\stackrel{G}{\rightarrow} V_R F^R_{\mu\nu} V_R^\dagger$ and
$F^L_{\mu\nu}\stackrel{G}{\rightarrow} V_L F^L_{\mu\nu} V_L^\dagger$.
Furthermore, we follow Ref.\ \cite{Gasser3} and introduce the linear
combination $\chi\equiv 2 B_0(s+ip)$, with $\chi\stackrel{G}{\rightarrow}
V_R \chi V_L^{\dagger}$, where $B_0$ is related to the
vacuum expectation value $<0|\bar{q}q|0>$.

The effective Lagrangian is constructed in terms of $U$, $U^\dagger$,
$\chi$, $\chi^\dagger$ and the field strength tensors
$F^{R}_{\mu\nu}$, $F^{L}_{\mu\nu}$
as well as covariant derivatives of these objects.
These covariant derivatives involve the gauge fields $R_\mu$ and $L_\mu$ and
transform in the same way under the group as the quantities they act upon.
Given the transformation properties of $R_\mu$ and $L_\mu$ of Eq.\ (\ref{sg})
we define the covariant derivatives as
\begin{eqnarray}
\label{covder}
A\stackrel{G}{\rightarrow}V_RAV_L^\dagger:
&\quad&D_\mu A\equiv\partial_\mu A-iR_\mu A+iA L_\mu,\nonumber\\
B\stackrel{G}{\rightarrow}V_LBV_R^\dagger:
&\quad&D_\mu B\equiv\partial_\mu B+iBR_\mu -iL_\mu B,\nonumber\\
C\stackrel{G}{\rightarrow}V_RCV_R^\dagger:
&\quad&D_\mu C\equiv\partial_\mu C-iR_\mu C+iCR_\mu,\nonumber\\
D\stackrel{G}{\rightarrow}V_LDV_L^\dagger:
&\quad&D_\mu D\equiv\partial_\mu D-iL_\mu D+iDL_\mu,\nonumber\\
E\stackrel{G}{\rightarrow}E:
&\quad&D_\mu E\equiv\partial_\mu E.
\end{eqnarray}
Note that we use the same symbol $D_\mu$ for the covariant derivative,
independent of the transformation property of the object it acts upon.
The advantage of this convention is that a chain rule analogous to ordinary
derivatives holds.
Given the product $Z=XY$ where $X,Y,Z$ have, according to Eq.\ (\ref{covder}),
well--defined but not necessarily the same transformation behavior,
the chain rule applies
\beq
D_\mu Z = D_\mu (XY)=(D_\mu X) Y+X( D_\mu Y),
\eeq{chr}
which is straightforward to verify using the definitions of
Eq.\ (\ref{covder}).

This chain rule is valuable as an intermediate step in a number of the
derivations of various relations. In essentially all cases however the
final results will be expressed solely in terms of quantities
transforming as $U$ and their covariant derivatives and thus require
only the covariant derivative defined in the first line
of the definitions of Eq.\ (\ref{covder}).

It will be very useful in the subsequent derivations to have all of the
building blocks transform in the same way and also to be able to handle the
external field terms $\chi, F^{R}_{\mu\nu}$, $F^{L}_{\mu\nu}$ in the same way.
To that end we define
\begin{eqnarray}
\label{chidef}
&& G^{\mu\nu} \equiv F^{\mu\nu}_R U + U F^{\mu\nu}_L, \nonumber \\
&& H^{\mu\nu} \equiv F^{\mu\nu}_R U - U F^{\mu\nu}_L.
\end{eqnarray}
We then let $\chi^{\mu\mu}=\chi$ and let $\chi^{\mu\nu}$, with $\mu \neq
\nu$, stand for $G^{\mu\nu}$ or $H^{\mu\nu}$. Thus we can consider
$\chi^{\mu\nu}$ with $\mu,\nu$ unspecified to stand for, collectively, the
set ($\chi$, $G^{\mu\nu}$, $H^{\mu\nu}$).

With these definitions, we have only two basic building blocks $U$,
$\chi^{\mu\nu}$ and their adjoints and covariant derivatives acting on them.
Note that by virtue of the chain rule, we do not need derivatives acting on
products of these basic terms.
All building blocks then transform as $U$ (or $U^\dagger$). In terms of the
momentum expansion $U$ is of order $1$, $\chi^{\mu\nu}$ is of order $p^2$ and
each covariant derivative $D_\mu$ is of order $p$ \cite{Gasser3}.

Finally, in order to construct terms which are invariant, we must put these
basic pieces together so as to get quantities which transform as
$V_R \dots V_R^\dagger$. Thus define for any $A$
transforming as $V_R A V_L^\dagger$, i.e. as $U$, the hermitian (or
antihermitian) combinations
\beq
\cpm{A} \equiv \frac{1}{2} (AU^\dagger \pm U A^\dagger).
\eeq{comdef}
We can then take as the basic building blocks $\cpm{A}$ with $A$ taken
as $\chi^{\mu\nu} $ or $D_\mu U$, or as some number of covariant derivatives
acting on $\chi^{\mu\nu} $ or $D_\mu U$.
The trace of a string of such quantities, or the product of such traces,
will then be invariant under the
group. Observe also that by defining the building blocks this way we include
each operator and its adjoint. By using building blocks transforming as
$V_R \dots V_R^\dagger$ we also solve the problem
of determining whether or not extra $U$'s and $U^\dagger$'s are
necessary. In the usual approach, starting with a string of quantities
transforming as $V_R \dots V_L^\dagger$ or $V_L \dots V_R^\dagger$,
one can always insert extra $U$'s and
$U^\dagger$'s in the string, replace some terms by their
adjoints, and get another invariant string. One then must check in each
case whether the resulting string is in fact independent of the original
one. However by using the forms of Eq.\ (\ref{comdef}) the only possible
combination which can be inserted is $U U^\dagger$ which is unity and of
course unnecessary. Observe also that if we were to insert $U U^\dagger$
between each term and after the last term in a trace of a string of
terms of the form of Eq.\ (\ref{comdef})
and use the relation $U^\dagger(AU^\dagger \pm U A^\dagger)U = U^\dagger
A \pm A^\dagger U$  we obtain a trace of the same form as the original,
but constructed of building blocks transforming as $V_L \dots V_L^\dagger$
which we could have defined in a fashion analogous to Eq.\
(\ref{comdef}).
This shows that such terms are not independent and that terms of the
form of Eq.\ (\ref{comdef}) are sufficient.

To summarize, we have seen that the basic building blocks necessary to
construct the most general chirally invariant Lagrangian are of the form
$\cpm{A}$, where $A$ is $D_\mu U$, $\chi^{\mu\nu}$ or (multiple)
covariant derivatives of such quantities. The problem now,
to be addressed in the next
section, is how to put these building blocks together in an efficient way
to get all possible terms in the Lagrangian.

\subsection{Strategy}
\label{ssstrat}
It is possible to generate a very large number of terms using just the
building blocks defined in the previous section. The most difficult part
of obtaining ${\cal L}_6$ is developing a strategy which obtains all of
the independent terms without generating a lot of extraneous terms which
have to be eliminated by hand.

To do this it is convenient to first define a hierarchy of terms. At
each level we will then find the most general set of terms. First
however, various relations can be used to eliminate some terms in
favor of those
lower down in the hierarchy. Since at each level we always find the most
general set, it is not necessary to actually work out what these (often
extremely complicated) relations are. To eliminate a term one must only
show that there exists a relation expressing it in terms of quantities
being kept at the same level and others lower down in the hierarchy.

We thus order the various classes of terms as follows:
\begin{itemize}
\item[I.] Those terms with six $D_\mu$'s.
\item[II.] Those terms with four $D_\mu$'s and one $\chi^{\mu\nu}$.
\item[III.] Those terms with two $D_\mu$'s and two $\chi^{\mu\nu}$'s.
\item[IV.] Those terms with three  $\chi^{\mu\nu}$'s.
\end{itemize}
At each major level terms which involve a single trace are considered
higher up than those with multiple traces. Likewise the ordering ensures
that the addition of an $F^{\mu\nu}$
always results in a term lower in the hierarchy. Since the result must be
a Lorentz scalar and since the only available tensors $\chi^{\mu\nu}$ and
$\epsilon_{\alpha\beta\gamma\delta}$ have an even number of indices there
are no terms with an odd number of $D_\mu$'s.

For each level one first determines the form of the possible structures.
For this it often is easier to start initially with the
original building blocks
$U,\chi^{\mu\nu}$ and their covariant derivatives. A number of
simplifications and tricks to be discussed in the next section are then
applied to reduce the number of possible terms and ensure that each term
satisfies parity and charge conjugation. The remaining structures are then
written
in terms of the final building blocks $\cpm{A}$. The result is a set of
general forms. Explicit results can then be read off directly, usually by
simply considering all possible permutations of the indices, and perhaps
of the order of the terms, and all possible ways single traces can be
broken up into multiple traces.

Terms involving $D_\mu D^\mu U$ are of particular interest as this factor
appears in the classical equation of motion and terms
proportional to the equation
of motion can, as we shall see in a later section (see also Ref.\
\cite{Scherer}),
be transformed away by a transformation
on, or a redefinition of, the fields. Thus our strategy will be to extract
in so far as it is possible such terms, which will be referred to as
'equation of motion terms'. At the end we can make such terms proportional
to the
full ${\cal L}_2$ equation of motion by simply adding in the rest of the terms
in the equation of motion, as these terms will come from a lower level in
the hierarchy.

\subsection{Simplifications-General Results}
\label{ssimpl}
\subsubsection{Total derivative arguments}
\label{sstd}
It is convenient to make use of the fact that a total derivative in
the Lagrangian density does not change the equation of motion.
This `total derivative argument' can be applied as follows.
For any pair of operators $A$ and $B$ which under the group either transform as
$A'=V_R A V_L^\dagger$ and $B'=V_L B V_R^\dagger$ or as
$A'=V_R A V_R^\dagger$ and $B'=V_R B V_R^\dagger$ the product $AB$ transforms
as $V_R AB V_R^\dagger$ and the chain rule and definitions of the
covariant derivatives yield
\beq
D_\mu(A B) = (D_\mu A) B+A (D_\mu B) = \partial_\mu (A B)
-i[R_\mu,A B],
\eeq{pr}
and thus using the fact that the trace of a commutator vanishes,
\beq
\Tr{(D_\mu A) B} + \Tr{A (D_\mu B)} = \partial_\mu \Tr{A B}.
\eeq{rm}
The same result is obtained for combinations $AB$ transforming as $V_L AB
V_L^\dagger$.

Clearly this result generalizes to products of terms of the
form $B_1 \dots B_n$,
and one gets
\begin{eqnarray}
\label{rmgen}
\Tr{(D_\mu B_1) B_2 \dots B_n} &+& \Tr{B_1 (D_\mu B_2) B_3 \dots
B_n}\nonumber \\ &+& \dots + \Tr{B_1 \dots (D_\mu B_n)}
= \partial_\mu \lb \Tr{B_1 \dots B_n} \rb.
\end{eqnarray}

It is also straightforward to extend the `total derivative argument'
to products of traces.
As an illustration, we find applying Eqs.\ (\ref{rm}) and (\ref{chr}),
\begin{eqnarray}
\label{tdap}
\Tr{(D_\mu A_1)A_2}\Tr{A_3A_4}&+&\Tr{A_1(D_\mu A_2)}\Tr{A_3A_4} \nonumber\\
+ \Tr{A_1A_2}\Tr{(D_\mu A_3) A_4}&+&\Tr{A_1A_2}\Tr{A_3 (D_\mu A_4)} \nonumber\\
&=&\partial_\mu\lb \Tr{A_1A_2}\Tr{A_3A_4}\rb.
\end{eqnarray}
As a consequence of these arguments we may move covariant derivatives
from one factor to another either within a single trace or across
multiple traces in accord with Eqs.\ (\ref{rm}),(\ref{rmgen}),(\ref{tdap}).
Terms involving total derivatives will result, but these
will not contribute to the equation of motion and thus can be  dropped
from the Lagrangian.

\subsubsection{Symmetrization of multiple covariant derivatives }
\label{sssym}
For any operator $A$ transforming as $U$ we have, using the definition of
covariant derivatives Eq.\ (\ref{covder}) and of the field strength
tensors Eq.\ (\ref{Fdef})
\beq
(D_\mu D_\nu-D_\nu D_\mu)A = i A F^L_{\mu\nu}- i F^R_{\mu\nu}A.
\eeq{ddtof}
Any term of the form $ D_\mu D_\nu A$ can be written as a term symmetric
and one antisymmetric in $\mu \leftrightarrow \nu$. Thus Eq.\
(\ref{ddtof})
implies that at any given level in the hierarchy $ D_\mu D_\nu A$ can
always be assumed to be symmetric in its indices, since the antisymmetric
part can be expressed in terms of the field strength tensors and thus
contributes only at a lower level in the
hierarchy. Clearly this can be generalized to more than two covariant
derivatives. Hence we will always assume that a multiple covariant
derivative, $D_{\mu_1} \dots D_{\mu_n} A$, is symmetric in the interchange
of any of its indices. Formally this means that we will always interpret
$ D_\mu D_\nu A$ as
\beq
D_\mu D_\nu A \rightarrow  \frac{1}{2} (D_\mu D_\nu+D_\nu D_\mu)A,
\eeq{symder}
with an analogous definition when there are more than two covariant
derivatives.

\subsubsection{Index exchange}
\label{ssinex}
In keeping with the strategy of extracting as many 'equation of motion'
factors as possible, one would like to arrange things so that covariant
derivatives with summed indices, e.g. $D_\alpha D^\alpha$, act on the
same factor $U$. For example consider $\Tr{(D_\alpha D_\beta A) (D^\alpha
D_\gamma B) C}$, where $A,B,C$ do not contain any covariant derivatives.
Using the results of Secs.\ \ref{sstd} and \ref{sssym} one can move the
$D_\beta$ off of $A$ and on to $B$ and $C$ and then move the $D_\alpha$
off of $B$ and on to $A$ and $C$. The result is
\begin{eqnarray}
\label{inex}
\Tr{(D_\alpha D_\beta A) (D^\alpha D_\gamma B) C}
&=& \Tr{(D_\alpha D^\alpha A) (D_\beta D_\gamma B) C}\nonumber\\
&&+\mbox{\rm terms with multiple $D$'s on only one of $A,B,C$} \nonumber\\
&&+\mbox{\rm total derivatives},
\end{eqnarray}
where we have dropped the terms arising from the commutation of the $D$'s at
various stages which can be expressed via terms lower in the hierarchy
using Eq.\ (\ref{ddtof}).

\subsubsection{Building blocks involving multiple covariant derivatives}
\label{ssbbmcd}
Consider a term of the form $\cp{ D_\alpha D_\beta U}$. By starting with
the trivial identity $0=D_\alpha D_\beta (UU^\dagger)$ and using Eq.\
(\ref{chr}) one finds
\beq
2 \cp{ D_\alpha D_\beta U} = - (D_\alpha U)(D_\beta U)^\dagger -
(D_\beta U)  (D_\alpha U)^\dagger
\eeq{d2u}
which means that $\cp{ D_\alpha D_\beta U}$ can always be expressed in
terms of quantities involving at most a single covariant derivative
acting on $U$. This can be generalized, starting with $0=D_{\mu_1} \dots
D_{\mu_n} (UU^\dagger)$, to show that $\cp{D_{\mu_1} \dots
D_{\mu_n} U}$ can always be expressed as a sum of products of $\cpm{A}$
where A has at most one less covariant derivative. Symbolically, $\cp{D^n U}
\sim (D^iU)(D^{n-i}U)^\dagger \sim \cpm{D^i U}\cpm{D^{n-i} U}$ for all
$(1\leq i \leq n-1)$. By iteration, one then shows that all terms of the
form $\cp{D_{\mu_1} \dots
D_{\mu_n} U}$ can be expressed via the combinations with relative
negative sign $\cm{A}$ and the right (or left) hand side of
Eq.\ (\ref{d2u}). Note also that the simplest case $\cp{D_\mu U} = 0$, so that
the right hand side of Eq.\ (\ref{d2u}) can also be expressed solely in terms
of $\cm{A}$ as well. This means in practice that we never need to include
factors of the form $\cp{D_{\mu_1} \dots D_{\mu_n} U}$.

\subsubsection{Symmetries, parity and charge conjugation}
\label{ssparity}
Each term in the final Lagrangian must be hermitian and must be invariant
under parity and charge conjugation. Hermiticity is normally trivial as
all terms come out to be either hermitian or antihermitian once parity
and charge conjugation invariance is imposed. The strategy for ensuring
P and C invariance is to take each candidate term and add to it the
parity transform of the term, thus getting a result which is trivially P
invariant. Similarly one then adds to this result its C transform, and so
generates a C and P invariant term.

Consider first parity. We will be interested in candidate terms for the
Lagrangian which are of the form of a trace of a string of factors of the
canonical form $\cpm{A}$, or alternatively a product of such traces,
where $A$ can be $D_\mu U$ or $\chi^{\mu\nu}$ or their (perhaps multiple)
covariant derivatives. Table \ref{trafprop} shows the
transformation properties
of the various factors. We see that under parity all allowed $A$'s
transform as $A \rightarrow (-1)^p A^\dagger$ , where in addition all
Lorentz indices in $A^\dagger$ have been raised (or lowered) with
respect to those in $A$, and $\vec{x} \rightarrow -\vec{x}$. The
quantity $p$ is an `intrinsic parity' which
is 1 for $H^{\mu\nu}$  and its covariant derivatives and 0 otherwise.
In most cases Lorentz indices from one factor will be contracted with
those of another factor, so that the raising or lowering of indices under
P compensates in the two factors, and can be ignored. The
exception to this is the case when
the indices are contracted not with another term of the set of $A$'s but
with the antisymmetric tensor $\epsilon_{\alpha\beta\gamma\delta}$. Parity
does not affect this tensor, so to get a Lorentz invariant quantity one
must raise or lower its indices by hand. This introduces a minus sign
since $\epsilon_{\alpha\beta\gamma\delta} = -
\epsilon^{\alpha\beta\gamma\delta}$. This can be summarized neatly by
including a $(-1)^\epsilon$ in the parity transform of a complete term,
where $\epsilon$ counts the number of
$\epsilon_{\alpha\beta\gamma\delta}$'s in the term.

Using these preliminaries, we see that under parity $\cpm{A} \rightarrow
\pm (-1)^p U^\dagger \cpm{A} U$. Under a trace the $U^\dagger$ and $U$
collapse to 1, using the cyclic properties of a trace. Thus we arrive at
the following operational method of making a candidate term in the
Lagrangian invariant under parity. Simply multiply the term by the factor
$1 + (-1)^{s+P+\epsilon}$ where $P$ is the sum of the intrinsic parities
$p$ of the various terms, $\epsilon$ is the number of
$\epsilon_{\alpha\beta\gamma\delta}$'s in the term, and where the
$(-1)^s$ is determined from the product of signs coming from the individual
factors in the term, counting $+$ for $\cp{A}$ and $-$ for $\cm{A}$.
Clearly this holds for a single trace or a product of several traces.

Note also that by extracting this factor, we make it easy to see which
terms are not allowed by parity conservation, namely those for which
$(-1)^{s+P+\epsilon}$ is $-1$.

Consider now charge conjugation and proceed in exactly the same way. From
Table \ref{trafprop} it is easy to see that the necessary $A$'s transform
under C as $(-1)^c A^T$ where $c$ is an `intrinsic charge conjugation
quantum number' which is $1$ for $G^{\mu\nu}$ and its covariant derivatives
and 0 otherwise and $T$ is the transpose. We then
have under charge conjugation $\cpm{A} \rightarrow (-1)^c
(U^\dagger \cpm{A} U )^T$.
Using the fact that the trace of a quantity is equal to the trace of its
transpose, we then find that under C the trace of a string of basic
factors $\cpm{A}$ goes simply to $(-1)^C$ times the trace with the basic
factors in reverse order, where here C is simply the sum of the intrinsic
charge conjugation quantum numbers of the various terms. Clearly in the
case of multiple traces the factors are reversed in each trace
individually.

Thus to summarize, the operational way to get a term in the Lagrangian to
be invariant under charge conjugation is to add to it a term $(-1)^C$
times  the term with the
major factors reversed in each trace. If all of the traces have at most
two factors, the cyclic property of a trace implies that the second term
has the same structure as the first, and that effectively one is simply
multiplying by the overall factor $(1+(-1)^C)$.

As noted ensuring hermiticity is relatively simple since the basic
building blocks satisfy $ \cpm{A}^\dagger = \pm \cpm{A}$. Thus the
adjoint of a string of such terms is just $(-1)^s$ times
the string in reversed order,
where $s$ is calculated as in the parity discussion above. But, by virtue of C
invariance, reversing the order produces an overall phase $(-1)^C$, so that
taking the adjoint of a term simply generates an overall factor $(-1)^{s+C} =
(-1)^{P+C+\epsilon}$.  Thus to get a
hermitian term we need only multiply by a factor of $i$ when
this factor is $-1$.
This happens for terms with an even (odd) number of $G^{\mu\nu}$'s and
$H^{\mu\nu}$'s (or their covariant derivatives) combined
and one (no) $\epsilon_{\al\be\ga\de}$ factor.
We will do this only in the final listing of results.

\subsubsection{Trace relations}
\label{sstrace}

In the construction of ${\cal L}_4$ a relation between $SU(3)$ generators was
used to eliminate one term \cite{Gasser3}.
That relation can be generalized and used to
eliminate a number of possible terms of order $p^6$. A proof of the
relation and some discussion is given in Appendix \ref{aprt}. For the present
purposes we need only the result in the following form.  For any set of  $3
\times 3$ matrices $B_i$, $i=1, \dots ,n$ with $n \geq 4$, we can express
\beq
\sum_{All\, perm.}\Tr{B_1 \dots B_n}
\eeq{trrel}
as a sum of products of traces, each trace containing no more than $n-1$
of the $B_i$. Since we are interested only in terms invariant under the
group, and since we take the $B_i$ in all permutations, we need $B_i$ to
transform as $V_R B_i V_R^\dagger$. Thus the $B_i$ of
interest will be the basic
building blocks $\cpm{A}$. Since a product of traces is always lower
down in the hierarchy than a single trace, this relation allows us to
discard one  single trace term from each set of terms made up of
all permutations of a particular group of building blocks.

\subsubsection{Epsilon relations}
\label{ssepsrel}
There is also a set of relations which relate various terms involving the
tensor $\epsilon_{\alpha\beta\gamma\delta}$, and allow one to eliminate
some in favor of others \cite{Akhoury}. These relations originate from the
observation
that a tensor antisymmetric in five Lorentz indices must be zero since
there are only four possible different indices. Such a tensor is
\beq
g^{\alpha\beta}\epsilon^{\gamma\rho\tau\eta} -
g^{\alpha\gamma}\epsilon^{\beta\rho\tau\eta} -
g^{\alpha\rho}\epsilon^{\gamma\beta\tau\eta} -
g^{\alpha\tau}\epsilon^{\gamma\rho\beta\eta} -
g^{\alpha\eta}\epsilon^{\gamma\rho\tau\beta} = 0.
\eeq{antisymten}

Now consider a further tensor $Q_{\al\be\ga\rh\ta\et}$ which
will actually be the trace, or product of traces, of the basic building
blocks $\cpm{A}$ and contract this tensor with that of
Eq.\ (\ref{antisymten}) in all possible ways. This leads to the following six
equations:
\beq
(\phantom{+{Q_{\ga\al\rh\ta\et}}^\al}-{{Q_{\al}}^\al}_{\ga\rh\ta\et} +
{{Q_{\al\ga}}^\al}_{\rh\ta\et} -
{{Q_{\al\ga\rh}}^\al}_{\ta\et}
+{{Q_{\al\ga\rh\ta}}^\al}_\et -{Q_{\al\ga\rh\ta\et}}^\al)
\epsilon^{\ga\rh\ta\et}=0,
\eeq{eps1}
\beq
(+{{Q_{\al}}^\al}_{\ga\rh\ta\et} \phantom{+{Q_{\ga\al\rh\ta\et}}^\al} +
{{Q_{\ga\al}}^\al}_{\rh\ta\et}
+{{Q_{\ga\al\rh}}^\al}_{\ta\et}
-{{Q_{\ga\al\rh\ta}}^\al}_\et +{Q_{\ga\al\rh\ta\et}}^\al)
\epsilon^{\ga\rh\ta\et}=0,
\eeq{eps2}
\beq
(-{{Q_{\al\ga}}^\al}_{\rh\ta\et} +{{Q_{\ga\al}}^\al}_{\rh\ta\et}
\phantom{+{Q_{\ga\al\rh\ta\et}}^\al} -
{{Q_{\ga\rh\al}}^\al}_{\ta\et}
+{{Q_{\ga\rh\al\ta}}^\al}_\et -{Q_{\ga\rh\al\ta\et}}^\al)
\epsilon^{\ga\rh\ta\et}=0,
\eeq{eps3}
\beq
(+{{Q_{\al\ga\rh}}^\al}_{\ta\et} -{{Q_{\ga\al\rh}}^\al}_{\ta\et}
+{{Q_{\ga\rh_\al}}^\al}_{\ta\et}
\phantom{+{Q_{\ga\al\rh\ta\et}}^\al}-{{Q_{\ga\rh\ta\al}}^\al}_\et +
{Q_{\ga\rh\ta\al\et}}^\al)
\epsilon^{\ga\rh\ta\et}=0,
\eeq{eps4}
\beq
(-{{Q_{\al\ga\rh\ta}}^\al}_\et +{{Q_{\ga\al\rh\ta}}^\al}_\et
-{{Q_{\ga\rh\al\ta}}^\al}_\et
+{{Q_{\ga\rh\ta\al}}^\al}_\et \phantom{+{Q_{\ga\al\rh\ta\et}}^\al} -
{Q_{\ga\rh\ta\et\al}}^\al)
\epsilon^{\ga\rh\ta\et}=0,
\eeq{eps5}
\beq
(+{Q_{\al\ga\rh\ta\et}}^\al -{Q_{\ga\al\rh\ta\et}}^\al
+{Q_{\ga\rh\al\ta\et}}^\al
-{Q_{\ga\rh\ta\al\et}}^\al +
{Q_{\ga\rh\ta\et\al}}^\al\phantom{+{Q_{\ga\al\rh\ta\et}}^\al})
\epsilon^{\ga\rh\ta\et}=0.
\eeq{eps6}

Observe the symmetry of the $Q$'s across the diagonal and the fact that
Eq.\ (\ref{eps6}) is just the sum of the first five. Thus in general there
are five independent equations which generate for each specific
$Q_{\al\be\ga\rh\ta\et}$ five relations among the
terms involving $\epsilon^{\ga\rh\ta\et}$. However if
$Q_{\al\be\ga\rh\ta\et}$ has some symmetry in its indices, not
all of these five relations are independent. For example if
$Q_{\al\be\ga\rh\ta\et}$ is symmetric in the first two indices
then Eqs.\ (\ref{eps3}), (\ref{eps4}), and (\ref{eps5}) are identities
and Eqs.\ (\ref{eps1}) and (\ref{eps2}) are not independent
and so there is only one independent 'epsilon relation'.

\section{Construction of the most general Lagrangian ${\cal L}_6$ -
Explicit Calculation}
\label{calc}
Having completed these preliminaries we now proceed to the explicit
calculation of ${\cal L}_6$. As noted earlier, the strategy will be to
start with the highest level in our hierarchy and work down. At each
level we first itemize the types of terms possible. At this stage it is
useful to adopt a schematic notation. Thus an expression like
$(D^2U)(DU)$ will mean the class of terms which have a $(D_\mu D_\nu U)$
or its adjoint and a $D_\alpha U$ or its adjoint, in any order, and with
any contraction of indices which lead to a chirally invariant term when
the trace is taken. We then use the results of Sec. \ref{ssimpl} to eliminate
as many classes of terms as possible and to force invariance under parity
and charge conjugation.  The results are a set of general expressions
which satisfy the various constraints. One can then read directly from
these expressions all of the allowed terms by simply taking all possible
index contractions, and in some cases all possible orders.

At a given level in the hierarchy the manipulations required to reduce terms to
lower levels and to ensure P and C invariance are essentially
the same for multiple trace terms as for
single trace ones. Hence we will generally work, somewhat
symbolically, with just a string of factors
and only at the end take the single trace and all possible
multiple traces.

We will always attempt to extract a factor $\cm{D_\mu D^\mu U}$ as the
terms containing this factor can be
made proportional to the classical equation of motion, and as discussed below,
transformed away via a transformation of the fields.

\subsection{Terms with six $D_\mu$'s and no $\chi^{\mu\nu}$'s}
\label{ss6d0chi}
Terms in this class will have six covariant derivatives acting singly or
in multiples on $U$. Since there are six Lorentz indices, two must be
contracted and the other four can be contracted pairwise in two pairs, or
contracted with the indices of the antisymmetric tensor
$\epsilon_{\al\be\ga\de}$. Using the total derivative
arguments of Sec. \ref{sstd} the covariant derivatives can always be
moved back and forth, so as to get a factor $(D_\mu D_\nu U)$ which will be
pulled to the first of the string of terms which eventually
will become an argument of a trace.
Either $\mu = \nu$ in this factor or, if not,  no other multiple
derivative has a
contracted index in it, since if it did we could move all but the two
contracted $D$'s onto another term and pull the contracted pair out as
the first
term instead. Note that by virtue of Sec. \ref{sssym} all multiple
derivatives can be taken as symmetric in their indices. This leads to
structures of the following types:
\begin{eqnarray}
\label{eq6ds1}
1) &\quad & (D_\mu D_\nu U) (D^4U), \nonumber \\
2) &\quad & (D_\mu D_\nu U) (D^3U) (DU) U, \nonumber \\
3) &\quad & (D_\mu D_\nu U) (D^2U) (D^2U) U, \nonumber \\
4) &\quad & (D_\mu D_\nu U) (D^2U) (DU) (DU), \nonumber \\
5) &\quad & (D_\mu D_\nu U) (DU) (DU) (DU) (DU) U.
\end{eqnarray}

To get an expression transforming as $V_R \dots V_R^\dagger$ so that the
trace will be invariant we use Sec. \ref{ssbbmcd} to write each of the
structures in terms of products of factors $\cpm{A}$. Each
combination $\cp{A}$, again by the
results of Sec. \ref{ssbbmcd}, can be expressed via terms with one fewer
covariant derivative, and hence via structures lower down in the list of
Eq.\ (\ref{eq6ds1}), and so does not need to be kept. However at the
bottom of the list $\cp{D_\mu D_\nu U}$ is expressed in terms of
$(DU)(DU)$ and so generates a sixth structure involving six $\cm{D U}$'s.

Requiring parity invariance introduces an overall factor
$1 + (-1)^{s+P+\epsilon}$ in accord with Sec. \ref{ssparity}. The
intrinsic parities $p$ are all zero so we find that the parity conserving
terms must have an even number of $\cm{A}$ factors and no
$\epsilon_{\al\be\ga\de}$
or an odd number with an $\epsilon_{\al\be\ga\de}$. Finally
charge conjugation invariance simply adds in a term with the $\cm{A}$
factors in reverse order.

Thus we obtain terms of the following forms:
\begin{eqnarray}
\label{eq6ds2}
1) &\quad & \cm{D^2U} \cm{D^4U}, \nonumber \\
2) &\quad & \cm{D^2U} \cm{D^3U} \cm{DU} \epsilon_{\al\be\ga\de}, \nonumber \\
3) &\quad & \cm{D^2U} \cm{D^2U} \cm{D^2U} \epsilon_{\al\be\ga\de}, \nonumber \\
4) &\quad & \cm{D^2U} \cm{D^2U} \cm{DU} \cm{DU}, \nonumber \\
5) &\quad & \cm{D^2U} \cm{DU}   \cm{DU} \cm{DU} \cm{DU}
\epsilon_{\al\be\ga\de}, \nonumber \\
6) &\quad & \cm{DU} \cm{DU} \cm{DU}  \cm{DU} \cm{DU} \cm{DU},
\end{eqnarray}
where in each case we must add a piece with all major factors in reverse
order to ensure charge conjugation invariance. We must also take all
possible starting orders, which in this case, once C is taken into account,
is relevant only for 4) and generates two different structures there.

Note that all manipulations so far hold for single or multiple traces,
except that for multiple traces we reverse the terms in each trace
separately.

Finally we put in indices  and the order reversal explicitly to obtain
the most general result for single trace expressions involving six
covariant derivatives. It consists of all possible index contractions of
\begin{eqnarray}
\label{eq6dres}
1) &\quad & \cm{D_\mu D_\nu U} \cm{D_\al D_\be D_\ga D_\de U},
\nonumber \\
2) &\quad & \cm{D_\mu D_\nu U} \lb \cm{D_\al D_\be D_\ga U}\cm{D_\de U}+
\mbox{rev}\rb \epsilon_{\rh\ta\et\si},
\nonumber \\
3) &\quad & \cm{D_\mu D_\nu U} \lb \cm{D_\al D_\be U} \cm{D_\ga D_\de U}+
\mbox{rev} \rb \epsilon_{\rh\ta\et\si},
\nonumber \\
4a) &\quad & \cm{D_\mu D_\nu U} \lb \cm{D_\al D_\be U} \cm{D_\ga U}
\cm{D_\de U}+\mbox{rev} \rb, \nonumber \\
4b) &\quad & \cm{D_\mu D_\nu U} \lb \cm{D_\al U} \cm{D_\be D_\ga U}
\cm{D_\de U}+\mbox{rev} \rb, \nonumber \\
5) &\quad & \cm{D_\mu D_\nu U}\lb \cm{D_\al U} \cm{D_\be U}
\cm{D_\ga U}\cm{D_\de U}+\mbox{rev} \rb \epsilon_{\rh\ta\et\si}, \nonumber \\
6) &\quad & \cm{D_\mu U}\cm{D_\nu U}\cm{D_\al U}  \cm{D_\be U}
\cm{D_\ga U}\cm{D_\de U} + \mbox{rev}.
\end{eqnarray}
Here 'rev' stands for the terms in reverse order and it is understood that
we must take all possible index contractions and in the end a trace.

The general result for multiple traces will be essentially the same,
except that we need to take all possible multiple traces, as well as all
index contractions, and except that the reversed term in each of the above is
modified so as to reverse the terms in each trace separately.

We are now in a position to simply read off the results from
Eq.\ (\ref{eq6dres}), as the only thing that really needs to be done is to be
sure that all independent index contractions are included. Note
that a number of
the possible contractions vanish because the multiple derivatives are
symmetric in their indices while $\epsilon_{\al\be\ga\de}$ is
antisymmetric.
Also we are interested only in the structure, so overall factors will be
neglected.

For $\mu = \nu$ we get from 1)
\beq \Tr{\cm{D_\mu D^\mu U}\cm{D_\al D^\al D_\be D^\be U}},
\eeq{6d:1a}
from 4a)
\beq \Tr{\cm{D_\mu D^\mu U}\cm{D_\al D^\al U}\cm{D_\be U}\cm{D^\be
U}}, \eeq{6d:4aa}
\beq \Tr{\cm{D_\mu D^\mu U} (\cm{D_\al D_\be U}\cm{D^\al U}
\cm{D^\be U} + \cm{D^\be U}\cm{D^\al U}\cm{D_\al D_\be U})},
\eeq{6d:4ab}
from 4b)
\beq \Tr{\cm{D_\mu D^\mu U} \cm{D_\be U} \cm{D_\al D^\al U}
\cm{D^\be U}}, \eeq{6d:4ba}
\beq \Tr{\cm{D_\mu D^\mu U} \cm{D^\al U} \cm{D_\al D_\be U}
\cm{D^\be U}}, \eeq{6d:4bb}
and from 5)
\beq \Tr{\cm{D_\mu D^\mu U} \cm{D^\al U}
\cm{D^\be U} \cm{D^\ga U}  \cm{D^\de U}}
\epsilon_{\al\be\ga\de} . \eeq{6d:5a}
Terms from 2) or 3) all vanish, using the symmetry of multiple covariant
derivatives.

For $\mu \neq \nu$, recall that we cannot have repeated indices in the
multiple derivatives. Thus we obtain from 4a)
\beq \Tr{\cm{D_\mu D_\nu U} \cm{D^\mu D^\nu U} \cm{D_\al U}
\cm{D^\al U}}, \eeq{6d:4ac}
\beq \Tr{\cm{D_\mu D_\nu U} \cm{D^\mu D^\al U} \cm{D^\nu U}
\cm{D_\al U}}, \eeq{6d:4ad}
\beq \Tr{\cm{D_\mu D_\nu U} \cm{D^\mu D^\al U} \cm{D_\al U}
\cm{D^\nu U}}, \eeq{6d:4ae}
from 4b)
\beq \Tr{\cm{D_\mu D_\nu U} \cm{D_\al U} \cm{D^\mu D^\nu U}
\cm{D^\al U}}, \eeq{6d:4bc}
\beq \Tr{\cm{D_\mu D_\nu U} (\cm{D^\nu U} \cm{D^\mu D^\al U}
\cm{D_\al U} + \cm{D_\al U} \cm{D^\mu D^\al U} \cm{D^\nu
U})}, \eeq{6d:4bd}
and from 5)
\beq \Tr{\cm{D^\mu D^\nu U}(\cm{D_\nu U} \cm{D^\be U} \cm{D^\ga
U} \cm{D^\de U} + \cm{D^\de U} \cm{D^\ga U} \cm{D^\be
U} \cm{D_\nu U} )} \epsilon_{\mu\be\ga\de}, \eeq{6d:5b}
\beq \Tr{\cm{D^\mu D^\nu U}(\cm{D^\be U} \cm{D_\nu U} \cm{D^\ga
U} \cm{D^\de U} + \cm{D^\de U} \cm{D^\ga U} \cm{D_\nu U}
\cm{D^\be U} )} \epsilon_{\mu\be\ga\de}. \eeq{6d:5c}

Finally from 6) we get
\beq \Tr{\cm{D_\al U} \cm{D^\al U} \cm{D_\be U}
\cm{D^\be U} \cm{D_\ga U} \cm{D^\ga U}}, \eeq{6d:6a}
\beq \Tr{\cm{D_\al U} \cm{D^\al U} \cm{D_\be U}
\cm{D_\ga U} \cm{D^\be U} \cm{D^\ga U}}, \eeq{6d:6b}
\beq \Tr{\cm{D_\al U} \cm{D^\al U} \cm{D_\be U}
\cm{D_\ga U} \cm{D^\ga U} \cm{D^\be U}}, \eeq{6d:6c}
\beq \Tr{\cm{D_\al U} \cm{D_\be U} \cm{D_\ga U}
\cm{D^\al U} \cm{D^\be U} \cm{D^\ga U}}, \eeq{6d:6d}
\beq \Tr{\cm{D_\al U} \cm{D_\be U} \cm{D_\ga U}
\cm{D^\be U} \cm{D^\al U} \cm{D^\ga U}}. \eeq{6d:6e}

A number of these terms differ from others only by a reordering of terms
and thus we might expect that the trace relations of Sec. \ref{sstrace} might
connect some of them. This is in fact the case. The sets Eqs.\
(\ref{6d:4aa}) and (\ref{6d:4ba}), Eqs.\ (\ref{6d:4ab}) and
(\ref{6d:4bb}),
Eqs.\ (\ref{6d:4ac}) and (\ref{6d:4bc}), Eqs.\ (\ref{6d:4ad}),
(\ref{6d:4ae}), and (\ref{6d:4bd}), and Eqs.\ (\ref{6d:6a}),
(\ref{6d:6b}),
(\ref{6d:6c}), (\ref{6d:6d}), and (\ref{6d:6e}) in each case
contain all of the
independent permutations. Hence one of each set can be eliminated using
the trace relation, say Eqs.\ (\ref{6d:4ba}), (\ref{6d:4ab}),
(\ref{6d:4bc}), (\ref{6d:4bd}), and (\ref{6d:6e}).
Likewise Eqs.\ (\ref{6d:5a}), (\ref{6d:5b}), and (\ref{6d:5c}) are
related by the epsilon
relation of Sec. \ref{ssepsrel} and we can use it to eliminate say
Eq.\ (\ref{6d:5c}).

It is interesting to note that this epsilon relation connects a
term involving the equation of motion, Eq.\ (\ref{6d:5a}), with two that do
not obviously contain the equation of motion, Eqs.\ (\ref{6d:5b}) and
(\ref{6d:5c}). As we discuss later, the
equation of motion terms can be transformed away by an appropriate
transformation of the fields, and so do not contribute. Had we chosen to
use the epsilon relation to eliminate the equation of motion term
Eq.\ (\ref{6d:5a}) rather than one of the others we would have ended up with
two terms from this set instead of one, and it would have been extremely
non obvious, from just looking at the terms, that one was in fact
redundant.

Consider now the multiple trace terms arising from the general result of
Eq.\ (\ref{eq6dres}). We simply have to construct all possible combinations of
traces, plus all possible contractions of indices. The terms arising from
charge conjugation have to be modified slightly, as we take the terms in
reverse order within each separate trace. The result is simplified very
much by the fact that $\Tr{\cm{D_\mu U}}=\Tr{\cm{D_\mu D_\nu U}} =0$.
That fact, plus the symmetry of multiple derivatives, means that there
are no contributions from lines 1), 2), 3) or 5) of Eq.\ (\ref{eq6dres}).

With $\mu = \nu$ we get from 4a)
\beq \Tr{\cm{D_\mu D^\mu U} \cm{D_\al D^\al U}} \Tr{\cm{D_\be U}
\cm{D^\be U}}, \eeq{6dm:4aa}
\beq \Tr{\cm{D_\mu D^\mu U} \cm{D_\al D_\be U}} \Tr{\cm{D^\al U}
\cm{D^\be U}}, \eeq{6dm:4ab}
and from 4b)
\beq \Tr{\cm{D_\mu D^\mu U} \cm{D_\al U}}  \Tr{\cm{D_\be D^\be U}
\cm{D^\al U}}. \eeq{6dm:4ba}
\beq \Tr{\cm{D_\mu D^\mu U} \cm{D_\al U}}  \Tr{\cm{D^\al D^\be U}
\cm{D_\be U}}, \eeq{6dm:4bb}

With $\mu \neq \nu$ we get from 4a)
\beq \Tr{\cm{D_\mu D_\nu U} \cm{D^\mu D^\nu U}} \Tr{\cm{D_\al U}
\cm{D^\al U}}, \eeq{6dm:4ac}
\beq \Tr{\cm{D_\mu D_\nu U} \cm{D^\mu D^\al U}} \Tr{\cm{D^\nu U}
\cm{D_\al U}}, \eeq{6dm:4ad}
and from 4b)
\beq \Tr{\cm{D_\mu D_\nu U} \cm{D_\al U}}  \Tr{\cm{D^\mu D^\nu U}
\cm{D^\al U}}, \eeq{6dm:4bc}
\beq \Tr{\cm{D_\mu D_\nu U} \cm{D^\nu U}}  \Tr{\cm{D^\mu D^\al U}
\cm{D_\al U}}, \eeq{6dm:4bd}
\beq \Tr{\cm{D_\mu D_\nu U} \cm{D_\al U}}  \Tr{\cm{D^\mu D^\al U}
\cm{D^\nu U}}. \eeq{6dm:4be}

Finally from 6) we get
\beq \Tr{\cm{D_\mu U} \cm{D^\mu U}} \Tr{\cm{D_\al U}
\cm{D^\al U} \cm{D_\be U} \cm{D^\be U}}, \eeq{6dm:6a}
\beq \Tr{\cm{D_\mu U} \cm{D^\mu U}} \Tr{\cm{D_\al U}
\cm{D_\be U} \cm{D^\al U} \cm{D^\be U}}, \eeq{6dm:6b}
\beq \Tr{\cm{D_\mu U} \cm{D_\nu U}} \Tr{\cm{D^\mu U}
\cm{D^\nu U} \cm{D_\al U} \cm{D^\al U}}, \eeq{6dm:6c}
\beq \Tr{\cm{D_\mu U} \cm{D_\nu U}} \Tr{\cm{D^\mu U}
\cm{D_\al U} \cm{D^\nu U} \cm{D^\al U}}, \eeq{6dm:6d}
\beq \Tr{\cm{D_\mu U} \cm{D^\mu U}\cm{D_\al U}}
\Tr{\cm{D_\be U} \cm{D^\be U}\cm{D^\al U}}, \eeq{6dm:6e}
\beq \Tr{\cm{D_\mu U} \cm{D_\nu U}\cm{D_\al U}}
\Tr{\cm{D^\mu U} \cm{D^\nu U}\cm{D^\al U}}, \eeq{6dm:6f}
\beq \Tr{\cm{D_\mu U} \cm{D_\nu U}\cm{D_\al U}}
\Tr{\cm{D^\mu U} \cm{D^\al U}\cm{D^\nu U}}, \eeq{6dm:6g}
\beq \Tr{\cm{D_\mu U} \cm{D^\mu U}} \Tr{\cm{D_\al U} \cm{D^\al U}}
\Tr{\cm{D_\be U} \cm{D^\be U}}, \eeq{6dm:6h}
\beq \Tr{\cm{D_\mu U} \cm{D^\mu U}} \Tr{\cm{D_\al U} \cm{D_\be U}}
\Tr{\cm{D^\al U} \cm{D^\be U}}, \eeq{6dm:6i}
\beq \Tr{\cm{D_\mu U} \cm{D_\nu U}} \Tr{\cm{D^\mu U} \cm{D^\al U}}
\Tr{\cm{D^\nu U} \cm{D_\al U}}. \eeq{6dm:6j}

There are two trace relations for the multiple trace terms, one relating
Eqs.\ (\ref{6dm:6a}) and (\ref{6dm:6b}) and one relating Eqs.\
(\ref{6dm:6c}) and (\ref{6dm:6d}). We use these to eliminate
Eqs.\ (\ref{6dm:6b}) and (\ref{6dm:6d}).

\subsection{Terms with four $D_\mu$'s and one $\chi^{\mu\nu}$}
\label{ss4d1chi}
Terms at this level will have four $D_\al$'s and one $\chi^{\mu\nu}$,
which may be a simple $\chi$ if $\mu = \nu$ or may be $G^{\mu\nu}$ or
$H^{\mu\nu}$ constructed from the $F^{\mu\nu}$'s according to Eq.\
(\ref{chidef}) if $\mu \neq \nu$. In the former case there will be four
Lorentz indices which may be contracted in two pairs or with an
$\epsilon_{\al\be\ga\de}$ symbol. In the second case there will be six
indices and the types of contractions will be the same as in the
preceding section, except that  $\chi^{\mu\nu}$ is antisymmetric in its
indices.

Proceeding as before we write down schematically the types of terms
which are possible, using the initial building blocks, with the
understanding that
eventually traces of these terms will be taken. We find
\begin{eqnarray}
\label{eq4ds1}
1) &\quad & (D^2U)(D^2 \chi^{\mu\nu}), \nonumber \\
2) &\quad & (D^2U) (D^2U) \chi^{\mu\nu} U, \nonumber \\
3) &\quad & (D^2U) (DU) (DU) \chi^{\mu\nu}, \nonumber \\
4) &\quad & (D^2U) (DU)  (D \chi^{\mu\nu}) U, \nonumber \\
5) &\quad & (D^2 \chi^{\mu\nu}) (DU) (DU) U,  \nonumber \\
6) &\quad & (D \chi^{\mu\nu}) (DU) (DU) (DU), \nonumber \\
7) &\quad & \chi^{\mu\nu} U (DU) (DU) (DU) (DU).
\end{eqnarray}

In writing these down we have already used the total derivative arguments
of Sec. \ref{sstd} to eliminate terms involving $D^4$ and $D^3$. In fact
it is a general result that one needs at most half of the covariant
derivatives in the term acting on a single factor. We have included
explicitly the minimum number of $U$'s necessary to give an even number of
factors  so that it is possible to get  the right overall
transformation properties. Note also that at the
present stage there may in principle be additional nonadjacent
$U$ and $U^\dagger$
pairs distributed through some of the terms. Using again the total
derivative argument one of the $D$'s  in 5) can be moved off of the
$\chi$ to give terms like 6) and 4). Similarly in 6) the $D$ can be moved
from the $\chi$ to give terms like 3) and, if there are extra
$U$,$U^\dagger$ pairs, like 7). Thus 5) and 6) need not be considered.

Using Sec. \ref{sssym} all multiple derivatives can be taken to be
symmetric in their indices and this allows us to simplify 1) and 2). In
these terms, to preserve Lorentz invariance,  at least two $D$'s must
be contracted together, or contracted with an $\epsilon_{\al\be\ga\de}$.
Using the index exchange result of Sec. \ref{ssinex} these two $D$'s can
be brought together acting on a $U$, with the additional terms
produced being either
total derivatives or like some of the other terms. By symmetry
the $\epsilon_{\al\be\ga\de}$ term must
vanish.  Likewise by symmetry, the remaining two $D$'s
 cannot be contracted with the $\chi^{\mu\nu}$, and so must also be
contracted together. Thus $\chi^{\mu\nu}$ must be simply $\chi$ and we find
for 1) and 2)
\begin{eqnarray}
1) &\rightarrow& (D_\al D^\al U) (D_\be D^\be \chi), \nonumber \\
2) &\rightarrow& (D_\al D^\al U) (D_\be D^\be U) \chi U.
\end{eqnarray}

As before to get an expression transforming as $V_R \dots V_R^\dagger$,
so that the trace will be invariant, we write each factor in
Eq.\ (\ref{eq4ds1}) in the form $\cpm{A}$. Then Sec. \ref{ssbbmcd} and
particularly Eq.\ (\ref{d2u}) allows us to eliminate $\cp{D^2U}$ in
favor of terms we already have.

The result of all these manipulations is the set of general forms
\begin{eqnarray}
\label{eq4ds2}
1) &\quad & \cm{D_\al D^\al U} \cpm{D_\be D^\be \chi}, \nonumber \\
2) &\quad & \cm{D_\al D^\al U} \cm{D_\be D^\be U} \cpm{\chi}, \nonumber \\
3) &\quad & \cm{D_\al D_\be U} \cm{D_\ga U} \cm{D_\de U}
\cpm{\chi^{\mu\nu}}, \nonumber \\
4) &\quad & \cm{D_\al D_\be U} \cm{D_\ga U} \cpm{D_\de \chi^{\mu\nu}},
\nonumber \\
7) &\quad & \cpm{\chi^{\mu\nu}} \cm{D_\al U} \cm{D_\be U} \cm{D_\ga U}
\cm{D_\de U},
\end{eqnarray}
where it is understood that we still must take all possible contractions
of indices, perhaps with $\epsilon_{\al\be\ga\de}$, and all
different independent orderings of terms.

The last step is to impose parity and charge conjugation invariance. In
accord with Sec. \ref{ssparity} requiring parity invariance simply leads
to the overall factor $1 + (-1)^{s+P+\epsilon}$ just as before. P is the
sum of intrinsic parities $p$, but in effect here just counts the number of
$H^{\mu\nu}$'s, as all other quantities have $p=0$. For the
five terms in Eq.\ (\ref{eq4ds2}) these factors are $(1 \mp 1)$, $(1 \pm
1)$, $(1 \mp (-1)^{P+\epsilon})$, $(1 \pm (-1)^{P+\epsilon})$, and
$(1 \pm (-1)^{P+\epsilon})$, respectively, where the upper (lower) signs
correspond just to the factors $\cp{A}$ ($\cm{A}$) involving $\chi$ or
$\chi^{\mu\nu}$. Similarly to get charge
conjugation invariance we must add a term $(-1)^C \times$ the original
term with all the major factors $\cpm{A}$ in reverse order.

We can now put all of this together to get the general result  for the
four $D$ one $\chi^{\mu\nu}$ terms analogous to the six $D$ result of
Eq.\ (\ref{eq6dres}). It is most convenient to separate the $\mu = \nu$,
i.e. $\chi$, terms from the $\mu \neq \nu$ terms as there are some
'accidental' simplifications which are different in the two cases, and
since the values of the intrinsic parity and charge conjugation quantum
numbers are different. For $\mu = \nu$ observe that the symmetry of the
multiple derivatives ensures that there will be no  $\epsilon_{\al\be\ga\de}$
terms arising from 3) or 4) in Eq.\ (\ref{eq4ds2}). For $\mu \neq \nu$ we have
by explicit calculation $\cm{\chi^{\mu\nu}} = 0$. Furthermore we can
show, by taking the covariant derivative of $\cm{\chi^{\mu\nu}}$ and
using the chain rule Eq.\ (\ref{chr}) that $\cm{D_\al \chi^{\mu\nu}}$can
be expressed in terms of other factors we have kept. Hence it is not
independent and can be dropped.

We thus find the most general form for single trace expressions with
four $D$'s and one $\chi^{\mu\nu}$ to be, for $\mu = \nu$
\begin{eqnarray}
\label{eq4dreschi}
1) &\quad & \cm{D_\al D^\al U} \cm{D_\be D^\be \chi}, \nonumber \\
2) &\quad & \cm{D_\al D^\al U} \cm{D_\be D^\be U} \cp{\chi} + \mbox{rev},
\nonumber \\
3) &\quad & \cm{D_\al D_\be U} \cm{D_\ga U} \cm{D_\de U} \cm{\chi} +
\mbox{rev}, \nonumber \\
4) &\quad & \cm{D_\al D_\be U} \cm{D_\ga U} \cp{D_\de \chi} +
\mbox{rev}, \nonumber \\
7a) &\quad & \cp{\chi} \cm{D_\al U} \cm{D_\be U} \cm{D_\ga U} \cm{D_\de
U} + \mbox{rev}, \nonumber \\
7b) &\quad & \cm{\chi} \cm{D_\al U} \cm{D_\be U} \cm{D_\ga U} \cm{D_\de U}
\epsilon^{\al\be\ga\de} + \mbox{rev},
\end{eqnarray}
and for $\mu \neq \nu$
\begin{eqnarray}
\label{eq4dresf}
3a) &\quad &  \cm{D_\al D_\be U} \cm{D_\ga U} \cm{D_\de U} \cp{G^{\mu\nu}}
\epsilon_{\rh\ta\et\si} - \mbox{rev}, \nonumber \\
3b) &\quad &  \cm{D_\al D_\be U} \cm{D_\ga U} \cm{D_\de U}
\cp{H^{\mu\nu}} + \mbox{rev}, \nonumber \\
4a) &\quad &  \cm{D_\al D_\be U} \cm{D_\ga U} \cp{D_\de G^{\mu\nu}} -
\mbox{rev}, \nonumber \\
4b) &\quad &  \cm{D_\al D_\be U} \cm{D_\ga U} \cp{D_\de H^{\mu\nu}}
\epsilon_{\rh\ta\et\si}  + \mbox{rev}, \nonumber \\
7a) &\quad &  \cp{G^{\mu\nu}} \cm{D_\al U} \cm{D_\be U} \cm{D_\ga U}
\cm{D_\de U} - \mbox{rev}, \nonumber \\
7b) &\quad &  \cp{H^{\mu\nu}} \cm{D_\al U} \cm{D_\be U} \cm{D_\ga U}
\cm{D_\de U} \epsilon_{\rh\ta\et\si} + \mbox{rev}.
\end{eqnarray}
In both of these results 'rev' stands for a piece with the factors in
reverse order and it is understood that we still must take all possible
contractions of free indices and all possible independent orderings of
factors and, to get the final contribution to the Lagrangian, take the
trace.

Everything done to derive these results holds also essentially unchanged
for multiple traces, by virtue of the
total derivative arguments of
Eqs.\ (\ref{rmgen}) and (\ref{tdap}). We simply must take all
possible combinations
of traces, instead of a single trace, and must reverse the order of the
terms in each individual trace separately.

It is now straightforward to tabulate the terms arising from
Eqs.\ (\ref{eq4dreschi}) and (\ref{eq4dresf}). Starting with
Eq.\ (\ref{eq4dreschi}) we get
for the terms involving $\chi$
from 1)
\beq \Tr{\cm{D_\al D^\al U} \cm{D_\be D^\be \chi}}, \eeq{4dc:1}
from 2)
\beq \Tr{\cm{D_\al D^\al U} \cm{D_\be D^\be U} \cp{\chi}}, \eeq{4dc:2}
from 3)
\beq \Tr{\cm{D_\al D^\al U}(\cm{D_\be U}\cm{D^\be U} \cm{\chi} +
\cm{\chi} \cm{D_\be U}\cm{D^\be U})}, \eeq{4dc:3a}
\beq \Tr{\cm{D_\al D^\al U} \cm{D_\be U} \cm{\chi} \cm{D^\be U}},
\eeq{4dc:3b}
\beq \Tr{\cm{D_\al D_\be U}(\cm{D^\al U}\cm{D^\be U} \cm{\chi} +
\cm{\chi} \cm{D^\be U}\cm{D^\al U})}, \eeq{4dc:3c}
\beq \Tr{\cm{D_\al D_\be U} \cm{D^\al U} \cm{\chi} \cm{D^\be U}},
\eeq{4dc:3d}
from 4)
\beq \Tr{\cm{D_\al D^\al U}(\cm{D_\be U}\cp{D^\be \chi} + \cp{D^\be
\chi}\cm{D_\be U})}, \eeq{4dc:4a}
\beq \Tr{\cm{D_\al D_\be U}(\cm{D^\al U}\cp{D^\be \chi} + \cp{D^\be
\chi}\cm{D^\al U})}, \eeq{4dc:4b}
and from 7)
\beq \Tr{\cp{\chi} \cm{D_\al U}\cm{D^\al U} \cm{D_\be U}
\cm{D^\be U}}, \eeq{4dc:7a}
\beq \Tr{\cp{\chi} \cm{D_\al U}\cm{D_\be U} \cm{D^\al U}
\cm{D^\be U}}, \eeq{4dc:7b}
\beq \Tr{\cp{\chi} \cm{D_\al U}\cm{D_\be U} \cm{D^\be U}
\cm{D^\al U}}, \eeq{4dc:7c}
\beq \Tr{\cm{\chi} \cm{D_\al U}\cm{D_\be U} \cm{D_\ga U}\cm{D_\de U}}
\epsilon^{\al\be\ga\de}. \eeq{4dc:7d}

It is equally straightforward to read off the multiple trace terms
originating from Eq.\ (\ref{eq4dreschi}). Recall that $\Tr{\cm{D_\al D_\be
U}} = \Tr{\cm{D_\al U}} = 0$. Thus we have no terms from 1) and get
from 2)
\beq \Tr{\cm{D_\al D^\al U} \cm{D_\be D^\be U}} \Tr{\cp{\chi}},
\eeq{4dmc:2}
from 3)
\beq \Tr{\cm{D_\al D^\al U}\cm{D_\be U}} \Tr{\cm{D^\be U} \cm{\chi}},
\eeq{4dmc:3a}
\beq \Tr{\cm{D_\al D^\al U}\cm{\chi}} \Tr{\cm{D_\be U}\cm{D^\be U} },
\eeq{4dmc:3b}
\beq \Tr{\cm{D_\al D^\al U}\cm{D_\be U}\cm{D^\be U}} \Tr{\cm{\chi}},
\eeq{4dmc:3c}
\beq \Tr{\cm{D_\al D_\be U}\cm{D^\al U}} \Tr{\cm{D^\be U} \cm{\chi}},
\eeq{4dmc:3d}
\beq \Tr{\cm{D_\al D_\be U}\cm{\chi}} \Tr{\cm{D^\al U}\cm{D^\be U} },
\eeq{4dmc:3e}
\beq \Tr{\cm{D_\al D_\be U}\cm{D^\al U}\cm{D^\be U}} \Tr{\cm{\chi}},
\eeq{4dmc:3f}
from 4)
\beq \Tr{\cm{D_\al D^\al U}\cm{D_\be U}} \Tr{\cp{D^\be \chi}},
\eeq{4dmc:4a}
\beq \Tr{\cm{D_\al D_\be U}\cm{D^\al U}} \Tr{\cp{D^\be \chi}},
\eeq{4dmc:4b}
and from 7)
\beq \Tr{\cp{\chi}} \Tr{\cm{D_\al U}\cm{D^\al U}\cm{D_\be U}\cm{D^\be
U}}, \eeq{4dmc:7a}
\beq \Tr{\cp{\chi}} \Tr{\cm{D_\al U}\cm{D_\be U}\cm{D^\al U}\cm{D^\be
U}}, \eeq{4dmc:7b}
\beq \Tr{\cp{\chi}\cm{D_\al U}} \Tr{\cm{D^\al U}\cm{D_\be U}\cm{D^\be
U}}, \eeq{4dmc:7c}
\beq \Tr{\cp{\chi}\cm{D_\al U}\cm{D^\al U}} \Tr{\cm{D_\be U}\cm{D^\be
U}}, \eeq{4dmc:7d}
\beq \Tr{\cp{\chi}\cm{D_\al U}\cm{D_\be U}} \Tr{\cm{D^\al U}\cm{D^\be
U}}, \eeq{4dmc:7e}
\beq \Tr{\cp{\chi}} \Tr{\cm{D_\al U}\cm{D^\al U}}\Tr{\cm{D_\be U}\cm{D^\be
U}}, \eeq{4dmc:7f}
\beq \Tr{\cp{\chi}} \Tr{\cm{D_\al U}\cm{D_\be U}}\Tr{\cm{D^\al U}\cm{D^\be
U}}. \eeq{4dmc:7g}

Like in the previous section there are here also several sets of terms
consisting of all permutations of the factors. Thus we can use the
trace relations of Sec. \ref{sstrace} to express one of each set in terms
of the others and of terms with more traces which are lower down in the
hierarchy.  In particular Eqs.\ (\ref{4dc:3a}) and (\ref{4dc:3b}),
Eqs.\ (\ref{4dc:3c}) and (\ref{4dc:3d}), Eqs.\ (\ref{4dc:7a}),
(\ref{4dc:7b}), and (\ref{4dc:7c}) and
Eqs.\ (\ref{4dmc:7a}) and (\ref{4dmc:7b}) form such sets and
we choose to eliminate Eqs.\ (\ref{4dc:3a}), (\ref{4dc:3c}),
(\ref{4dc:7b}), and (\ref{4dmc:7b}) using the trace relations.

We now evaluate the terms coming from Eq.\ (\ref{eq4dresf}) which contain
$G^{\mu\nu}$ and $H^{\mu\nu}$. We obtain for the single trace terms from 3a)
\beq \Tr{\cm{D_\al D^\al U}(\cm{D^\be U}\cm{D^\ga U}\cp{G^{\mu\nu}} -
\cp{G^{\mu\nu}}\cm{D^\ga U}\cm{D^\be U})} \epsilon_{\be\ga\mu\nu},
\eeq{4df:3aa}
\beq \Tr{\cm{D_\al D^\al U}\cm{D^\be U}\cp{G^{\mu\nu}}\cm{D^\ga U}}
\epsilon_{\be\ga\mu\nu}, \eeq{4df:3ab}
\beq \Tr{\cm{D^\al D^\be U}(\cm{D_\be U}\cm{D^\ga U}\cp{G^{\mu\nu}} -
\cp{G^{\mu\nu}}\cm{D^\ga U}\cm{D_\be U})} \epsilon_{\al\ga\mu\nu},
\eeq{4df:3ac}
\beq \Tr{\cm{D^\al D^\be U}(\cm{D_\be U}\cp{G^{\mu\nu}}\cm{D^\ga U} -
\cm{D^\ga U}\cp{G^{\mu\nu}}\cm{D_\be U})} \epsilon_{\al\ga\mu\nu},
\eeq{4df:3ad}
\beq \Tr{\cm{D^\al D^\be U}(\cm{D^\ga U}\cm{D_\be U}\cp{G^{\mu\nu}} -
\cp{G^{\mu\nu}}\cm{D_\be U}\cm{D^\ga U})} \epsilon_{\al\ga\mu\nu},
\eeq{4df:3ae}
\beq \Tr{\cm{D^\al D_\be U}(\cm{D^\ga U}\cm{D^\de U}\cp{G^{\be\nu}} -
\cp{G^{\be\nu}}\cm{D^\de U}\cm{D^\ga U})} \epsilon_{\al\ga\de\nu},
\eeq{4df:3af}
\beq \Tr{\cm{D^\al D_\be U}\cm{D^\ga U}\cp{G^{\be\nu}}\cm{D^\de U} }
\epsilon_{\al\ga\de\nu}, \eeq{4df:3ag}
from 3b)
\beq \Tr{\cm{D_\al D^\al U}(\cm{D_\mu U}\cm{D_\nu U}\cp{H^{\mu\nu}} +
\cp{H^{\mu\nu}}\cm{D_\nu U}\cm{D_\mu U})}, \eeq{4df:3ba}
\beq \Tr{\cm{D_\mu D_\be U}(\cm{D^\be U}\cm{D_\nu U}\cp{H^{\mu\nu}} +
\cp{H^{\mu\nu}}\cm{D_\nu U}\cm{D^\be U})}, \eeq{4df:3bb}
\beq \Tr{\cm{D_\mu D_\be U}(\cm{D_\nu U}\cm{D^\be U}\cp{H^{\mu\nu}} +
\cp{H^{\mu\nu}}\cm{D^\be U}\cm{D_\nu U})}, \eeq{4df:3bc}
\beq \Tr{\cm{D_\mu D_\be U}(\cm{D_\nu U}\cp{H^{\mu\nu}}\cm{D^\be U} +
\cm{D^\be U}\cp{H^{\mu\nu}}\cm{D_\nu U})}, \eeq{4df:3bd}
from 4a)
\beq \Tr{\cm{D_\al D^\al U}(\cm{D_\mu U}\cp{D_\nu G^{\mu\nu}} -
\cp{D_\nu G^{\mu\nu}}\cm{D_\mu U})}, \eeq{4df:4aa}
\beq \Tr{\cm{D_\mu D^\be U}(\cm{D_\be U}\cp{D_\nu G^{\mu\nu}} -
\cp{D_\nu G^{\mu\nu}}\cm{D_\be U})}, \eeq{4df:4ab}
\beq \Tr{\cm{D_\mu D^\be U}(\cm{D_\nu U}\cp{D_\be G^{\mu\nu}} -
\cp{D_\be G^{\mu\nu}}\cm{D_\nu U})}, \eeq{4df:4ac}
from 4b)
\beq \Tr{\cm{D_\al D^\al U}(\cm{D^\be U}\cp{D^\ga H^{\mu\nu}} +
\cp{D^\ga H^{\mu\nu}}\cm{D^\be U})} \epsilon_{\be\ga\mu\nu},
\eeq{4df:4ba}
\beq \Tr{\cm{D^\al D^\be U}(\cm{D_\be U}\cp{D^\ga H^{\mu\nu}} +
\cp{D^\ga H^{\mu\nu}}\cm{D_\be U})} \epsilon_{\al\ga\mu\nu},
\eeq{4df:4bb}
\beq \Tr{\cm{D^\al D^\be U}(\cm{D^\ga U}\cp{D_\be H^{\mu\nu}} +
\cp{D_\be H^{\mu\nu}}\cm{D^\ga U})} \epsilon_{\al\ga\mu\nu},
\eeq{4df:4bc}
\beq \Tr{\cm{D^\al D_\be U}(\cm{D^\ga U}\cp{D^\de H^{\be\nu}} +
\cp{D^\de H^{\be\nu}}\cm{D^\ga U})} \epsilon_{\al\ga\de\nu},
\eeq{4df:4bd}
from 7a)
\beq \Tr{\cp{G^{\mu\nu}}(\cm{D_\mu U}\cm{D_\nu U}\cm{D_\al
U}\cm{D^\al U} -
\cm{D_\al U}\cm{D^\al U}\cm{D_\nu U}\cm{D_\mu U})}, \eeq{4df:7aa}
\beq \Tr{\cp{G^{\mu\nu}}(\cm{D_\mu U}\cm{D_\al U}\cm{D_\nu
U}\cm{D^\al U} -
\cm{D_\al U}\cm{D_\nu U}\cm{D^\al U}\cm{D_\mu U})}, \eeq{4df:7ab}
\beq \Tr{\cp{G^{\mu\nu}}\cm{D_\mu U}\cm{D_\al U}\cm{D^\al U}
\cm{D_\nu U}},\eeq{4df:7ac}
\beq \Tr{\cp{G^{\mu\nu}}\cm{D_\al U}\cm{D_\mu U}\cm{D_\nu U}
\cm{D^\al U}},\eeq{4df:7ad}
and from 7b)
\beq \Tr{\cp{H^{\mu\nu}}(\cm{D^\al U}\cm{D^\be U}\cm{D_\ga
U}\cm{D^\ga U} +
\cm{D_\ga U}\cm{D^\ga U}\cm{D^\be U}\cm{D^\al U})}\epsilon_{\mu\nu\al\be},
\eeq{4df:7ba}
\beq \Tr{\cp{H^{\mu\nu}}(\cm{D^\al U}\cm{D_\ga U}\cm{D^\be
U}\cm{D^\ga U} +
\cm{D_\ga U}\cm{D^\be U}\cm{D^\ga U}\cm{D^\al U})}\epsilon_{\mu\nu\al\be},
\eeq{4df:7bb}
\beq \Tr{\cp{H^{\mu\nu}}(\cm{D_\nu U}\cm{D^\be U}\cm{D^\ga
U}\cm{D^\de U} +
\cm{D^\de U}\cm{D^\ga U}\cm{D^\be U}\cm{D_\nu U})}\epsilon_{\mu\be\ga\de},
\eeq{4df:7bc}
\beq \Tr{\cp{H^{\mu\nu}}(\cm{D^\be U}\cm{D_\nu U}\cm{D^\ga
U}\cm{D^\de U} +
\cm{D^\de U}\cm{D^\ga U}\cm{D_\nu U}\cm{D^\be U})}\epsilon_{\mu\be\ga\de}.
\eeq{4df:7bd}

In this group the set Eqs.\ (\ref{4df:3bb}), (\ref{4df:3bc}), and
(\ref{4df:3bd}) are related
by the trace relations of Sec. \ref{sstrace}, and we use that result to
eliminate Eq.\ (\ref{4df:3bd}). Also a number of terms are related by the
epsilon relations of Sec. \ref{ssepsrel}. In particular there is one
relation among terms of each of the sets
Eqs.\ (\ref{4df:3ab}), (\ref{4df:3ad}), and (\ref{4df:3ag}),
Eqs.\ (\ref{4df:3aa}), (\ref{4df:3ac}), (\ref{4df:3ae}), and
(\ref{4df:3af}) and
Eqs.\ (\ref{4df:4ba}), (\ref{4df:4bb}), (\ref{4df:4bc}), and
(\ref{4df:4bd}) which we use to eliminate say
Eqs.\ (\ref{4df:3ad}), (\ref{4df:3af}), and (\ref{4df:4bd}). Just as in
the discussion following Eq.\ (\ref{6d:6e}) we here also keep the
equation of motion terms, since they can be eliminated later via field
transformations, and use these relations to eliminate other terms.
There are two relations among the
set Eqs.\ (\ref{4df:7ba}), (\ref{4df:7bb}), (\ref{4df:7bc}), and
(\ref{4df:7bd}) which we use to
eliminate Eqs.\ (\ref{4df:7bc}) and (\ref{4df:7bd}).

Again we can write down directly from Eq.\ (\ref{eq4dresf}) the multiple
trace terms. Many possible terms vanish because for $\mu \neq \nu$ we have
$\Tr{\cpm{\chi^{\mu\nu}}} = \Tr{\cp{D_\al \chi^{\mu\nu}}} =0$, as can be
shown by explicit evaluation using the definitions Eq.\ (\ref{chidef}).
Because of these relations, and also $\Tr{\cm{D_\al D_\be U}} =
\Tr{\cm{D_\al U}} = 0$, we obtain no terms from 3a), 4a), or 4b),
and we get from 3b)
\beq \Tr{\cm{D_\al D^\al U}\cm{D_\mu U}}\Tr{\cm{D_\nu U}
\cp{H^{\mu\nu}}}, \eeq{4dmf:3ba}
\beq \Tr{\cm{D_\mu D_\be U}\cm{D^\be U}}\Tr{\cm{D_\nu U}
\cp{H^{\mu\nu}}}, \eeq{4dmf:3bb}
\beq \Tr{\cm{D_\mu D_\be U}\cm{D_\nu U}}\Tr{\cm{D^\be U}
\cp{H^{\mu\nu}}}, \eeq{4dmf:3bc}
\beq \Tr{\cm{D_\mu D_\be U}\cp{H^{\mu\nu}}}\Tr{\cm{D^\be U}
\cm{D_\nu U}}, \eeq{4dmf:3bd}
from 7a)
\beq
\Tr{\cp{G^{\mu\nu}}\cm{D_\al U}}\Tr{\cm{D^\al U}\cm{D_\mu U}
\cm{D_\nu U}}, \eeq{4dmf:7aa}
\beq \Tr{\cp{G^{\mu\nu}}\cm{D_\mu U}\cm{D_\nu U}}\Tr{\cm{D_\al U}
\cm{D^\al U}}, \eeq{4dmf:7ab}
\beq \Tr{\cp{G^{\mu\nu}}(\cm{D_\mu U}\cm{D_\al U}-
\cm{D_\al U}\cm{D_\mu U})}
\Tr{\cm{D^\al U}\cm{D_\nu U}}, \eeq{4dmf:7ac}
and from 7b)
\beq \Tr{\cp{H^{\mu\nu}}\cm{D^\al U}}\Tr{\cm{D^\be U}
\cm{D_\ga U}\cm{D^\ga U}}\epsilon_{\mu\nu\al\be}, \eeq{4dmf:7ba}
\beq \Tr{\cp{H^{\mu\nu}}(\cm{D^\al U}\cm{D^\ga U}+
\cm{D^\ga U}\cm{D^\al U})}
\Tr{\cm{D^\be U}\cm{D_\ga U}}\epsilon_{\mu\nu\al\be}. \eeq{4dmf:7bb}

\subsection{Terms with two $D_\mu$'s and two $\chi^{\mu\nu}$'s}
\label{ss2d2chi}
Consider now terms of the third level in the hierarchy, those containing
two $D_\mu$'s and two $\chi^{\mu\nu}$'s. The $\chi^{\mu\nu}$'s can be
just $\chi$ if $\mu = \nu$, or can be the full $\chi^{\mu\nu}$ involving
the $F^{\mu\nu}$'s if $\mu \neq \nu$, or there can be one of each. Because
of these different possibilities, and the fact that each $\chi^{\mu\nu}$
has its own parity and charge conjugation quantum numbers there are so
many options that it is  much more difficult in this case to develop a
general result and simply evaluate it, as was done in the previous two
sections. We will however proceed as far as possible in general, but will
be forced in the end to simply enumerate the possibilities.

Note that we will have to distinguish between the two $\chi^{\mu\nu}$'s.
To do this we will call one $\chi$ and the other $\tchi$. To simplify
the notation we will also drop the Lorentz indices on $\chi$ until the end.

We can then write the possible terms directly in terms of the
building blocks $\cpm{A}$ as
\begin{eqnarray}
\label{eq2ds1}
1) &\quad & \cpm{D \chi} \cpm{D \tchi}, \nonumber \\
2a) &\quad & \cm{D U} \cpm{D \chi} \cpm{\tchi}, \nonumber \\
2b) &\quad & \cm{D U} \cpm{D \tchi} \cpm{\chi}, \nonumber \\
3) &\quad &  \cm{D U} \cm{D U} \cpm{\chi} \cpm{\tchi}.
\end{eqnarray}
In these expressions we must take all independent orders, and of course
eventually put the Lorentz indices back in and take all possible
contractions. The $\pm$ signs on the $\chi$ and $\tchi$ terms are
uncorrelated, so that all combinations must be included.

To obtain these expressions we have already systematically used the
total derivative argument of Sec. \ref{sstd} to remove any $D^2$'s.
However this procedure may have hidden some equation of motion
terms in this set, which we
want to extract. For example, moving the $D$ acting on $\chi$ in 2a) over
to the $DU$ using the total derivative argument could generate a
$\cm{D_\al D^\al U}$ term. There are some subtleties in such a
transformation however. So far we have eliminated terms high in the
hierarchy in favor of those lower down, i.e. in this case we
eliminate those with more
$D$'s acting on a single factor in favor of those with fewer $D$'s. Since
we always keep the most general set of terms at the lower level, it is
never necessary to actually work out explicitly the relation used to
eliminate the higher level term. Here however we are asking whether a
specific lower level term can be eliminated in favor of a higher level
one. Thus to be sure one is not over or under counting the independent
terms it will be necessary here to work out the transformation
explicitly.

To do this consider the total derivative
$\partial_\mu \lb \cm{D_\nu U} \cpm{\chi} \cpm{\tchi} \rb$, where the
argument of the derivative is understood to be a single or multiple trace.
Since total
derivatives can be dropped in the Lagrangian, this can be treated as
if it were effectively zero. Thus using the results of Sec. \ref{sstd}
and particularly Eq.\ (\ref{rmgen}) we find
\begin{eqnarray}
\label{tdappl}
0 & = & \cm{D_\mu D_\nu U} \cpm{\chi} \cpm{\tchi} \nonumber \\
  &   & \mbox{ } + \cm{D_\nu U} \cpm{D_\mu \chi} \cpm{\tchi} +
\cm{D_\nu U} \cpm{\chi} \cpm{D_\mu \tchi} \nonumber \\
  &   & \mbox{ } + \mbox{terms of the class 3)},
\end{eqnarray}
where again the traces have not been put in explicitly.
The terms of the class 3) arise from the covariant
derivative acting on the $U$ and $U^\dagger$ in the various factors $\cpm{A}$
together with some algebraic rearrangement to express everything in terms
of the standard factors $\cpm{A}$.

Thus we see that we can express a particular sum of the terms
from the classes 2a) and 2b) in terms of a class containing a double
derivative term
$\cm{D_\mu D_\nu U}$. To get this particular sum we must take a new
basis for the terms of 2a) and 2b) consisting of the sums and differences
of terms. When this is done we can use this double derivative term instead of
the sum if we wish, but we must keep the remaining (difference)
terms from 2a) and 2b). Note that we could have started this discussion
with alternative ordering, $\chi \leftrightarrow \tchi$, but since we
always include all orders of the factors, this gives the same results.

This means that in the general result of Eq.\ (\ref{eq2ds1}) we may
use instead of 2a) and 2b)
\begin{eqnarray}
\label{eq2ds2}
2a)^\prime &\quad & \cm{D U} \lb \cpm{D \chi} \cpm{\tchi} - \cpm{\chi}
\cpm{D \tchi} \rb, \nonumber \\
2b)^\prime &\quad & \cm{D_\mu D_\nu U} \cpm{\chi} \cpm{\tchi},
\end{eqnarray}
where as usual, all orderings must be taken and the $\pm$ signs are not
correlated.

Observe that everything done so far applies to multiple traces just as
to single traces. One sees, by following through the arguments leading to
Eq.\ (\ref{eq2ds2}), that if we start in 2a) and 2b) with say, the trace
of the first two factors times the trace of the third, then we will
also get the trace of the first two
factors times the trace of the third in each of the terms making up
2a)$^\prime$ and 2b)$^\prime$.

Finally it is possible to apply essentially the same argument to 1) of
Eq.\ (\ref{eq2ds1}) and show that we can use either of the alternative
forms
\begin{eqnarray}
\label{eq2ds3}
1)^\prime &\quad & \cpm{\chi} \cpm{D_\mu D_\nu \tchi}, \nonumber \\
1)^{\prime\prime} &\quad & \cpm{D_\mu D_\nu \chi} \cpm{\tchi}.
\end{eqnarray}

Next we must ensure that parity and charge conjugation are satisfied.
Recall that effectively under parity $\cpm{A} \rightarrow \pm (-1)^p U^\dagger
\cpm{A} U$ with $p$ being the intrinsic parity of the operator $A$.
There is also the extra factor of $(-1)^\epsilon$, where $\epsilon$
counts the number of $\epsilon_{\al\be\ga\de}$'s in the term. Let $p$ and
$\tilde{p}$ be the intrinsic parities of $\chi$ and $\tchi$ and let $(-
1)^s$ here account for the product of $\pm$ signs coming from the
$\cpm{\chi}$ and $\cpm{\tchi}$ factors only, e.\ g.\, $s=0$ for
$\cp{\chi}\cp{\tchi}$ and $s=1$ for $\cp{\chi}\cm{\tchi}$.
Thus we get a parity invariant term by multiplying the forms 1) and
3) by $(1+(-1)^{s+p+\tilde{p}+\epsilon})$ and the forms 2a)$^\prime$ and
2b)$^\prime$ by $(1-(-1)^{s+p+\tilde{p}+\epsilon})$.

To ensure charge conjugation invariance we must add $(-1)^{c+\tilde{c}}$
times the major factors in reverse order, where $c$ and $\tilde{c}$ are
the intrinsic charge conjugation quantum numbers of $\chi$ and $\tchi$.
For terms of type 1) and the multiple trace terms from type 2a)$^\prime$
and 2b)$^\prime$ with only two factors in the trace the cyclic property of the
trace means that this amounts to multiplying by the overall factor
$(1+(-1)^{c+\tilde{c}})$.  For the others the reversed term has to be
added explicitly, though in many cases adding such a term makes different
initial starting orders for the factors give the same result.

We can now summarize the general form for the two $D_\mu$ and two
$\chi^{\mu\nu}$ terms as follows.
\begin{eqnarray}
\label{eq2dsres}
1) &\quad & (1+(-1)^{s+p+\tilde{p}+\epsilon})(1+(-1)^{c+\tilde{c}})
\cpm{D_\mu \chi^{\al\be}} \cpm{D_\nu \tchi^{\ga\de}}, \nonumber \\
2a)^\prime &\quad & (1-(-1)^{s+p+\tilde{p}+\epsilon}) \lb \cm{D_\mu U}
(\cpm{D_\nu \chi^{\al\be}}\cpm{\tchi^{\ga\de}} -
\cpm{\chi^{\al\be}}\cpm{D_\nu \tchi^{\ga\de}}) + (-1)^{c+\tilde{c}}
\  \mbox{(rev)} \rb, \nonumber \\
2b)^\prime &\quad & (1-(-1)^{s+p+\tilde{p}+\epsilon}) \lb
\cm{D_\mu D_\nu U} \cpm{\chi^{\al\be}}
\cpm{\tchi^{\ga\de}} + (-1)^{c+\tilde{c}} \  \mbox{(rev)} \rb, \nonumber \\
3) &\quad & (1+(-1)^{s+p+\tilde{p}+\epsilon}) \lb \cm{D_\mu U}
\cm{D_\nu U} \cpm{\chi^{\al\be}}
\cpm{\tchi^{\ga\de}} + (-1)^{c+\tilde{c}} \  \mbox{(rev)} \rb.
\end{eqnarray}
In these expressions one must contract the Lorentz indices in all
possible ways, including perhaps contracting with an
$\epsilon_{\al\be\ga\de}$, must take all possible independent orders
for the factors, and in the end must take a trace.
The $\pm$ signs are not correlated, though the $(-1)^s$ in the parity
factor will enforce a correlation for given values of the other quantum
numbers. The indication 'rev' means to take the major factors in reverse
order. For multiple traces, one must in addition take all
possible combinations
of different traces. Just as in the previous section the evaluation of
the result,
particularly the multiple trace part, is simplified by the relations
$\Tr{\cm{D_\mu D_\nu U}}=\Tr{\cm{D_\mu U}}=0$. Also for $\mu \neq \nu$
$\Tr{\cpm{\chi^{\mu\nu}}}=\cm{\chi^{\mu\nu}}=\Tr{\cp{D_\al
\chi^{\mu\nu}}}=0$. Furthermore $\cm{D_\al \chi^{\mu\nu}}$ can be expressed in
terms of other quantities lower in the hierarchy which have been kept,
and so can be dropped.

We now proceed to evaluate this general expression. There are three main
cases, namely, I) $\chi$ and $\tchi$ both simple $\chi$'s, II) one a $\chi$
and the other a $\chi^{\mu\nu}$ with $\mu \neq \nu$, and III) both
$\chi^{\mu\nu}$'s. Furthermore each $\chi^{\mu\nu}$ can be a $G^{\mu\nu}$
or an $H^{\mu\nu}$. We will evaluate the single and multiple traces
together, as that seems simplest here, and will always drop irrelevant
overall numerical factors.

Consider case I) where both $\chi$ and $\tchi$ are simple $\chi$'s. The
quantum numbers $p=\tilde{p}=c=\tilde{c}=\epsilon=0$ so that $s=0$
for terms from 1) and 3) and $s=1$ for those from 2a)$^\prime$ and
2b)$^\prime$. We obtain from 1)
\beq \Tr{\cp{D_\mu \chi} \cp{D^\mu \chi} },    \eeq{2dcc:1a}
\beq \Tr{\cm{D_\mu \chi} \cm{D^\mu \chi} },    \eeq{2dcc:1b}
\beq \Tr{\cp{D_\mu \chi}} \Tr{\cp{D^\mu \chi} },    \eeq{2dmcc:1a}
\beq \Tr{\cm{D_\mu \chi}} \Tr{\cm{D^\mu \chi} },    \eeq{2dmcc:1b}
from 2a)$^\prime$
\beq \Tr{\cm{D_\mu U}(\cp{D^\mu \chi} \cm{\chi} - \cp{\chi} \cm{D^\mu \chi}
+ \cm{\chi} \cp{D^\mu \chi} - \cm{D^\mu \chi} \cp{\chi})}, \eeq{2dcc:2a}
\beq \Tr{\cm{D_\mu U}\cp{D^\mu \chi}} \Tr{\cm{\chi}} - \Tr{\cm{D_\mu
U}\cp{\chi}} \Tr{\cm{D^\mu \chi}}, \eeq{2dmcc:2aa}
\beq \Tr{\cm{D_\mu U}\cm{D^\mu \chi}} \Tr{\cp{\chi}} - \Tr{\cm{D_\mu
U}\cm{\chi}} \Tr{\cp{D^\mu \chi}}, \eeq{2dmcc:2ab}
from 2b)$^\prime$
\beq \Tr{\cm{D_\mu D^\mu U}(\cp{\chi}\cm{\chi} + \cm{\chi}\cp{\chi})},
\eeq{2dcc:2b}
\beq \Tr{\cm{D_\mu D^\mu U}\cp{\chi}} \Tr{\cm{\chi}}, \eeq{2dmcc:2ba}
\beq \Tr{\cm{D_\mu D^\mu U}\cm{\chi}} \Tr{\cp{\chi}}, \eeq{2dmcc:2bb}
and from 3)
\beq \Tr{\cm{D_\mu U}\cm{D^\mu U}\cp{\chi}\cp{\chi}},  \eeq{2dcc:3a}
\beq \Tr{\cm{D_\mu U}\cp{\chi}\cm{D^\mu U}\cp{\chi}},  \eeq{2dcc:3b}
\beq \Tr{\cm{D_\mu U}\cm{D^\mu U}\cm{\chi}\cm{\chi}},  \eeq{2dcc:3c}
\beq \Tr{\cm{D_\mu U}\cm{\chi}\cm{D^\mu U}\cm{\chi}},  \eeq{2dcc:3d}
\beq \Tr{\cm{D_\mu U}\cm{D^\mu U}\cp{\chi}}\Tr{\cp{\chi}},
\eeq{2dmcc:3a}
\beq \Tr{\cm{D_\mu U}\cm{D^\mu U}\cm{\chi}}\Tr{\cm{\chi}},
\eeq{2dmcc:3b}
\beq \Tr{\cm{D_\mu U}\cm{D^\mu U}}\Tr{\cp{\chi}\cp{\chi}},
\eeq{2dmcc:3c}
\beq \Tr{\cm{D_\mu U}\cm{D^\mu U}}\Tr{\cm{\chi}\cm{\chi}},
\eeq{2dmcc:3d}
\beq \Tr{\cm{D_\mu U}\cp{\chi}}\Tr{\cm{D^\mu U}\cp{\chi}},
\eeq{2dmcc:3e}
\beq \Tr{\cm{D_\mu U}\cm{\chi}}\Tr{\cm{D^\mu U}\cm{\chi}},
\eeq{2dmcc:3f}
\beq \Tr{\cm{D_\mu U}\cm{D^\mu U}}\Tr{\cp{\chi}}\Tr{\cp{\chi}},
\eeq{2dmcc:3g}
\beq \Tr{\cm{D_\mu U}\cm{D^\mu U}}\Tr{\cm{\chi}}\Tr{\cm{\chi}}.
\eeq{2dmcc:3h}
Of these there are two sets of terms which consist of all permutations
of factors, Eqs.\ (\ref{2dcc:3a}) and (\ref{2dcc:3b}) and Eqs.\
(\ref{2dcc:3c}) and (\ref{2dcc:3d}). Within each set the terms are thus
related by the trace relations of Sec. \ref{sstrace} which we use
to eliminate Eqs.\ (\ref{2dcc:3b}) and (\ref{2dcc:3d}).

Now consider case II) where say $\chi$ is the simple $\chi$ and $\tchi$
is one of $G^{\mu\nu}$ or $H^{\mu\nu}$. Now $p=c=0$, with $\tilde{p}=0$ and
$\tilde{c}=1$ for $G^{\ga\de}$ or $\tilde{p}=1$ and
$\tilde{c}=0$ for $H^{\ga\de}$. There are now four indices so it is
possible to have an $\epsilon_{\mu\nu\ga\de}$ factor. Observe that
2b)$^\prime$, and also 1), if we use the equivalent form 1)$^\prime$ of
Eq.\ (\ref{eq2ds3}), are symmetric in the interchange $\mu \leftrightarrow
\nu$ and hence they will vanish when contracted either with
$\tchi^{\mu\nu}$ or $\epsilon_{\mu\nu\ga\de}$. Thus non zero terms come
only from 2a)$^\prime$ and 3).

We then obtain from 2a)$^\prime$ without an $\epsilon_{\mu\nu\ga\de}$
factor
\beq \Tr{\cm{D_\mu U}(\cm{D_\nu \chi}\cp{G^{\mu\nu}}-
\cm{\chi}\cp{D_\nu G^{\mu\nu}} - \cp{G^{\mu\nu}}\cm{D_\nu \chi}
+ \cp{D_\nu G^{\mu\nu}}\cm{\chi})}, \eeq{2dcg:2aa}
\beq \Tr{\cm{D_\mu U}(\cp{D_\nu \chi}\cp{H^{\mu\nu}}-
\cp{\chi}\cp{D_\nu H^{\mu\nu}} + \cp{H^{\mu\nu}}\cp{D_\nu \chi}
- \cp{D_\nu H^{\mu\nu}}\cp{\chi})}, \eeq{2dch:2aa}
\beq \Tr{\cm{D_\mu U}\cp{H^{\mu\nu}}}\Tr{\cp{D_\nu \chi}} -
\Tr{\cm{D_\mu U}\cp{D_\nu H^{\mu\nu}}}\Tr{\cp{\chi}}, \eeq{2dmch:2aa}
and with an $\epsilon_{\mu\nu\ga\de}$ factor
\beq \Tr{\cm{D^\mu U}(\cp{D^\nu \chi}\cp{G^{\ga\de}}-
\cp{\chi}\cp{D^\nu G^{\ga\de}} - \cp{G^{\ga\de}}\cp{D^\nu \chi}
+ \cp{D^\nu G^{\ga\de}}\cp{\chi})}\epsilon_{\mu\nu\ga\de},
\eeq{2dcg:2ab}
\beq \Tr{\cm{D^\mu U}(\cm{D^\nu \chi}\cp{H^{\ga\de}}-
\cm{\chi}\cp{D^\nu H^{\ga\de}} + \cp{H^{\ga\de}}\cm{D^\nu \chi}
- \cp{D^\nu H^{\ga\de}}\cm{\chi})}\epsilon_{\mu\nu\ga\de},
\eeq{2dch:2ab}
\beq \lb \Tr{\cm{D^\mu U}\cp{H^{\ga\de}}}\Tr{\cm{D^\nu \chi}} -
\Tr{\cm{D^\mu U}\cp{D^\nu H^{\ga\de}}}\Tr{\cm{\chi}} \rb
\epsilon_{\mu\nu\ga\de}. \eeq{2dmch:2ab}

{}From 3) without an $\epsilon_{\mu\nu\ga\de}$ factor we get
\beq \Tr{\cm{D_\mu U}\cm{D_\nu U}(\cp{\chi}\cp{G^{\mu\nu}} +
\cp{G^{\mu\nu}}\cp{\chi})}, \eeq{2dcg:3a}
\beq \Tr{\cm{D_\mu U}\cp{\chi}\cm{D_\nu U}\cp{G^{\mu\nu}}},
\eeq{2dcg:3b}
\beq \Tr{\cm{D_\mu U}\cm{D_\nu U}\cp{G^{\mu\nu}}}\Tr{\cp{\chi}},
\eeq{2dmcg:3a}
\beq \Tr{\cm{D_\mu U}\cm{D_\nu U}(\cm{\chi}\cp{H^{\mu\nu}} -
\cp{H^{\mu\nu}}\cm{\chi})}, \eeq{2dch:3a}
\beq \Tr{\cm{D_\mu U}\cp{H^{\mu\nu}}}\Tr{\cm{D_\nu U}\cm{\chi}},
\eeq{2dmch:3a}
and with an $\epsilon_{\mu\nu\ga\de}$ factor
\beq \Tr{\cm{D^\mu U}\cm{D^\nu U}(\cm{\chi}\cp{G^{\ga\de}} +
\cp{G^{\ga\de}}\cm{\chi})} \epsilon_{\mu\nu\ga\de}, \eeq{2dcg:3c}
\beq \Tr{\cm{D^\mu U}\cm{\chi}\cm{D^\nu U}\cp{G^{\ga\de}}}
\epsilon_{\mu\nu\ga\de}, \eeq{2dcg:3d}
\beq \Tr{\cm{D^\mu U}\cm{D^\nu U}\cp{G^{\ga\de}}}\Tr{\cm{\chi}}
\epsilon_{\mu\nu\ga\de}, \eeq{2dmcg:3b}
\beq \Tr{\cm{D^\mu U}\cm{D^\nu U}(\cp{\chi}\cp{H^{\ga\de}} -
\cp{H^{\ga\de}}\cp{\chi})}\epsilon_{\mu\nu\ga\de}, \eeq{2dch:3b}
\beq \Tr{\cm{D^\mu U}\cp{H^{\ga\de}}}\Tr{\cm{D^\nu U}\cp{\chi}}
\epsilon_{\mu\nu\ga\de}. \eeq{2dmch:3b}

Consider now the third case in which both $\chi$ and $\tchi$ are
$\chi^{\mu\nu}$ with $\mu \neq \nu$. Observe first that $s=0$ since all
factors involving $\chi^{\mu\nu}$'s are of the form $\cp{\dots \chi^{\mu\nu}
\dots}$. This means that when $p+\tilde{p}+\epsilon$ is even only 1) and
3) contribute while when it is odd only 2a)$^\prime$ and 2b)$^\prime$
contribute. Furthermore for the $HH$ or $GG$ terms $p+\tilde{p}+\epsilon
\rightarrow \epsilon$ and $c+\tilde{c}$ is even whereas for the $GH$
terms $p+\tilde{p}+\epsilon \rightarrow 1 + \epsilon$ and $c+\tilde{c}$ is odd.
All of the individual factors are traceless, so the only multiple traces
possible must involve two factors and thus can come only from 3). Finally
by considering various permutations we see that we get fewer terms at
this level if we use the 1)$^\prime$ form instead
of the 1) form.

We thus obtain from 1)$^\prime$
\beq \Tr{\cp{G^{\al\be}}\cp{D_\mu D^\mu G_{\al\be}}},  \eeq{2dgg:1a}
\beq \Tr{\cp{G^{\al\be}}\cp{D_\al D^\ga G_{\be\ga}}},  \eeq{2dgg:1b}
\beq \Tr{\cp{H^{\al\be}}\cp{D_\mu D^\mu H_{\al\be}}},  \eeq{2dhh:1a}
\beq \Tr{\cp{H^{\al\be}}\cp{D_\al D^\ga H_{\be\ga}}}.  \eeq{2dhh:1b}

{}From 3) we get the following terms without an $\epsilon_{\al\be\ga\de}$
\beq \Tr{\cm{D_\mu U}\cm{D^\mu U}\cp{G^{\al\be}}\cp{G_{\al\be}}},
\eeq{2dgg:3a}
\beq \Tr{\cm{D_\mu U}\cp{G^{\al\be}}\cm{D^\mu U}\cp{G_{\al\be}}},
\eeq{2dgg:3b}
\beq \Tr{\cm{D_\al U}\cm{D^\be U}\cp{G^{\al\ga}}\cp{G_{\be\ga}}},
\eeq{2dgg:3c}
\beq \Tr{\cm{D_\al U}(\cp{G^{\al\ga}}\cm{D^\be U}\cp{G_{\be\ga}} +
\cp{G_{\be\ga}}\cm{D^\be U}\cp{G^{\al\ga}})}, \eeq{2dgg:3d}
\beq \Tr{\cm{D_\al U}\cm{D^\be U}\cp{G_{\be\ga}}\cp{G^{\al\ga}}},
\eeq{2dgg:3e}
\beq \Tr{\cm{D_\mu U}\cm{D^\mu U}}\Tr{\cp{G^{\al\be}}\cp{G_{\al\be}}},
\eeq{2dmgg:3a}
\beq \Tr{\cm{D_\mu U}\cp{G^{\al\be}}}\Tr{\cm{D^\mu U}\cp{G_{\al\be}}},
\eeq{2dmgg:3b}
\beq \Tr{\cm{D_\al U}\cm{D^\be U}}\Tr{\cp{G^{\al\ga}}\cp{G_{\be\ga}}},
\eeq{2dmgg:3c}
\beq \Tr{\cm{D_\al U}\cp{G^{\al\ga}}}\Tr{\cm{D^\be U}\cp{G_{\be\ga}}},
\eeq{2dmgg:3d}
\beq \Tr{\cm{D^\al U}\cp{G^{\be\ga}}}\Tr{\cm{D_\be U}\cp{G_{\al\ga}}},
\eeq{2dmgg:3e}
\beq \Tr{\cm{D_\mu U}\cm{D^\mu U}\cp{H^{\al\be}}\cp{H_{\al\be}}},
\eeq{2dhh:3a}
\beq \Tr{\cm{D_\mu U}\cp{H^{\al\be}}\cm{D^\mu U}\cp{H_{\al\be}}},
\eeq{2dhh:3b}
\beq \Tr{\cm{D_\al U}\cm{D^\be U}\cp{H^{\al\ga}}\cp{H_{\be\ga}}},
\eeq{2dhh:3c}
\beq \Tr{\cm{D_\al U}(\cp{H^{\al\ga}}\cm{D^\be U}\cp{H_{\be\ga}} +
\cp{H_{\be\ga}}\cm{D^\be U}\cp{H^{\al\ga}})}, \eeq{2dhh:3d}
\beq \Tr{\cm{D_\al U}\cm{D^\be U}\cp{H_{\be\ga}}\cp{H^{\al\ga}}},
\eeq{2dhh:3e}
\beq \Tr{\cm{D_\mu U}\cm{D^\mu U}}\Tr{\cp{H^{\al\be}}\cp{H_{\al\be}}},
\eeq{2dmhh:3a}
\beq \Tr{\cm{D_\mu U}\cp{H^{\al\be}}}\Tr{\cm{D^\mu U}\cp{H_{\al\be}}},
\eeq{2dmhh:3b}
\beq \Tr{\cm{D_\al U}\cm{D^\be U}}\Tr{\cp{H^{\al\ga}}\cp{H_{\be\ga}}},
\eeq{2dmhh:3c}
\beq \Tr{\cm{D_\al U}\cp{H^{\al\ga}}}\Tr{\cm{D^\be U}\cp{H_{\be\ga}}},
\eeq{2dmhh:3d}
\beq \Tr{\cm{D^\al U}\cp{H^{\be\ga}}}\Tr{\cm{D_\be U}\cp{H_{\al\ga}}}.
\eeq{2dmhh:3e}
Of these there are four sets of terms whose elements are related by the
trace relations of Sec. \ref{sstrace}, namely,
Eqs.\ (\ref{2dgg:3a}) and (\ref{2dgg:3b}),
Eqs.\ (\ref{2dgg:3c}), (\ref{2dgg:3d}), and (\ref{2dgg:3e}),
Eqs.\ (\ref{2dhh:3a}) and (\ref{2dhh:3b}), and
Eqs.\ (\ref{2dhh:3c}), (\ref{2dhh:3d}), and (\ref{2dhh:3e}).
We use those relations to eliminate one term of each set,
Eqs.\ (\ref{2dgg:3b}), (\ref{2dgg:3d}), (\ref{2dhh:3b}), and
(\ref{2dhh:3d}).

The $\epsilon_{\al\be\ga\de}$ terms from 3) are
\beq \Tr{\cm{D_\mu U}\cm{D^\mu U}(\cp{G^{\al\be}}\cp{H^{\ga\de}} -
\cp{H^{\ga\de}}\cp{G^{\al\be}})}\epsilon_{\al\be\ga\de},  \eeq{2dgh:3a}
\beq \Tr{\cm{D^\mu U}\cm{D^\nu U}(\cp{G^{\al\be}}\cp{{H_\al}^\ga} +
\cp{{H_\al}^\ga}\cp{G^{\al\be}})}\epsilon_{\mu\nu\be\ga},  \eeq{2dgh:3b}
\beq \Tr{\cm{D^\mu U}\cp{G^{\al\be}}\cm{D^\nu U}\cp{{H_\al}^\ga}}
\epsilon_{\mu\nu\be\ga},  \eeq{2dgh:3c}
\beq \Tr{\cm{D^\mu U}(\cm{D_\al U}\cp{G^{\al\be}}\cp{H^{\ga\de}} -
\cp{H^{\ga\de}}\cp{G^{\al\be}}\cm{D_\al U})} \epsilon_{\mu\be\ga\de},
\eeq{2dgh:3d}
\beq \Tr{\cm{D^\mu U}(\cp{G^{\al\be}}\cm{D_\al U}\cp{H^{\ga\de}} -
\cp{H^{\ga\de}}\cm{D_\al U}\cp{G^{\al\be}})} \epsilon_{\mu\be\ga\de},
\eeq{2dgh:3e}
\beq \Tr{\cm{D^\mu U}(\cp{G^{\al\be}}\cp{H^{\ga\de}}\cm{D_\al U} -
\cm{D_\al U}\cp{H^{\ga\de}}\cp{G^{\al\be}})} \epsilon_{\mu\be\ga\de},
\eeq{2dgh:3f}
\beq \Tr{\cm{D^\mu U}(\cm{D_\al U}\cp{H^{\al\be}}\cp{G^{\ga\de}} -
\cp{G^{\ga\de}}\cp{H^{\al\be}}\cm{D_\al U})} \epsilon_{\mu\be\ga\de},
\eeq{2dgh:3g}
\beq \Tr{\cm{D^\mu U}(\cp{H^{\al\be}}\cm{D_\al U}\cp{G^{\ga\de}} -
\cp{G^{\ga\de}}\cm{D_\al U}\cp{H^{\al\be}})} \epsilon_{\mu\be\ga\de},
\eeq{2dgh:3h}
\beq \Tr{\cm{D^\mu U}(\cp{H^{\al\be}}\cp{G^{\ga\de}}\cm{D_\al U} -
\cm{D_\al U}\cp{G^{\ga\de}}\cp{H^{\al\be}})} \epsilon_{\mu\be\ga\de}.
\eeq{2dgh:3i}

There are two sets of epsilon relations, as described in Sec.
\ref{ssepsrel}, among this group of terms. One consists of two relations
among the set Eqs.\ (\ref{2dgh:3c}), (\ref{2dgh:3e}), and (\ref{2dgh:3h})
and will be used to eliminate Eqs.\ (\ref{2dgh:3e}) and (\ref{2dgh:3h}).
The other involves three relations among the set Eqs.\ (\ref{2dgh:3a}),
(\ref{2dgh:3b}), (\ref{2dgh:3d}), (\ref{2dgh:3f}), (\ref{2dgh:3g}), and
(\ref{2dgh:3i}) and will be used to eliminate Eqs.\ (\ref{2dgh:3f}),
(\ref{2dgh:3g}), and (\ref{2dgh:3i}).

{}From 2a)$^\prime$ we obtain
\beq \Tr{\cm{D_\mu U}(\cp{D^\mu G^{\al\be}}\cp{H_{\al\be}} -
\cp{G_{\al\be}}\cp{D^\mu H^{\al\be}} - \cp{H_{\al\be}}\cp{D^\mu G^{\al\be}}
+ \cp{D^\mu H^{\al\be}}\cp{G_{\al\be}})}, \eeq{2dgh:2aa}
\beq \Tr{\cm{D_\al U}(\cp{D^\ga G^{\al\be}}\cp{H_{\ga\be}} -
\cp{G^{\al\be}}\cp{D^\ga H_{\ga\be}} - \cp{H_{\ga\be}}\cp{D^\ga G^{\al\be}}
+ \cp{D^\ga H_{\ga\be}}\cp{G^{\al\be}})}, \eeq{2dgh:2ab}
\beq \Tr{\cm{D^\ga U}(\cp{D_\al G^{\al\be}}\cp{H_{\ga\be}} -
\cp{G^{\al\be}}\cp{D_\al H_{\ga\be}} - \cp{H_{\ga\be}}\cp{D_\al G^{\al\be}}
+ \cp{D_\al H_{\ga\be}}\cp{G^{\al\be}})}, \eeq{2dgh:2ac}
\beq \Tr{\cm{D^\mu U}(\cp{D^\nu H^{\al\be}}\cp{{H_\al}^\ga} -
\cp{H^{\al\be}}\cp{D^\nu {H_\al}^\ga})}\epsilon_{\mu\nu\be\ga},
\eeq{2dhh:2aa}
\beq \Tr{\cm{D^\mu U}(\cp{D_\al H^{\al\be}}\cp{H^{\ga\de}} -
\cp{H^{\al\be}}\cp{D_\al H^{\ga\de}} -
\cp{D_\al H^{\ga\de}}\cp{H^{\al\be}} + \cp{H^{\ga\de}}\cp{D_\al H^{\al\be}})}
\epsilon_{\mu\be\ga\de}, \eeq{2dhh:2ab}
\beq \Tr{\cm{D_\al U}(\cp{D^\mu H^{\al\be}}\cp{H^{\ga\de}} -
\cp{H^{\al\be}}\cp{D^\mu H^{\ga\de}} -
\cp{D^\mu H^{\ga\de}}\cp{H^{\al\be}} + \cp{H^{\ga\de}}\cp{D^\mu H^{\al\be}})}
\epsilon_{\mu\be\ga\de}, \eeq{2dhh:2ac}
\beq \Tr{\cm{D^\mu U}(\cp{D^\nu G^{\al\be}}\cp{{G_\al}^\ga} -
\cp{G^{\al\be}}\cp{D^\nu {G_\al}^\ga})}\epsilon_{\mu\nu\be\ga},
\eeq{2dgg:2aa}
\beq \Tr{\cm{D^\mu U}(\cp{D_\al G^{\al\be}}\cp{G^{\ga\de}} -
\cp{G^{\al\be}}\cp{D_\al G^{\ga\de}} -
\cp{D_\al G^{\ga\de}}\cp{G^{\al\be}} + \cp{G^{\ga\de}}\cp{D_\al G^{\al\be}})}
\epsilon_{\mu\be\ga\de}, \eeq{2dgg:2ab}
\beq \Tr{\cm{D_\al U}(\cp{D^\mu G^{\al\be}}\cp{G^{\ga\de}} -
\cp{G^{\al\be}}\cp{D^\mu G^{\ga\de}} -
\cp{D^\mu G^{\ga\de}}\cp{G^{\al\be}} + \cp{G^{\ga\de}}\cp{D^\mu G^{\al\be}})}
\epsilon_{\mu\be\ga\de}. \eeq{2dgg:2ac}
There are two sets in this group,
Eqs.\ (\ref{2dhh:2aa}), (\ref{2dhh:2ab}), and (\ref{2dhh:2ac}) and
Eqs.\ (\ref{2dgg:2aa}), (\ref{2dgg:2ab}), and (\ref{2dgg:2ac}), each
related by a
single epsilon relation which we use to eliminate Eqs.\ (\ref{2dhh:2ac})
and (\ref{2dgg:2ac}).

Finally from 2b)$^\prime$ we get
\beq \Tr{\cm{D_\mu D^\mu U}(\cp{G^{\al\be}}\cp{H_{\al\be}} -
\cp{H_{\al\be}}\cp{G^{\al\be}})}, \eeq{2dgh:2ba}
\beq \Tr{\cm{D_\al D^\ga U}(\cp{G^{\al\be}}\cp{H_{\ga\be}} -
\cp{H_{\ga\be}}\cp{G^{\al\be}})}, \eeq{2dgh:2bb}
\beq \Tr{\cm{D_\mu D^\mu U}\cp{H^{\al\be}}\cp{H^{\ga\de}}}
\epsilon_{\al\be\ga\de}, \eeq{2dhh:2ba}
\beq \Tr{\cm{D^\mu D_\al U}(\cp{H^{\al\be}}\cp{H^{\ga\de}} +
\cp{H^{\ga\de}}\cp{H^{\al\be}})} \epsilon_{\mu\be\ga\de}, \eeq{2dhh:2bb}
\beq \Tr{\cm{D_\mu D^\mu U}\cp{G^{\al\be}}\cp{G^{\ga\de}}}
\epsilon_{\al\be\ga\de}, \eeq{2dgg:2ba}
\beq \Tr{\cm{D^\mu D_\al U}(\cp{G^{\al\be}}\cp{G^{\ga\de}} +
\cp{G^{\ga\de}}\cp{G^{\al\be}})} \epsilon_{\mu\be\ga\de}. \eeq{2dgg:2bb}
There are two epsilon relations for this group, which allow us to
eliminate Eq.\ (\ref{2dhh:2bb}) in favor of Eq.\ (\ref{2dhh:2ba}) and
Eq.\ (\ref{2dgg:2bb}) in favor of Eq.\ (\ref{2dgg:2ba}). Note that this
is again a case where it is important, if we are to minimize the number
of terms, to use the relations to eliminate the non equation of motion
terms, as the equation of motion terms can be eliminated in a different way.

\subsection{Terms with no $D_\mu$'s and three $\chi^{\mu\nu}$'s}
\label{ss0d3chi}
Consider now the final case with three $\chi^{\mu\nu}$'s and no covariant
derivatives. This is relatively simple compared with the previous cases
and we can write the general case directly,
\beq (1+(-1)^{s+p_1+p_2+p_3+\epsilon})\cpm{\chi_1^{\mu\nu}}
(\cpm{\chi_2^{\al\be}}\cpm{\chi_3^{\ga\de}} +
(-1)^{c_1+c_2+c_3}\cpm{\chi_3^{\ga\de}}\cpm{\chi_2^{\al\be}}).
\eeq{eq0ds1}
Here $p_i$ and $c_i$ are the intrinsic parity and charge conjugation
quantum numbers of the three $\chi$'s. The $\pm$ signs are not correlated
and as before $(-1)^s$ is the product of the signs coming from the parity
transformation on the individual $\cp{A}$'s, counting $+$ for $\cp{A}$
and $-$ for $\cm{A}$. By virtue of the second term with factors in
reverse order, which comes from charge conjugation, the interchange
$\chi_2$ and $\chi_3$ gives, up to a sign, the original expression.
Thus the three $\chi$'s can be treated as distinguishable from the beginning.

For the simplest case with three simple $\chi$'s all of the $p_i$ and
$c_i$ as well as $\epsilon$ are zero, which requires $s=0$ also. Thus we get
\beq \Tr{\cp{\chi}\cp{\chi}\cp{\chi}}, \eeq{0dccc:a}
\beq \Tr{\cm{\chi}\cm{\chi}\cp{\chi}}, \eeq{0dccc:b}
\beq \Tr{\cp{\chi}}\Tr{\cp{\chi}\cp{\chi}}, \eeq{0dmccc:a}
\beq \Tr{\cp{\chi}}\Tr{\cm{\chi}\cm{\chi}}, \eeq{0dmccc:b}
\beq \Tr{\cm{\chi}}\Tr{\cm{\chi}\cp{\chi}}, \eeq{0dmccc:c}
\beq \Tr{\cp{\chi}}\Tr{\cp{\chi}}\Tr{\cp{\chi}}, \eeq{0dmccc:d}
\beq \Tr{\cp{\chi}}\Tr{\cm{\chi}}\Tr{\cm{\chi}}. \eeq{0dmccc:e}

There are no terms with two $\chi$'s and one $\chi^{\mu\nu}$ as there is
nothing with which to contract the indices. With only one simple $\chi$ the
general form of Eq.\ (\ref{eq0ds1}) reduces to
\beq (1+(-1)^{s+p_2+p_3+\epsilon})\cpm{\chi}
(\cp{\chi_2^{\al\be}}\cp{\chi_3^{\ga\de}} +
(-1)^{c_2+c_3}\cp{\chi_3^{\ga\de}}\cp{\chi_2^{\al\be}})
\eeq{eq0ds2}
since $\cm{\chi^{\mu\nu}}=0$ for $\mu \neq \nu$. We see from this that if
$\chi_2 = \chi_3$, so that both are $H^{\mu\nu}$'s or both are
$G^{\mu\nu}$'s, then
$s=1$ if there is an epsilon term and $s=0$ if not. This is reversed if
$\chi_2 \neq \chi_3$. The number of multiple traces will be quite limited
because $\Tr{\cpm{\chi^{\mu\nu}}} =0$ when $\mu \neq \nu$. Thus we obtain
\beq \Tr{\cp{\chi}\cp{H^{\mu\nu}}\cp{H_{\mu\nu}}},  \eeq{0dchh:a}
\beq \Tr{\cp{\chi}\cp{G^{\mu\nu}}\cp{G_{\mu\nu}}},  \eeq{0dcgg:a}
\beq \Tr{\cm{\chi}(\cp{H^{\mu\nu}}\cp{G_{\mu\nu}} -
\cp{G_{\mu\nu}}\cp{H^{\mu\nu}})},  \eeq{0dcgh:a}
\beq \Tr{\cp{\chi}}\Tr{\cp{H^{\mu\nu}}\cp{H_{\mu\nu}}},  \eeq{0dmchh:a}
\beq \Tr{\cp{\chi}}\Tr{\cp{G^{\mu\nu}}\cp{G_{\mu\nu}}},  \eeq{0dmcgg:a}
\beq \Tr{\cm{\chi}\cp{H^{\al\be}}\cp{H^{\ga\de}}}
\epsilon_{\al\be\ga\de},  \eeq{0dchh:b}
\beq \Tr{\cm{\chi}\cp{G^{\al\be}}\cp{G^{\ga\de}}}
\epsilon_{\al\be\ga\de}, \eeq{0dcgg:b}
\beq \Tr{\cp{\chi}(\cp{H^{\al\be}}\cp{G^{\ga\de}} -
\cp{G^{\ga\de}}\cp{H^{\al\be}})}\epsilon_{\al\be\ga\de},  \eeq{0dcgh:b}
\beq \Tr{\cm{\chi}}\Tr{\cp{H^{\al\be}}\cp{H^{\ga\de}}}
\epsilon_{\al\be\ga\de}, \eeq{0dmchh:b}
\beq \Tr{\cm{\chi}}\Tr{\cp{G^{\al\be}}\cp{G^{\ga\de}}}
\epsilon_{\al\be\ga\de}. \eeq{0dmcgg:b}

Finally when all three $\chi$'s are $\chi^{\mu\nu}$'s the number of terms
is very limited because of the parity and charge conjugation factors and
because all factors are traceless. We find
\beq \Tr{\cp{G^{\mu\nu}}\cp{G_{\mu\al}}\cp{{G_\nu}^\al}}, \eeq{0dggg:a}
\beq \Tr{\cp{G^{\mu\nu}}\cp{H_{\mu\al}}\cp{{H_\nu}^\al}}, \eeq{0dghh:a}
\beq \Tr{\cp{G^{\mu\nu}}(\cp{G^{\al\be}}\cp{{H_\be}^\ga} -
\cp{H^{\al\be}}\cp{{G_\be}^\ga})} \epsilon_{\mu\nu\al\ga}. \eeq{0dggh:a}
The last of these, Eq.\ (\ref{0dggh:a}), is identically zero by virtue of
the epsilon relations of Sec.\ \ref{ssepsrel}.

\section{Simplification and Reorganization of Terms in the Lagrangian
- Final Results}
\label{results}
We have now derived in Secs. \ref{ss6d0chi} - \ref{ss0d3chi} the complete
set of terms
contributing to the order $p^6$ Lagrangian ${\cal L}_6$. The results are
scattered through these sections in the order they were derived. We now
want to collect those results in one place in
a form of a Lagrangian with effective coefficients analogous to the
standard Gasser-Leutwyler Lagrangian. In the course of doing this we want
to simplify the forms as much as possible and to reorganize them so as to
select out those terms which are likely to be most immediately useful.

\subsection{Equation of motion terms}
\label{sseqofmot}
In the course of the derivation we have extracted as many terms as
possible which are proportional to the factor $\cm{D_\mu D^\mu U}$
which we have called the 'equation of motion terms' and have stated, but
not proved, that these can be transformed away. Details of this
transformation procedure are given with respect to ${\cal L}_6$ in
\cite{Scherer} and with respect to lower orders in a number of earlier
papers \cite{Gasser3,Leutwyler1,Donoghue2}. Here we just outline
the general idea, as it
produces a large reduction in the final number of terms which need to be
considered for the general Lagrangian.

The lowest order Lagrangian is given by
\beq
{\cal L}_2 = \frac{F_0^2}{4} \Tr{D_\mu U (D^\mu U)^{\dagger}}
+\frac{F_0^2}{4} \Tr{\chi U^\dagger+ U \chi^\dagger}. \eeq{l2}
{}From this one can obtain the lowest order or classical equation of motion
${\cal O}_{EOM}^{(2)} = 0$ where
\beq
{\cal O}_{EOM}^{(2)} = 2\cm{D_\mu D^\mu U} - 2\cm{\chi} +\frac{2}{3}
\Tr{\cm{\chi}}. \eeq{eomop}

We observe first that we can make the replacement $\cm{D_\mu D^\mu U}
\rightarrow {\cal O}_{EOM}^{(2)}$ in each term where it appears, where as
always we have dropped the irrelevant numerical factor. Since we have
the most general form the extra terms added and subtracted to get
${\cal O}_{EOM}^{(2)}$ are just terms we already have.

Now make a transformation on the fields of the form
\beq
U \rightarrow U^\prime = \exp(iS)U \eeq{utrans}
where $S=S^\dagger$ and $\Tr{S}=0$. It is a general result
\cite{Haag,Kamefuchi,Coleman1,Georgi,Arzt,Grosse} that such a
transformation does not affect
measurable quantities such as the S-matrix.
Applied to ${\cal L}_2$ it generates a correction to lowest
order in $S$ of the form \cite{Scherer}
\beq
\delta {\cal L}_2=\frac{F^2_0}{4}
\Tr{iS{\cal O}_{EOM}^{(2)}} \eeq{delt2}
and we can choose an $S$ of order $p^2$ so that this term cancels the
equation of motion terms in ${\cal L}_4$. This transformation generates
corrections at order $p^6$ as well \cite{Scherer}, both from the second order
in $S$
correction to ${\cal L}_2$ and the first order in $S$ correction to
${\cal L}_4$. For our purposes these corrections can simply be absorbed in
the terms of ${\cal L}_6$ since we have the most general form. Finally we
make a second transformation on ${\cal L}_2$ using an $S$ of order $p^4$
and thus generate a correction term analogous to that of Eq.\
(\ref{delt2}) which is of order $p^6$ and proportional to
${\cal O}_{EOM}^{(2)}$. By choosing $S$ properly we can eliminate
those terms of
${\cal L}_6$ which contain ${\cal O}_{EOM}^{(2)}$.

The conclusion one draws from this discussion is that in order to
generate the most general ${\cal L}_6$ in its simplest form, we can just
drop all terms proportional to ${\cal O}_{EOM}^{(2)}$. This allows us to
eliminate the 23 terms given by Eqs.\
(\ref{6d:1a}), (\ref{6d:4aa}), (\ref{6d:4bb}),
(\ref{6dm:4aa}), (\ref{6dm:4ab}),
(\ref{6dm:4ba}), (\ref{6dm:4bb}), (\ref{4dc:1}), (\ref{4dc:2}), (\ref{4dc:3b}),
(\ref{4dc:4a}), (\ref{4dmc:2}), (\ref{4dmc:3a}), (\ref{4dmc:3b}),
(\ref{4dmc:3c}),
(\ref{4dmc:4a}), (\ref{4df:3ba}), (\ref{4df:4aa}), (\ref{4dmf:3ba}),
(\ref{2dcc:2b}), (\ref{2dmcc:2ba}), (\ref{2dmcc:2bb}), (\ref{2dgh:2ba})
and the additional 6
terms containing an $\epsilon_{\al\be\ga\de}$ given by Eqs.\
(\ref{6d:5a}), (\ref{4df:3aa}), (\ref{4df:3ab}), (\ref{4df:4ba}),
(\ref{2dhh:2ba}), (\ref{2dgg:2ba}).
For completeness, these terms are given explicitly in Table \ref{eom}
in Appendix \ref{apeom}. Recall that three additional equation of motion terms,
Eqs.\ (\ref{6d:4ab}), (\ref{6d:4ba}), and (\ref{4dc:3a}), had
previously been eliminated using the trace relations.

It is important to note that this argument must be applied with extreme
care if one works in the other direction. Thus if one starts with a
particular ${\cal L}_6$ generated from some model which contains terms
proportional to ${\cal O}_{EOM}^{(2)}$ and tries to put it into the general
form we have derived it will be necessary to keep track of the changes
in the coefficients generated by the successive transformations,
particularly the second order correction to ${\cal L}_2$ which would be
missed by just dropping the ${\cal O}_{EOM}^{(2)}$ terms. This is explained in
more detail in \cite{Scherer}.

\subsection{Reorganization}
\label{ssreorg}
In the course of the derivations of the preceding sections we devoted a
great deal of effort to getting all of the terms of the Lagrangian in a
systematic way, without any thought at all as to which terms would be of
most practical importance. It is clear however that there will be some
terms which will be of immediate importance for simple processes. In fact
there have been already recent calculations, e.g. Ref.\ \cite{Bellucci},
which in the absence of
the general ${\cal L}_6$, have included a few ad hoc $p^6$ terms,
motivated by the need to cancel infinities arising from the loops
involving ${\cal L}_4$.  On the other hand
some terms in ${\cal L}_6$ contribute only to processes which are so
complicated that they probably will not be of practical interest for a
long time. For example, naively the factor $\cm{D_\mu U}$ goes like
$\partial_\mu \phi$ in leading order and so it would seem that a term
like that of Eq.\ (\ref{6d:6a}) would contribute at tree
level only to a process involving six boson fields, which is probably
not of much immediate interest.

One concludes from the preceding discussion that it would be useful to
organize the terms in the Lagrangian in such a way as to separate out
those contributing to simple processes. Such an organization is more
subtle than it might seem however. Consider for example the simplest
factor appearing in many terms of the Lagrangian, $\cm{D_\mu U}$.
If we start with Eq.\ (\ref{u}) for $U$ and expand in powers of $\phi$ we
get symbolically $U \sim 1 + i \phi + {\cal O}(\phi^2)$, where we have
absorbed the $F_0$ into $\phi$ for the purposes of this section. Now
using the first line of Eq.\ (\ref{covder}) we expand $\cm{D_\mu U}$ as
\beq
\cm{D_\mu U} \sim   i \partial_\mu \phi - i (R_\mu - L_\mu)
+[L_\mu,\phi]  + {\cal O}(\phi^2). \eeq{dexp}
If $R_\mu = L_\mu =0$, corresponding to pure QCD with no external
fields,
then $\cm{D_\mu U} \sim \partial_\mu \phi$  and a
term like Eq.\ (\ref{6d:6a}) which involves six $D_\mu$'s does contribute
only to a process with six bosons in accordance with our naive
expectation. In general however when $R_\mu \neq
L_\mu$ this term could contribute to anything with $n$ bosons and $6-n$
external fields, where $0 \leq n \leq 6$. Thus for the general case
it doesn't appear possible to sort the terms in a useful way.

However if we limit the external fields to the electromagnetic field,
which is a useful physical situation, then we can take $R_\mu = L_\mu = -
e A_\mu Q$, where $A_\mu$ is the electromagnetic field, $e > 0$ is the
electric charge, and $Q$ is the diagonal quark charge matrix
$3Q=\mbox{diag}(2,-1,-1)$.  This allows us to classify the various
terms according to the number of boson and external electromagnetic
fields required in the process in order that the term
produce a non zero contribution in this limit. The explicit term of
course remains perfectly general, but the sorting into various groups
depends on this special assumption.

With the assumption of only external electromagnetic fields, i.e.\ $R_\mu =
L_\mu = -eA_\mu Q$, and with
$\chi = \chi^\dagger$ the various
building blocks contribute to order $\phi$ and lower as:
\begin{eqnarray}
\cm{D_\mu U} & \sim & \phi, A \phi, \nonumber \\
\cm{D_\mu D_\nu U} & \sim & \phi, A \phi, A^2 \phi, \nonumber \\
\cp{\chi} & \sim & \chi, \chi \phi,\nonumber \\
\cp{D_\mu \chi} & \sim & \chi, \chi \phi, A \chi, A \chi \phi,\nonumber \\
\cm{\chi} & \sim & \chi \phi, \nonumber \\
\cm{D_\mu \chi} & \sim & \chi \phi, A \chi \phi, \nonumber\\
\cm{D_\mu D_\nu \chi} & \sim & \chi \phi, A \chi \phi, A^2 \chi \phi,
\nonumber\\
\cp{G^{\mu\nu}}& \sim & A, A \phi, \nonumber \\
\cp{D_\al G^{\mu\nu}}& \sim & A, A \phi, \nonumber \\
\cp{D_\al D_\be G^{\mu\nu}}& \sim & A, A \phi, \nonumber \\
\cp{H^{\mu\nu}} & \sim & A \phi, \nonumber \\
\cp{D_\al H^{\mu\nu}} & \sim & A \phi, A^2 \phi, \nonumber \\
\cp{D_\al D_\be H^{\mu\nu}} & \sim & A \phi, A^2 \phi, A^3 \phi.
\end{eqnarray}
Note that for the electromagnetic case and in addition the usual
choice of $\chi = 2 B_0 M$, with $M$ the diagonal quark mass matrix,
$M = \mbox{diag}(m_u,m_d,m_s)$, the quantities
$\cpm{D_\mu \chi}$ and $\cpm{D_\mu D_\nu \chi}$ actually vanish
since $[Q,\chi]=0$ implies $D_\mu \chi \rightarrow \partial_\mu \chi$.
We keep them, however, to preserve a bit more generality.

There is another useful simplification we can use also.
Under parity, using the results of Sec. \ref{ssparity}, we
have in effect for each term $U \rightarrow U^\dagger$, $\chi \rightarrow
\chi^\dagger$, $R_\mu
\leftrightarrow L^\mu$ and $F^{\mu\nu}_R \leftrightarrow F_{\mu\nu}^L$,
with an extra minus sign for the $\epsilon_{\al\be\ga\de}$ terms.
This means that for the electromagnetic or pure QCD case,
with $R_\mu = L_\mu$ and $F^{\mu\nu}_R = F^{\mu\nu}_L$, and
with $\chi = \chi^\dagger$, we can use the fact that $U
\rightarrow U^\dagger$ is equivalent to $\phi \rightarrow -\phi$
to show that
parity invariance implies that terms without an $\epsilon_{\al\be\ga\de}$
will have only even powers of $\phi$ while those with an
$\epsilon_{\al\be\ga\de}$ will have odd powers \cite{Witten,Bijnens2}.

We have used both of these simplifications to group the terms in the
final result of Tables \ref{2phi} -\ref{eps5phi} according to
the smallest number
of $\phi$'s and $A_\mu$'s which a term can have. There can be more than
the minimum number of course, if one goes past leading order in the
expansion in powers of $\phi$. Also it is possible that there may be accidental
cancellations which make the leading term vanish so that a given structure may
in fact have more $\phi$'s or $A$'s than indicated in the tables. In a few
cases the leading behavior of the structure contains no $\phi$'s. These terms
have been included with the terms having two $\phi$'s, since such non leading
contributions would seem most relevant for practical calculations.

To see how this sorting works, suppose we are interested in the contact
terms at order $p^6$ contributing to the  process
$\gamma + \gamma \rightarrow \pi + \pi$. We thus need two $\phi$'s and
two $A_\mu$'s in tree level. Thus none of the terms in Tables \ref{4phi} -
\ref{eps5phi} contribute as they require an odd number or too many $\phi$'s.
Of the terms in Table \ref{2phi} the first group has no $A_\mu$'s and the
last group has too many, so neither will contribute. Thus we need
consider only the middle three groups, and even some of those terms will
vanish because as noted above $\cpm{D_\mu \chi}$ and
$\cpm{D_\mu D_\nu \chi}$ vanish.

If one wishes to consider a general external interaction which has $R_\mu \neq
L_\mu$ then there seems to be no substitute for considering each term in
detail to determine which will contribute. We emphasize again that the
results listed in Tables \ref{2phi} - \ref{eps5phi} are completely general and
appropriate also for general external interactions. Only the
classification into groups depends on the assumption of just the
electromagnetic interaction.

\subsection{Simplifications}
\label{sssimp}
The notation used so far was developed to simplify the derivations. It has the
advantage of leading to a Lagrangian which depends on a relatively small
number of building blocks which have well defined and simple transformation
properties under parity and charge conjugation and hermitian conjugation.

For the purposes of calculation however there are a few simplifications which
will be collected here for reference, though most have been mentioned earlier.
Some may simplify the evaluations in specific cases.

The building block $\cm{D_\mu U}$ can be written as $\cm{D_\mu U} =  (D_\mu
U) U^\dagger = - U (D_\mu U)^\dagger$. Under the trace the $U$'s
and $U^\dagger$'s
for the most part commute through and collapse to unity so that
the effect is to
convert a string of $\cm{D_\mu U}$'s to a string of $(D_\mu U)$'s and
$(D_\mu U)^\dagger$'s.

The factors $\cp{G^{\mu\nu}}$ and $\cp{H^{\mu\nu}}$ can be expressed in terms
of the original $F^{\mu\nu}_R$ and $F^{\mu\nu}_L$ via
\beq \cp{G^{\mu\nu}} =  G^{\mu\nu} U^\dagger =  U {G^{\mu\nu}}^\dagger =
(F^{\mu\nu}_R + U F^{\mu\nu}_L U^\dagger), \eeq{gsimp}
\beq \cp{H^{\mu\nu}} =  H^{\mu\nu} U^\dagger =  U {H^{\mu\nu}}^\dagger =
(F^{\mu\nu}_R - U F^{\mu\nu}_L U^\dagger). \eeq{hsimp}

For purely electromagnetic external gauge fields, with $A_\mu$ proportional to
the (diagonal) quark charge matrix, or for pure QCD with no external gauge
fields, and for the usual choice for $\chi$ as a diagonal quark mass
matrix the covariant derivative
$D_\mu \chi \rightarrow \partial_\mu \chi \rightarrow 0$.  Thus in this
situation the terms containing $\cpm{D_\mu \chi}$ or
$\cpm{D_\mu D_\nu \chi}$ all vanish.

\subsection{Final Results}
\label{ssfnlres}
We have collected our final results for ${\cal L}_6$, the complete
Lagrangian to order
$p^6$, in Tables \ref{2phi} - \ref{eps5phi}, ordering the various
terms according to the
scheme described in Sec. \ref{ssreorg} above. Each term has been written
in such a way that it is chirally invariant and for real coefficients
is hermitian and invariant under parity and charge conjugation. There are a
total of 32 terms of odd intrinsic parity, corresponding to the
coefficients $A_i$ and involving an $\epsilon_{\al\be\ga\de}$. There are
111 terms of even intrinsic parity, corresponding to the coefficients
$B_i$.  In the course of the derivation we obtained 23 independent
equation of motion
terms proportional to the operator ${\cal O}_{EOM}^{(2)}$ of Eq.\
(\ref{eomop}). For completeness these have been listed in Table \ref{eom}
in Appendix \ref{apeom}, though we assume that for the simplest Lagrangian
these will have been transformed away via an appropriate field
transformation \cite{Leutwyler1,Scherer}.

Also in the course of the derivation we used trace relations to express
18 structures originally obtained in terms of others. Table
\ref{tracerel} of Appendix \ref{aprt} shows which of the original terms
are related and which we chose to eliminate. Likewise epsilon relations were
used to eliminate 16 dependent structures from the original set of odd
intrinsic parity terms. Table \ref{epsilonrel} of
Appendix \ref{apepsrel} indicates
which of the original equations were eliminated. Both of these tables,
though not strictly necessary for the final results, should make it easier
to compare our work with that of others.

We have tried to start with all possible structures and to eliminate those
which are not independent and to extract from
the remaining terms as many equation of motion terms as possible.  The
procedure for doing this requires several tricks, i.e. the trace and
epsilon relations, and depends in some cases on the way in which one writes
the various terms. We cannot prove in a general and rigorous way that the
resulting terms are all independent. Thus the reader should be aware that
it is possible that there may be additional tricks which have been missed
which could be used to express some of the structures in terms of
others and thus reduce the number of independent terms.

\subsection{Comparison with other results}
\label{sscompar}

The set of independent structures of $O(p^6)$ in the odd intrinsic parity
sector has already been discussed by Issler \cite{Issler} and by
Akhoury and Alfakih \cite{Akhoury}.
However, the number of independent terms we find does not agree with either
of the above references, which mutually disagree with each other.
In the following, we will try to locate potential sources of this discrepancy.
A direct term--by--term comparison is made difficult by the fact that, in
general, different conventions and, furthermore, different basic building
blocks are used.

Let us start with Ref.\ \cite{Issler} which quotes 49 independent terms.
This number is close to our starting number of 54 terms.
However, we made use of the epsilon relations to eliminate 16 terms.
It appears that these relations were not used in any derivation
prior to the work by Akhoury and Alfakih \cite{Akhoury}
(see for example \cite{Donoghue4}).
Thus one has to conclude that, in general, too many terms were found which,
in fact, are not independent.
Furthermore, there is no reference in \cite{Issler} to the use of the
equation of motion or field transformations to eliminate terms.
For example, of the 4 terms proportional to $k^{(0)}_1-k^{(0)}_4$ in
Ref.\ \cite{Issler}, the epsilon relation can be used to eliminate say the
structure proportional to $k^{(0)}_4$, and from the remaining 3 terms
the ones proportional to $k^{(0)}_1$ and $k^{(0)}_2$ can be related using
a field transformation, resulting in only two, instead of four, independent
structures.
We made use of 6 field transformations which reduces our final number to
32 as compared with 49 in Ref.\ \cite{Issler}.

However, even after taking these two observations into consideration, there
remain some discrepancies.
It appears that in Ref.\ \cite{Issler} not all independent orderings of
operators under the trace have been taken into account.
As an example, in our opinion, there should be another term similar to
the structure proportional to $k^{(1)}_{12}$ involving a different
contraction of indices.

Finally, it appears that the set of terms includes structures which can be
related using the total derivative argument resulting in a reduction of the
number of terms.
To be specific, let us consider as an example the structure proportional
to $k^{(1)}_{17}$.
It is straightforward but tedious to show that up to a total derivative
it is related to the terms proportional to $k^{(1)}_5,k^{(1)}_9,k^{(1)}_{14}$
and $k^{(2)}_{11}$.
For that purpose one has to take the covariant derivative off of the field
strength tensor in the term proportional to $k^{(1)}_{17}$ and use the total
derivative argument as outlined in Sec.\ \ref{sstd}.

A comparison with the work of Akhoury and Alfakih \cite{Akhoury} turns out to
be more difficult as their choice of the building blocks is very different
from ours.
The final number quoted in Ref.\ \cite{Akhoury} is 30 where 5 terms
have been eliminated using the equation of motion.
This has to be compared with our 32 terms using 6 equation of motion terms.

Even though the use of epsilon relations was first proposed in
Ref.\ \cite{Akhoury} it seems that their set still contains structures
which are not independent as a consequence of such relations.
To give an example, the terms proportional to $w_{11}-w_{13}$ can be
interpreted to originate from a tensor $Q_{\alpha\beta\mu\nu\rho\sigma}$
which is antisymmetric in the index pairs $(\alpha,\beta)$, $(\mu,\nu)$,
and $(\rho,\sigma)$, respectively.
Without the epsilon relation, one would naively expect 3 independent
contractions from such a tensor, of which only one independent term remains
after use of the epsilon relation.
In a similar fashion one can show that of the four terms proportional to
$w_7-w_{10}$ only three are independent.
Finally, the very first structure proportional to $w_1$ vanishes identically
(see our Eq.\ (\ref{0dggh:a})).

On the other hand, it appears that there are terms missing in
Ref.\ \cite{Akhoury}.
To be specific, there exists an additional independent contraction of indices
for the structure of the type proportional to $w_6$.
Furthermore, Ref.\ \cite{Akhoury} does not contain any terms involving
covariant derivatives of $\chi$ of which we find three independent terms.
Finally, note that in Ref.\ \cite{Akhoury} the equation of motion is used so
as to eliminate terms which are proportional to $[\chi]_-$ instead of the
structures proportional to $[D_\mu D^\mu U]_-$.
On the other hand this means that in Ref.\ \cite{Akhoury} terms
proportional to $[D_\mu D^\mu U]_-$ should have been kept.
In fact, we find six such terms (see Table \ref{eom}) whereas
Ref.\ \cite{Akhoury} quotes only three.

Thus it appears that in both of the previous cases where a systematic study of
the odd intrinsic parity terms was made there are terms in the resulting sets
which are not independent and terms which have been missed.

\section{Summary}
\label{concl}

In the preceding sections we have developed the complete chirally invariant
Lagrangian ${\cal L}_6$ for the meson sector to order $p^6$. This is intended
to be an extension of the order $p^4$ Lagrangian ${\cal L}_4$ of Gasser and
Leutwyler which has become the standard in chiral perturbation theory and has
been used in many applications. Such an extension is important at this time
because we are beginning to see two loop calculations of processes for which
the leading contributions are order $p^4$. Such calculations generate some
$p^6$ contributions, but the full ${\cal L}_6$ is needed to produce a
consistent result.

Throughout we have emphasized a careful and pedagogical development of the
steps leading to the full Lagrangian, since we feel that it is only via such an
approach that the reader can be confident that the extremely complicated final
result is complete and correct. To do this we have first outlined a
hierarchical strategy which allows us to eliminate terms in favor of ones lower
in the hierarchy. We then discussed a number of general results which allowed
us to simplify and reduce the number of terms. After imposing parity and charge
conjugation invariance we could obtain a set of general structures at each
level, which could then be evaluated to find the set of possible terms. Trace
relations and epsilon relations were then used to eliminate terms which were
not independent and we described how field transformations could eliminate
those terms proportional to the lowest order equation of motion.

The resulting set of terms was then sorted, for the usual QCD plus
electromagnetic case, according to the minimum number of boson and
electromagnetic fields appearing. The final result for ${\cal L}_6$ is given in
Tables \ref{2phi} - \ref{eps5phi}. It consists of 111 terms in the even
intrinsic parity sector and 32 terms in the odd intrinsic parity sector.

To our knowledge there have been no prior systematic studies of the even
intrinsic parity sector to this order, though isolated terms have been used in
a variety of calculations. In the odd sector however there have been two
previous analyses \cite{Issler,Akhoury}, which disagree in the number of terms
with our result and with each other. We have shown that in each of these
previous cases, terms have been missed and terms which are not independent have
been included.

It is clear that the coefficients of all of these terms will never be evaluated
from experiment. However a much smaller subset actually contributes to most
simple processes, and it may be possible to get information about some of them.

In any case we hope that our derivation of the complete and most general ${\cal
L}_6$ Lagrangian will stimulate systematic chiral perturbation theory studies
of processes at this order.

\acknowledgments
This work was supported in part by a grant from the
Natural Sciences and Engineering Research Council of Canada.

\appendix

\section{Relations between traces}
\label{aprt}

Following the lecture notes of Coleman \cite{Coleman2}, we will
derive relations
between traces of $3\times 3$ matrices. Let $A$ be any complex $3\times 3$
matrix with eigenvalues $a_1,a_2,$ and $a_3$ (possibly complex and identical).
The solution of the characteristic equation is then equivalent to
\begin{equation}
\label{ce}
(A-a_1I)(A-a_2I)(A-a_3I)=0,
\end{equation}
where $I$ is the $3\times 3$ identity matrix.
As $A$ is similar to a matrix $B$ of the form \cite{Finkbeiner},
\begin{equation}
\label{asimb}
A=TBT^{-1},\quad B=\left( \begin{array}{ccc} a_1&\delta&0\\
0&a_2&\delta'\\0&0&a_3\end{array}\right),
\quad \delta,\delta'=0 \, \mbox{or} \, 1,
\end{equation}
one finds
\begin{eqnarray}
Tr(A)=Tr(B)&=&a_1+a_2+a_3,\nonumber\\
Tr(A^2)=Tr(B^2)&=&a_2^2+a_2^2+a_3^2,\nonumber\\
det(A)=det(B)&=&a_1a_2a_3.
\end{eqnarray}
These relations may be used to rewrite Eq.\ (\ref{ce}) as
\begin{equation}
\label{mu}
A^3-Tr(A)A^2+\frac{1}{2}\left((Tr(A))^2-Tr(A^2)\right)A-det(A)=0.
\end{equation}
A first important observation is made
by taking the trace of Eq.\ (\ref{mu}), namely that the determinant
of a matrix can be expressed in terms of traces.
This is the justification for not considering determinants as separate
building blocks in the construction of the chiral Lagrangian
\cite{Chivukula}.
Taking the trace of Eq.\ (\ref{mu}) one eliminates the determinant
to obtain (see also Eq.\ (80) of Ref.\ \cite{Urech})
\begin{eqnarray}
\label{a3}
A^3-Tr(A)A^2+\frac{1}{2}(Tr(A))^2 A-\frac{1}{2}Tr(A^2)A&&\nonumber\\
-\frac{1}{3}Tr(A^3)+\frac{1}{2}Tr(A^2)Tr(A)-\frac{1}{6}(Tr(A))^3&=&0.
\end{eqnarray}
Starting from Eq.\ (\ref{a3}) we will derive various trace relations
for traces involving between four and six $3\times 3$ matrices.

Multiplying Eq.\ (\ref{a3}) by $A$ and taking the trace results in
\begin{equation}
\label{tra4}
Tr(A^4)-\frac{4}{3}Tr(A^3)Tr(A) -\frac{1}{2}(Tr(A^2))^2
+Tr(A^2)(Tr(A))^2-\frac{1}{6}(Tr(A))^4=0.
\end{equation}
Inserting $A=\lambda_1A_1+\lambda_2A_2+\lambda_3A_3+\lambda_4A_4$
into Eq.\ (\ref{tra4}) and comparing the coefficients of
$\lambda_1\lambda_2\lambda_3\lambda_4$ one finds (see also
Eq.\ (81) of Ref.\ \cite{Urech})
\begin{eqnarray}
\label{tra4a}
\sum_{6\, perm.}Tr(A_1A_2A_3A_4)
-\sum_{8\, perm.}Tr(A_1A_2A_3)Tr(A_4)
-\sum_{3\, perm.}Tr(A_1A_2)Tr(A_3A_4)
&&\nonumber\\
+\sum_{6\, perm.}Tr(A_1A_2)Tr(A_3)Tr(A_4)
-Tr(A_1)Tr(A_2)Tr(A_3)Tr(A_4)=0.&&\nonumber\\
\end{eqnarray}
In the following, we list special cases of Eq.\ (\ref{tra4a}) which we
used to relate different terms and thus eliminate redundant structures
in the chiral Lagrangian:
\begin{eqnarray}
\label{tra4:1}
\sum_{6\, perm.}Tr(A_1A_2A_3A_4)-Tr(A_1A_2A_3+A_1A_3A_2)Tr(A_4)&&\nonumber\\
-\sum_{3\, perm.}Tr(A_1A_2)Tr(A_3A_4)&=&0,
\end{eqnarray}
for $Tr(A_1)=Tr(A_2)=Tr(A_3)=0$, and $A_4$ arbitrary;
\begin{equation}
\label{tra4:2}
\sum_{6\, perm.}Tr(A_1A_2A_3A_4)-\sum_{3\, perm.}Tr(A_1A_2)Tr(A_3A_4)=0,
\end{equation}
for $Tr(A_i)=0$;
\begin{eqnarray}
\label{tra4:3}
2Tr((A^2B+ABA+BA^2)C)-2Tr(A^2B)Tr(C)&&\nonumber\\
-Tr(A^2)Tr(BC)-2Tr(AB)Tr(AC)&=&0,
\end{eqnarray}
for $Tr(A)=Tr(B)=0$, and $C$ arbitrary;
\begin{equation}
\label{tra4:4}
2Tr((A^2B+ABA+BA^2)C)-Tr(A^2)Tr(BC)-2Tr(AB)Tr(AC)=0,
\end{equation}
for $Tr(A)=Tr(B)=Tr(C)=0$;
\begin{eqnarray}
\label{tra4:5}
4 Tr(A^2B^2)+2Tr(ABAB)-4Tr(A^2B)Tr(B)&&\nonumber\\
-Tr(A^2)Tr(B^2)-2(Tr(AB))^2+Tr(A^2)(Tr(B))^2&=&0,
\end{eqnarray}
for $Tr(A)=0$, and $B$ arbitrary, and
\begin{equation}
\label{tra4:6}
4Tr(A^2B^2)+2Tr(ABAB)-Tr(A^2)Tr(B^2)-2(Tr(AB))^2=0,
\end{equation}
for $Tr(A)=Tr(B)=0$.

The last relation, Eq.\ (\ref{tra4:6}), was already used by
Gasser and Leutwyler
in the construction of the $p^4$ Lagrangian \cite{Gasser3}.
Note, however, that the matrices $A$ and $B$ do not have to be hermitian
for Eq.\ (\ref{tra4:6}) to hold, as is sometimes stated in the literature.
Furthermore, Eq.\ (\ref{tra4:2}) is the result obtained in
Ref.\ \cite{Coleman2}.

Next we multiply Eq.\ (\ref{a3}) by $A^2$, take the trace, and rewrite
$Tr(A^4)$ using Eq.\ (\ref{tra4}) to obtain
\begin{equation}
\label{tra5}
Tr(A^5)-\frac{5}{6}Tr(A^3)Tr(A^2)-\frac{5}{6}Tr(A^3)(Tr(A))^2
+\frac{5}{6}Tr(A^2)(Tr(A))^3-\frac{1}{6}(Tr(A))^5=0.
\end{equation}
Inserting $A=\sum_{i=1}^5 \lambda_iA_i$ into Eq.\ (\ref{tra5}) one finds
\begin{eqnarray}
\label{tra5:1}
\sum_{24\, perm.}Tr(A_1A_2A_3A_4A_5)
-\sum_{20\, perm.}Tr(A_1A_2A_3)Tr(A_4A_5)&&\nonumber\\
-\sum_{20\, perm.}Tr(A_1A_2A_3)Tr(A_4)Tr(A_5)
+2\sum_{10\, perm.}Tr(A_1A_2)Tr(A_3)Tr(A_4)Tr(A_5)&&\nonumber\\
-4Tr(A_1)Tr(A_2)Tr(A_3)Tr(A_4)Tr(A_5)=0.&&\nonumber\\
\end{eqnarray}
We applied the following special case of Eq.\ (\ref{tra5:1})
\begin{eqnarray}
\label{tra5b}
\sum_{6\,perm}Tr(AABBC)-\frac{1}{2}Tr(A^2C)Tr(B^2)
-\frac{1}{2}Tr(B^2C)Tr(A^2)
&&\nonumber\\
-Tr(ABC)Tr(AB)
-Tr(ACB)Tr(AB)
&&\nonumber\\
-Tr(A^2B)Tr(BC)-Tr(AB^2)Tr(AC)&=&0,
\end{eqnarray}
for $Tr(A)=Tr(B)=0$, and $C$ arbitrary.

For our final application we can restrict ourselves to $Tr(A)=0$.
Multiplying Eq.\ (\ref{a3}) by $A^3$, and taking the trace
one obtains
\begin{equation}
\label{tra6}
Tr(A^6)-\frac{1}{2}Tr(A^4)Tr(A^2)-\frac{1}{3}(Tr(A^3))^2=0,
\end{equation}
for $Tr(A)=0$.
Inserting $A=\sum_{i=1}^6 \lambda_i A_i$ with $Tr(A_i)=0$ into
Eq.\ (\ref{tra6}) yields
\begin{eqnarray}
\label{tra6:1}
\sum_{120\, perm.}Tr(A_1A_2A_3A_4A_5A_6)
-\frac{2}{3}\sum_{90\, perm.}Tr(A_1A_2A_3A_4)Tr(A_5A_6)&&\nonumber\\
-\sum_{40\, perm.}Tr(A_1A_2A_3)Tr(A_4A_5A_6)&=&0,
\end{eqnarray}
for $Tr(A_i)=0$.
As a special application we insert
$A_1=P_\alpha$, $A_2=P^\alpha$, $A_3=P_\beta$, $A_4=P^\beta$,
$A_5=P_\gamma$, and $A_6=P^\gamma$ into Eq.\ (\ref{tra6:1}) to obtain
\begin{eqnarray}
\label{tra6:2}
2 Tr((P\cdot P)^3)+3Tr(P\cdot P P_\mu P\cdot P P^\mu)
+6Tr(P\cdot P P_\mu P_\nu P^\mu P^\nu)&&\nonumber\\
+3Tr(P_\mu P_\nu P^\mu P_\rho P^\nu P^\rho)
+Tr(P_\mu P_\nu P_\rho P^\mu P^\nu P^\rho)&&\nonumber\\
-Tr(P\cdot P) Tr((P\cdot P)^2)
-\frac{1}{2}Tr(P\cdot P)Tr(P_\mu P_\nu P^\mu P^\nu)&&\nonumber\\
-4Tr(P_\mu P_\nu)Tr(P\cdot P P^\mu P^\nu)
-2Tr(P_\mu P_\nu)Tr(P^\mu P_\rho P^\nu P^\rho)&&\nonumber\\
-3Tr(P\cdot P P_\mu)Tr(P\cdot P P^\mu)
-Tr(P_\mu P_\nu P_\rho)Tr(P^\mu P^\nu P^\rho)&&\nonumber\\
-Tr(P_\mu P_\nu P_\rho)Tr(P^\mu P^\rho P^\nu)&=&0,
\end{eqnarray}
for $Tr(P_\alpha)=0$.

In Table \ref{tracerel} we summarize how we applied the trace relations.
It contains the equation numbers of structures which are not independent
due to trace relations as well as the specific trace relation which
connects them.
Furthermore, we list which structure we have chosen to eliminate.

\section{Equation of motion terms}
\label{apeom}
For the purposes of completeness we list in Table \ref{eom} in the same
form as our final results all of the
terms in the original Lagrangian which are proportional to the factor
$\cm{D_\mu D^\mu U}$, which has been replaced by ${\cal O}_{EOM}^{(2)}$.
As detailed in Sec.\ \ref{sseqofmot} these terms can simply be dropped from
the most general form.

\section{Epsilon relations}
\label{apepsrel}
We list in Table \ref{epsilonrel} the equation numbers of
the original structures
we derived which are related by the epsilon relations of Sec.
\ref{ssepsrel} and the ones which we chose to eliminate.

\mediumtext
\begin{table}
\caption[test]{\label{trafprop}
Transformation properties under the group (G), charge conjugation (C)
and parity (P). The expressions for adjoint matrices are trivially obtained
by taking the hermitian conjugate of each entry. In the parity transformed
expression it is understood that the argument is $(-\vec{x},t)$ and that
partial derivatives $\partial_{\mu}$ act with respect to $x$ and not with
respect to the argument of the corresponding function.}
\begin{center}
\begin{tabular}{cccc}

operator&G&C&P\\
\hline
$U$&$V_R U V_L^\dagger$&$U^T$&$U^\dagger$\\

$D_{\lambda_1}\dots D_{\lambda_n}U$&
$V_R D_{\lambda_1}\dots D_{\lambda_n}U V_L^\dagger$&
$(D_{\lambda_1}\dots D_{\lambda_n}U)^T$&
$(D^{\lambda_1}\dots D^{\lambda_n}U)^\dagger$\\

$\chi$&$V_R \chi V_L^\dagger$&$\chi^T$&$\chi^\dagger$\\

$D_{\lambda_1}\dots D_{\lambda_n}\chi$&
$V_R D_{\lambda_1}\dots D_{\lambda_n}\chi V_L^\dagger$&
$(D_{\lambda_1}\dots D_{\lambda_n}\chi)^T$&
$(D^{\lambda_1}\dots D^{\lambda_n}\chi)^\dagger$\\

$R_\mu$&$V_R R_\mu V_R^\dagger+iV_R\partial_\mu
V^\dagger_R$&$-L_\mu^T$&$L^\mu$\\

$L_\mu$&$V_L L_\mu V_L^\dagger+iV_L\partial_\mu
V^\dagger_L$&$-R_\mu^T$&$R^\mu$\\

$F^R_{\mu\nu}$&$V_R F^R_{\mu\nu}V_R^\dagger$&
$-(F_{\mu\nu}^L)^T$&$F_L^{\mu\nu}$\\


$F^L_{\mu\nu}$&$V_L F^L_{\mu\nu}V_L^\dagger$&
$-(F_{\mu\nu}^R)^T$&$F_R^{\mu\nu}$\\


$G_{\mu\nu}$ & $V_R G_{\mu\nu} V_L^\dagger$&$-{G_{\mu\nu}}^T$&
${G^{\mu\nu}}^\dagger$\\

$D_{\lambda_1}\dots D_{\lambda_n}G_{\mu\nu}$ & $V_R D_{\lambda_1}\dots
D_{\lambda_n}G_{\mu\nu} V_L^\dagger$&$-(D_{\lambda_1}\dots
D_{\lambda_n}G_{\mu\nu})^T$
&$(D^{\lambda_1}\dots D^{\lambda_n}G^{\mu\nu})^\dagger$\\

$H_{\mu\nu}$ & $V_R H_{\mu\nu} V_L^\dagger$&${H_{\mu\nu}}^T$&
$-{H^{\mu\nu}}^\dagger$\\

$D_{\lambda_1}\dots D_{\lambda_n}H_{\mu\nu}$ & $V_R D_{\lambda_1}\dots
D_{\lambda_n}H_{\mu\nu} V_L^\dagger$&$(D_{\lambda_1}\dots
D_{\lambda_n}H_{\mu\nu})^T$
&$-(D^{\lambda_1}\dots D^{\lambda_n}H^{\mu\nu})^\dagger$
\end{tabular}
\end{center}
\end{table}


\renewcommand{\beq}{$}
\renewcommand{\eeq}[1]{$ & \ref{#1} \\ }
\widetext
\begin{table} \squeezetable
\caption[test]{Terms in the Lagrangian with 2 or more $\phi$'s, sorted
according to the minimum number of electromagnetic fields $A_\mu$. The
formulas are general, but the sorting order and number of $\phi$'s
depends on the specific assumption of external electromagnetic fields only,
i.e.\ on the assumption $R_\mu = L_\mu = -eA_\mu Q$, and on $\chi =
\chi^\dagger$ as detailed in Sec.\ \ref{ssreorg}. The quantities $\cpm{D_\mu
\chi}$ and  $\cpm{D_\mu D_\nu \chi}$ vanish if one in addition makes the usual
choice $\chi = 2 B_0 M$, but have been kept here for generality. Terms with
leading behavior independent of $\phi$ have been included
with the two $\phi$ terms.
The double covariant derivative  $D_\mu D_\nu A$ is assumed to be symmetric
in its indices in accord with Eq.\ (\ref{symder}).
\label{2phi}}

\begin{tabular}[t]{ld}
\mbox{Terms in the Lagrangian with 2 or more $\phi$'s and
no $A_\mu$'s possible} & \mbox{Eq.} \\
\hline

\beq + B_{1} \Tr{\cp{\chi}\cp{\chi}\cp{\chi}} \eeq{0dccc:a}
\beq + B_{2} \Tr{\cm{\chi}\cm{\chi}\cp{\chi}} \eeq{0dccc:b}
\beq + B_{3} \Tr{\cp{\chi}}\Tr{\cp{\chi}\cp{\chi}} \eeq{0dmccc:a}
\beq + B_{4} \Tr{\cp{\chi}}\Tr{\cm{\chi}\cm{\chi}} \eeq{0dmccc:b}
\beq + B_{5} \Tr{\cm{\chi}}\Tr{\cm{\chi}\cp{\chi}} \eeq{0dmccc:c}
\beq + B_{6} \Tr{\cp{\chi}}\Tr{\cp{\chi}}\Tr{\cp{\chi}} \eeq{0dmccc:d}
\beq + B_{7} \Tr{\cp{\chi}}\Tr{\cm{\chi}}\Tr{\cm{\chi}} \eeq{0dmccc:e}
\end{tabular}
\end{table}
\vspace{-3ex}
\begin{table} \squeezetable
\begin{tabular}[t]{ld}
\mbox{Terms in the Lagrangian with 2 or more $\phi$'s, with an $A_\mu$ possible
but not necessary} & \mbox{Eq.} \\
\hline
\beq + B_{8} \Tr{\cm{D_\al D_\be U}(\cm{D^\al U}\cp{D^\be \chi} + \cp{D^\be
\chi}\cm{D^\al U})} \eeq{4dc:4b}

\beq + B_{9} \Tr{\cm{D_\al D_\be U}\cm{D^\al U}} \Tr{\cp{D^\be \chi}}
\eeq{4dmc:4b}

\beq + B_{10} \Tr{\cp{D_\mu \chi} \cp{D^\mu \chi} }    \eeq{2dcc:1a}
\beq + B_{11} \Tr{\cm{D_\mu \chi} \cm{D^\mu \chi} }    \eeq{2dcc:1b}
\beq + B_{12} \Tr{\cp{D_\mu \chi}} \Tr{\cp{D^\mu \chi} }    \eeq{2dmcc:1a}
\beq + B_{13} \Tr{\cm{D_\mu \chi}} \Tr{\cm{D^\mu \chi} }    \eeq{2dmcc:1b}

\beq + B_{14} \Tr{\cm{D_\mu U}(\cp{D^\mu \chi} \cm{\chi} -
\cp{\chi} \cm{D^\mu \chi}
+ \cm{\chi} \cp{D^\mu \chi} - \cm{D^\mu \chi} \cp{\chi})} \eeq{2dcc:2a}
\beq + B_{15} \Tr{\cm{D_\mu U}\cp{D^\mu \chi}} \Tr{\cm{\chi}} - \Tr{\cm{D_\mu
U}\cp{\chi}} \Tr{\cm{D^\mu \chi}} \eeq{2dmcc:2aa}
\beq + B_{16} \Tr{\cm{D_\mu U}\cm{D^\mu \chi}} \Tr{\cp{\chi}} - \Tr{\cm{D_\mu
U}\cm{\chi}} \Tr{\cp{D^\mu \chi}} \eeq{2dmcc:2ab}

\beq + B_{17} \Tr{\cm{D_\mu U}\cm{D^\mu U}\cp{\chi}\cp{\chi}}  \eeq{2dcc:3a}
\beq + B_{18} \Tr{\cm{D_\mu U}\cm{D^\mu U}\cp{\chi}}\Tr{\cp{\chi}}
\eeq{2dmcc:3a}
\beq + B_{19} \Tr{\cm{D_\mu U}\cm{D^\mu U}}\Tr{\cp{\chi}\cp{\chi}}
\eeq{2dmcc:3c}
\beq + B_{20} \Tr{\cm{D_\mu U}\cp{\chi}}\Tr{\cm{D^\mu U}\cp{\chi}}
\eeq{2dmcc:3e}
\beq + B_{21} \Tr{\cm{D_\mu U}\cm{D^\mu U}}\Tr{\cp{\chi}}\Tr{\cp{\chi}}
\eeq{2dmcc:3g}
\end{tabular}
\end{table}
\vspace{-3ex}
\begin{table} \squeezetable
\begin{tabular}[t]{ld}
\mbox{Terms in the Lagrangian with 2 or more $\phi$'s and at least 1
$A_\mu$} & \mbox{Eq.} \\
\hline
\beq + iB_{22} \Tr{\cm{D_\mu D^\be U}(\cm{D_\be U}\cp{D_\nu G^{\mu\nu}} -
\cp{D_\nu G^{\mu\nu}}\cm{D_\be U})} \eeq{4df:4ab}
\beq + iB_{23} \Tr{\cm{D_\mu D^\be U}(\cm{D_\nu U}\cp{D_\be G^{\mu\nu}} -
\cp{D_\be G^{\mu\nu}}\cm{D_\nu U})} \eeq{4df:4ac}

\beq + iB_{24} \Tr{\cm{D_\mu U}\cm{D_\nu U}(\cp{\chi}\cp{G^{\mu\nu}} +
\cp{G^{\mu\nu}}\cp{\chi})} \eeq{2dcg:3a}
\beq + iB_{25} \Tr{\cm{D_\mu U}\cp{\chi}\cm{D_\nu U}\cp{G^{\mu\nu}}}
\eeq{2dcg:3b}
\beq + iB_{26} \Tr{\cm{D_\mu U}\cm{D_\nu U}\cp{G^{\mu\nu}}}\Tr{\cp{\chi}}
\eeq{2dmcg:3a}

\beq + iB_{27} \Tr{\cm{D_\mu U}(\cm{D_\nu \chi}\cp{G^{\mu\nu}}-
\cm{\chi}\cp{D_\nu G^{\mu\nu}} - \cp{G^{\mu\nu}}\cm{D_\nu \chi}
+ \cp{D_\nu G^{\mu\nu}}\cm{\chi})} \eeq{2dcg:2aa}
\beq + iB_{28} \Tr{\cm{D_\mu U}(\cp{D_\nu \chi}\cp{H^{\mu\nu}}-
\cp{\chi}\cp{D_\nu H^{\mu\nu}} + \cp{H^{\mu\nu}}\cp{D_\nu \chi}
- \cp{D_\nu H^{\mu\nu}}\cp{\chi})} \eeq{2dch:2aa}
\beq + iB_{29} \Tr{\cm{D_\mu U}\cp{H^{\mu\nu}}}\Tr{\cp{D_\nu \chi}} -
\Tr{\cm{D_\mu U}\cp{D_\nu H^{\mu\nu}}}\Tr{\cp{\chi}} \eeq{2dmch:2aa}
\end{tabular}
\end{table}
\vspace{-3ex}
\begin{table} \squeezetable
\begin{tabular}[t]{ld}
\mbox{Terms in the Lagrangian with 2 or more $\phi$'s and at least 2
$A_\mu$'s} & \mbox{Eq.} \\
\hline
\beq + B_{30} \Tr{\cp{G^{\al\be}}\cp{D_\mu D^\mu G_{\al\be}}}  \eeq{2dgg:1a}
\beq + B_{31} \Tr{\cp{G^{\al\be}}\cp{D_\al D^\ga G_{\be\ga}}}  \eeq{2dgg:1b}
\beq + B_{32} \Tr{\cp{H^{\al\be}}\cp{D_\mu D^\mu H_{\al\be}}}  \eeq{2dhh:1a}
\beq + B_{33} \Tr{\cp{H^{\al\be}}\cp{D_\al D^\ga H_{\be\ga}}}  \eeq{2dhh:1b}

\beq + B_{34} \Tr{\cm{D_\mu U}\cm{D^\mu U}\cp{G^{\al\be}}\cp{G_{\al\be}}}
\eeq{2dgg:3a}
\beq + B_{35} \Tr{\cm{D_\al U}\cm{D^\be U}\cp{G^{\al\ga}}\cp{G_{\be\ga}}}
\eeq{2dgg:3c}
\beq + B_{36} \Tr{\cm{D_\al U}\cm{D^\be U}\cp{G_{\be\ga}}\cp{G^{\al\ga}}}
\eeq{2dgg:3e}
\beq + B_{37} \Tr{\cm{D_\mu U}\cm{D^\mu U}}\Tr{\cp{G^{\al\be}}\cp{G_{\al\be}}}
\eeq{2dmgg:3a}
\beq + B_{38} \Tr{\cm{D_\mu U}\cp{G^{\al\be}}}\Tr{\cm{D^\mu U}\cp{G_{\al\be}}}
\eeq{2dmgg:3b}
\beq + B_{39} \Tr{\cm{D_\al U}\cm{D^\be U}}\Tr{\cp{G^{\al\ga}}\cp{G_{\be\ga}}}
\eeq{2dmgg:3c}
\beq + B_{40} \Tr{\cm{D_\al U}\cp{G^{\al\ga}}}\Tr{\cm{D^\be U}\cp{G_{\be\ga}}}
\eeq{2dmgg:3d}
\beq + B_{41} \Tr{\cm{D^\al U}\cp{G^{\be\ga}}}\Tr{\cm{D_\be U}\cp{G_{\al\ga}}}
\eeq{2dmgg:3e}

\beq + B_{42} \Tr{\cm{D_\mu U}(\cp{D^\mu G^{\al\be}}\cp{H_{\al\be}} -
\cp{G_{\al\be}}\cp{D^\mu H^{\al\be}} - \cp{H_{\al\be}}\cp{D^\mu G^{\al\be}}
+ \cp{D^\mu H^{\al\be}}\cp{G_{\al\be}})} \eeq{2dgh:2aa}
\beq + B_{43} \Tr{\cm{D_\al U}(\cp{D^\ga G^{\al\be}}\cp{H_{\ga\be}} -
\cp{G^{\al\be}}\cp{D^\ga H_{\ga\be}} - \cp{H_{\ga\be}}\cp{D^\ga G^{\al\be}}
+ \cp{D^\ga H_{\ga\be}}\cp{G^{\al\be}})} \eeq{2dgh:2ab}
\beq + B_{44} \Tr{\cm{D^\ga U}(\cp{D_\al G^{\al\be}}\cp{H_{\ga\be}} -
\cp{G^{\al\be}}\cp{D_\al H_{\ga\be}} - \cp{H_{\ga\be}}\cp{D_\al G^{\al\be}}
+ \cp{D_\al H_{\ga\be}}\cp{G^{\al\be}})} \eeq{2dgh:2ac}
\beq + B_{45} \Tr{\cm{D_\al D^\ga U}(\cp{G^{\al\be}}\cp{H_{\ga\be}} -
\cp{H_{\ga\be}}\cp{G^{\al\be}})} \eeq{2dgh:2bb}

\beq + B_{46} \Tr{\cp{\chi}\cp{H^{\mu\nu}}\cp{H_{\mu\nu}}}  \eeq{0dchh:a}
\beq + B_{47} \Tr{\cp{\chi}\cp{G^{\mu\nu}}\cp{G_{\mu\nu}}}  \eeq{0dcgg:a}
\beq + B_{48} \Tr{\cm{\chi}(\cp{H^{\mu\nu}}\cp{G_{\mu\nu}} -
\cp{G_{\mu\nu}}\cp{H^{\mu\nu}})}  \eeq{0dcgh:a}
\beq + B_{49} \Tr{\cp{\chi}}\Tr{\cp{H^{\mu\nu}}\cp{H_{\mu\nu}}}  \eeq{0dmchh:a}
\beq + B_{50} \Tr{\cp{\chi}}\Tr{\cp{G^{\mu\nu}}\cp{G_{\mu\nu}}}  \eeq{0dmcgg:a}
\end{tabular}
\end{table}
\vspace{-3ex}
\begin{table} \squeezetable
\begin{tabular}[t]{ld}
\mbox{Terms in the Lagrangian with 2 or more $\phi$'s and 3
$A_\mu$'s} & \mbox{Eq.} \\
\hline
\beq + iB_{51} \Tr{\cp{G^{\mu\nu}}\cp{G_{\mu\al}}\cp{{G_\nu}^\al}}
\eeq{0dggg:a}
\beq + iB_{52} \Tr{\cp{G^{\mu\nu}}\cp{H_{\mu\al}}\cp{{H_\nu}^\al}}
\eeq{0dghh:a}
\end{tabular}
\end{table}
\narrowtext
\renewcommand{\beq}{\begin{equation}}
\renewcommand{\eeq}[1]{\label{#1} \end{equation}}

\renewcommand{\beq}{$}
\renewcommand{\eeq}[1]{$ & \ref{#1} \\ }
\widetext
\begin{table} \squeezetable
\caption[test]{Terms in the Lagrangian with 4 or more $\phi$'s, sorted
according to the minimum number of electromagnetic fields $A_\mu$. The
formulas are general, but the sorting order and number of $\phi$'s
depends on the specific assumptions described in Sec.\ \ref{ssreorg} and the
caption of Table \ref{2phi}.
The double covariant derivative  $D_\mu D_\nu A$ is assumed to be symmetric
in its indices in accord with Eq.\ (\ref{symder}).
\label{4phi}}

\begin{tabular}[t]{ld}
\mbox{Terms in the Lagrangian with 4 or more $\phi$'s, with an $A_\mu$ possible
but not necessary} & \mbox{Eq.} \\
\hline
\beq + B_{53} \Tr{\cm{D_\mu D_\nu U} \cm{D^\mu D^\nu U} \cm{D_\al U}
\cm{D^\al U}} \eeq{6d:4ac}
\beq + B_{54} \Tr{\cm{D_\mu D_\nu U} \cm{D^\mu D^\al U} \cm{D^\nu U}
\cm{D_\al U}} \eeq{6d:4ad}
\beq + B_{55} \Tr{\cm{D_\mu D_\nu U} \cm{D^\mu D^\al U} \cm{D_\al U}
\cm{D^\nu U}} \eeq{6d:4ae}

\beq + B_{56} \Tr{\cm{D_\mu D_\nu U} \cm{D^\mu D^\nu U}} \Tr{\cm{D_\al U}
\cm{D^\al U}} \eeq{6dm:4ac}
\beq + B_{57} \Tr{\cm{D_\mu D_\nu U} \cm{D^\mu D^\al U}} \Tr{\cm{D^\nu U}
\cm{D_\al U}} \eeq{6dm:4ad}

\beq + B_{58} \Tr{\cm{D_\mu D_\nu U} \cm{D_\al U}}  \Tr{\cm{D^\mu D^\nu U}
\cm{D^\al U}} \eeq{6dm:4bc}
\beq + B_{59} \Tr{\cm{D_\mu D_\nu U} \cm{D^\nu U}}  \Tr{\cm{D^\mu D^\al U}
\cm{D_\al U}} \eeq{6dm:4bd}
\beq + B_{60} \Tr{\cm{D_\mu D_\nu U} \cm{D_\al U}}  \Tr{\cm{D^\mu D^\al U}
\cm{D^\nu U}} \eeq{6dm:4be}

\beq + B_{61} \Tr{\cm{D_\al D_\be U} \cm{D^\al U} \cm{\chi} \cm{D^\be U}}
\eeq{4dc:3d}

\beq + B_{62} \Tr{\cp{\chi} \cm{D_\al U}\cm{D^\al U} \cm{D_\be U}
\cm{D^\be U}} \eeq{4dc:7a}
\beq + B_{63} \Tr{\cp{\chi} \cm{D_\al U}\cm{D_\be U} \cm{D^\be U}
\cm{D^\al U}} \eeq{4dc:7c}

\beq + B_{64} \Tr{\cm{D_\al D_\be U}\cm{D^\al U}} \Tr{\cm{D^\be U} \cm{\chi}}
\eeq{4dmc:3d}
\beq + B_{65} \Tr{\cm{D_\al D_\be U}\cm{\chi}} \Tr{\cm{D^\al U}\cm{D^\be U} }
\eeq{4dmc:3e}
\beq + B_{66} \Tr{\cm{D_\al D_\be U}\cm{D^\al U}\cm{D^\be U}} \Tr{\cm{\chi}}
\eeq{4dmc:3f}

\beq + B_{67} \Tr{\cp{\chi}} \Tr{\cm{D_\al U}\cm{D^\al U}
\cm{D_\be U}\cm{D^\be U}} \eeq{4dmc:7a}
\beq + B_{68} \Tr{\cp{\chi}\cm{D_\al U}} \Tr{\cm{D^\al U}\cm{D_\be U}\cm{D^\be
U}} \eeq{4dmc:7c}
\beq + B_{69} \Tr{\cp{\chi}\cm{D_\al U}\cm{D^\al U}} \Tr{\cm{D_\be U}\cm{D^\be
U}} \eeq{4dmc:7d}
\beq + B_{70} \Tr{\cp{\chi}\cm{D_\al U}\cm{D_\be U}} \Tr{\cm{D^\al U}\cm{D^\be
U}} \eeq{4dmc:7e}
\beq + B_{71} \Tr{\cp{\chi}} \Tr{\cm{D_\al U}\cm{D^\al U}}
\Tr{\cm{D_\be U}\cm{D^\be U}} \eeq{4dmc:7f}
\beq + B_{72} \Tr{\cp{\chi}} \Tr{\cm{D_\al U}\cm{D_\be U}}
\Tr{\cm{D^\al U}\cm{D^\be U}} \eeq{4dmc:7g}

\beq + B_{73} \Tr{\cm{D_\mu U}\cm{D^\mu U}\cm{\chi}\cm{\chi}}  \eeq{2dcc:3c}
\beq + B_{74} \Tr{\cm{D_\mu U}\cm{D^\mu U}\cm{\chi}}\Tr{\cm{\chi}}
\eeq{2dmcc:3b}
\beq + B_{75} \Tr{\cm{D_\mu U}\cm{D^\mu U}}\Tr{\cm{\chi}\cm{\chi}}
\eeq{2dmcc:3d}
\beq + B_{76} \Tr{\cm{D_\mu U}\cm{\chi}}\Tr{\cm{D^\mu U}\cm{\chi}}
\eeq{2dmcc:3f}
\beq + B_{77} \Tr{\cm{D_\mu U}\cm{D^\mu
U}}\Tr{\cm{\chi}}\Tr{\cm{\chi}}
\eeq{2dmcc:3h}
\end{tabular}
\end{table}
\vspace{-3ex}
\begin{table} \squeezetable
\begin{tabular}[t]{ld}
\mbox{Terms in the Lagrangian with 4 or more $\phi$'s and at least 1
$A_\mu$} & \mbox{Eq.} \\
\hline
\beq + iB_{78} \Tr{\cm{D_\mu D_\be U}(\cm{D^\be U}\cm{D_\nu U}\cp{H^{\mu\nu}}
+ \cp{H^{\mu\nu}}\cm{D_\nu U}\cm{D^\be U})} \eeq{4df:3bb}
\beq + iB_{79} \Tr{\cm{D_\mu D_\be U}(\cm{D_\nu U}\cm{D^\be U}\cp{H^{\mu\nu}}
+ \cp{H^{\mu\nu}}\cm{D^\be U}\cm{D_\nu U})} \eeq{4df:3bc}

\beq + iB_{80} \Tr{\cp{G^{\mu\nu}}(\cm{D_\mu U}\cm{D_\nu U}\cm{D_\al
U}\cm{D^\al U} -
\cm{D_\al U}\cm{D^\al U}\cm{D_\nu U}\cm{D_\mu U})} \eeq{4df:7aa}
\beq + iB_{81} \Tr{\cp{G^{\mu\nu}}(\cm{D_\mu U}\cm{D_\al U}\cm{D_\nu
U}\cm{D^\al U} -
\cm{D_\al U}\cm{D_\nu U}\cm{D^\al U}\cm{D_\mu U})} \eeq{4df:7ab}
\beq + iB_{82} \Tr{\cp{G^{\mu\nu}}\cm{D_\mu U}\cm{D_\al U}\cm{D^\al U}
\cm{D_\nu U}}\eeq{4df:7ac}
\beq + iB_{83} \Tr{\cp{G^{\mu\nu}}\cm{D_\al U}\cm{D_\mu U}\cm{D_\nu U}
\cm{D^\al U}}\eeq{4df:7ad}

\beq + iB_{84} \Tr{\cm{D_\mu D_\be U}\cm{D^\be U}}\Tr{\cm{D_\nu U}
\cp{H^{\mu\nu}}} \eeq{4dmf:3bb}
\beq + iB_{85} \Tr{\cm{D_\mu D_\be U}\cm{D_\nu U}}\Tr{\cm{D^\be U}
\cp{H^{\mu\nu}}} \eeq{4dmf:3bc}
\beq + iB_{86} \Tr{\cm{D_\mu D_\be U}\cp{H^{\mu\nu}}}\Tr{\cm{D^\be U}
\cm{D_\nu U}} \eeq{4dmf:3bd}

\beq + iB_{87} \Tr{\cp{G^{\mu\nu}}\cm{D_\al U}}\Tr{\cm{D^\al U}\cm{D_\mu U}
\cm{D_\nu U}} \eeq{4dmf:7aa}
\beq + iB_{88} \Tr{\cp{G^{\mu\nu}}\cm{D_\mu U}\cm{D_\nu U}}\Tr{\cm{D_\al U}
\cm{D^\al U}} \eeq{4dmf:7ab}
\beq + iB_{89} \Tr{\cp{G^{\mu\nu}}(\cm{D_\mu U}\cm{D_\al U}-
\cm{D_\al U}\cm{D_\mu U})}
\Tr{\cm{D^\al U}\cm{D_\nu U}} \eeq{4dmf:7ac}

\beq + iB_{90} \Tr{\cm{D_\mu U}\cm{D_\nu U}(\cm{\chi}\cp{H^{\mu\nu}} -
\cp{H^{\mu\nu}}\cm{\chi})} \eeq{2dch:3a}
\beq + iB_{91} \Tr{\cm{D_\mu U}\cp{H^{\mu\nu}}}\Tr{\cm{D_\nu U}\cm{\chi}}
\eeq{2dmch:3a}
\end{tabular}
\end{table}
\vspace{-3ex}
\begin{table} \squeezetable
\begin{tabular}[t]{ld}
\mbox{Terms in the Lagrangian with 4 or more $\phi$'s and at least 2
$A_\mu$'s} & \mbox{Eq.} \\
\hline
\beq + B_{92} \Tr{\cm{D_\mu U}\cm{D^\mu U}\cp{H^{\al\be}}\cp{H_{\al\be}}}
\eeq{2dhh:3a}
\beq + B_{93} \Tr{\cm{D_\al U}\cm{D^\be U}\cp{H^{\al\ga}}\cp{H_{\be\ga}}}
\eeq{2dhh:3c}
\beq + B_{94} \Tr{\cm{D_\al U}\cm{D^\be U}\cp{H_{\be\ga}}\cp{H^{\al\ga}}}
\eeq{2dhh:3e}
\beq + B_{95} \Tr{\cm{D_\mu U}\cm{D^\mu U}}\Tr{\cp{H^{\al\be}}\cp{H_{\al\be}}}
\eeq{2dmhh:3a}
\beq + B_{96} \Tr{\cm{D_\mu U}\cp{H^{\al\be}}}\Tr{\cm{D^\mu U}\cp{H_{\al\be}}}
\eeq{2dmhh:3b}
\beq + B_{97} \Tr{\cm{D_\al U}\cm{D^\be U}}\Tr{\cp{H^{\al\ga}}\cp{H_{\be\ga}}}
\eeq{2dmhh:3c}
\beq + B_{98} \Tr{\cm{D_\al U}\cp{H^{\al\ga}}}\Tr{\cm{D^\be U}\cp{H_{\be\ga}}}
\eeq{2dmhh:3d}
\beq + B_{99} \Tr{\cm{D^\al U}\cp{H^{\be\ga}}}\Tr{\cm{D_\be
U}\cp{H_{\al\ga}}}
\eeq{2dmhh:3e}
\end{tabular}
\end{table}
\narrowtext
\renewcommand{\beq}{\begin{equation}}
\renewcommand{\eeq}[1]{\label{#1} \end{equation}}

\clearpage

\renewcommand{\beq}{$}
\renewcommand{\eeq}[1]{$ & \ref{#1} \\ }
\widetext
\begin{table} \squeezetable
\caption[test]{Terms in the Lagrangian with 6 or more $\phi$'s. The
formulas are general, but the sorting order and number of $\phi$'s
depends on the specific assumptions described in Sec.\ \ref{ssreorg} and the
caption of Table \ref{2phi}.
\label{6phi}}

\begin{tabular}[t]{ld}
\mbox{Terms in the Lagrangian with 6 or more $\phi$'s, with an $A_\mu$ possible
but not necessary} & \mbox{Eq.} \\
\hline
\beq + B_{100} \Tr{\cm{D_\al U} \cm{D^\al U} \cm{D_\be U}
\cm{D^\be U} \cm{D_\ga U} \cm{D^\ga U}} \eeq{6d:6a}
\beq + B_{101} \Tr{\cm{D_\al U} \cm{D^\al U} \cm{D_\be U}
\cm{D_\ga U} \cm{D^\be U} \cm{D^\ga U}} \eeq{6d:6b}
\beq + B_{102} \Tr{\cm{D_\al U} \cm{D^\al U} \cm{D_\be U}
\cm{D_\ga U} \cm{D^\ga U} \cm{D^\be U}} \eeq{6d:6c}
\beq + B_{103} \Tr{\cm{D_\al U} \cm{D_\be U} \cm{D_\ga U}
\cm{D^\al U} \cm{D^\be U} \cm{D^\ga U}} \eeq{6d:6d}

\beq + B_{104} \Tr{\cm{D_\mu U} \cm{D^\mu U}} \Tr{\cm{D_\al U}
\cm{D^\al U} \cm{D_\be U} \cm{D^\be U}} \eeq{6dm:6a}
\beq + B_{105} \Tr{\cm{D_\mu U} \cm{D_\nu U}} \Tr{\cm{D^\mu U}
\cm{D^\nu U} \cm{D_\al U} \cm{D^\al U}} \eeq{6dm:6c}
\beq + B_{106} \Tr{\cm{D_\mu U} \cm{D^\mu U}\cm{D_\al U}}
\Tr{\cm{D_\be U} \cm{D^\be U}\cm{D^\al U}} \eeq{6dm:6e}
\beq + B_{107} \Tr{\cm{D_\mu U} \cm{D_\nu U}\cm{D_\al U}}
\Tr{\cm{D^\mu U} \cm{D^\nu U}\cm{D^\al U}} \eeq{6dm:6f}
\beq + B_{108} \Tr{\cm{D_\mu U} \cm{D_\nu U}\cm{D_\al U}}
\Tr{\cm{D^\mu U} \cm{D^\al U}\cm{D^\nu U}} \eeq{6dm:6g}
\beq + B_{109} \Tr{\cm{D_\mu U} \cm{D^\mu U}} \Tr{\cm{D_\al U} \cm{D^\al U}}
\Tr{\cm{D_\be U} \cm{D^\be U}} \eeq{6dm:6h}
\beq + B_{110} \Tr{\cm{D_\mu U} \cm{D^\mu U}} \Tr{\cm{D_\al U} \cm{D_\be U}}
\Tr{\cm{D^\al U} \cm{D^\be U}} \eeq{6dm:6i}
\beq + B_{111} \Tr{\cm{D_\mu U} \cm{D_\nu U}} \Tr{\cm{D^\mu U} \cm{D^\al U}}
\Tr{\cm{D^\nu U} \cm{D_\al U}} \eeq{6dm:6j}
\end{tabular}
\end{table}
\narrowtext
\renewcommand{\beq}{\begin{equation}}
\renewcommand{\eeq}[1]{\label{#1} \end{equation}}

\renewcommand{\beq}{$}
\renewcommand{\eeq}[1]{$ & \ref{#1} \\ }
\widetext
\begin{table} \squeezetable
\caption[test]{Terms in the Lagrangian with 1 or more $\phi$'s and an
$\epsilon_{\al\be\ga\de}$, ($\epsilon_{0123}=+1$), sorted
according to the minimum number of electromagnetic fields $A_\mu$. The
formulas are general, but the sorting order and number of $\phi$'s
depends on the specific assumptions described in Sec.\ \ref{ssreorg} and the
caption of Table \ref{2phi}.
\label{eps1phi}}

\begin{tabular}[t]{ld}
\mbox{Terms in the Lagrangian with 1 or more $\phi$'s, an
$\epsilon_{\al\be\ga\de}$}, and at least 1 $A_\mu$ & \mbox{Eq.} \\
\hline
\beq + A_{1} \Tr{\cm{D^\mu U}(\cp{D^\nu \chi}\cp{G^{\ga\de}}-
\cp{\chi}\cp{D^\nu G^{\ga\de}} - \cp{G^{\ga\de}}\cp{D^\nu \chi}
+ \cp{D^\nu G^{\ga\de}}\cp{\chi})}\epsilon_{\mu\nu\ga\de}
\eeq{2dcg:2ab}
\end{tabular}
\end{table}
\vspace{-3ex}
\begin{table} \squeezetable
\begin{tabular}[t]{ld}
\mbox{Terms in the Lagrangian with 1 or more $\phi$'s, an
$\epsilon_{\al\be\ga\de}$}, and at least 2 $A_\mu$'s & \mbox{Eq.} \\
\hline
\beq + iA_{2} \Tr{\cm{D^\mu U}(\cp{D^\nu G^{\al\be}}\cp{{G_\al}^\ga} -
\cp{G^{\al\be}}\cp{D^\nu {G_\al}^\ga})}\epsilon_{\mu\nu\be\ga}
\eeq{2dgg:2aa}
\beq + iA_{3} \Tr{\cm{D^\mu U}(\cp{D_\al G^{\al\be}}\cp{G^{\ga\de}} -
\cp{G^{\al\be}}\cp{D_\al G^{\ga\de}} -
\cp{D_\al G^{\ga\de}}\cp{G^{\al\be}} + \cp{G^{\ga\de}}\cp{D_\al G^{\al\be}})}
\epsilon_{\mu\be\ga\de} \eeq{2dgg:2ab}

\beq + iA_{4} \Tr{\cm{\chi}\cp{G^{\al\be}}\cp{G^{\ga\de}}}
\epsilon_{\al\be\ga\de} \eeq{0dcgg:b}
\beq + iA_{5} \Tr{\cp{\chi}(\cp{H^{\al\be}}\cp{G^{\ga\de}} -
\cp{G^{\ga\de}}\cp{H^{\al\be}})}\epsilon_{\al\be\ga\de}  \eeq{0dcgh:b}
\beq + iA_{6} \Tr{\cm{\chi}}\Tr{\cp{G^{\al\be}}\cp{G^{\ga\de}}}
\epsilon_{\al\be\ga\de} \eeq{0dmcgg:b}
\end{tabular}
\end{table}
\narrowtext
\renewcommand{\beq}{\begin{equation}}
\renewcommand{\eeq}[1]{\label{#1} \end{equation}}

\renewcommand{\beq}{$}
\renewcommand{\eeq}[1]{$ & \ref{#1} \\ }
\widetext
\begin{table} \squeezetable
\caption[test]{Terms in the Lagrangian with 3 or more $\phi$'s and an
$\epsilon_{\al\be\ga\de}$, ($\epsilon_{0123}=+1$), sorted
according to the minimum number of electromagnetic fields $A_\mu$. The
formulas are general, but the sorting order and number of $\phi$'s
depends on the specific assumptions described in Sec.\ \ref{ssreorg} and the
caption of Table \ref{2phi}.
The double covariant derivative  $D_\mu D_\nu A$ is assumed to be symmetric
in its indices in accord with Eq.\ (\ref{symder}).
\label{eps3phi}}

\begin{tabular}[t]{ld}
\mbox{Terms in the Lagrangian with 3 or more $\phi$'s, an
$\epsilon_{\al\be\ga\de}$}, and at least 1 $A_\mu$ & \mbox{Eq.} \\
\hline
\beq + A_{7} \Tr{\cm{D^\al D^\be U}(\cm{D_\be U}\cm{D^\ga U}\cp{G^{\mu\nu}} -
\cp{G^{\mu\nu}}\cm{D^\ga U}\cm{D_\be U})} \epsilon_{\al\ga\mu\nu}
\eeq{4df:3ac}
\beq + A_{8} \Tr{\cm{D^\al D^\be U}(\cm{D^\ga U}\cm{D_\be U}\cp{G^{\mu\nu}} -
\cp{G^{\mu\nu}}\cm{D_\be U}\cm{D^\ga U})} \epsilon_{\al\ga\mu\nu}
\eeq{4df:3ae}
\beq + A_{9} \Tr{\cm{D^\al D_\be U}\cm{D^\ga U}\cp{G^{\be\nu}}\cm{D^\de U} }
\epsilon_{\al\ga\de\nu} \eeq{4df:3ag}

\beq + A_{10} \Tr{\cm{D^\al D^\be U}(\cm{D_\be U}\cp{D^\ga H^{\mu\nu}} +
\cp{D^\ga H^{\mu\nu}}\cm{D_\be U})} \epsilon_{\al\ga\mu\nu}
\eeq{4df:4bb}
\beq + A_{11} \Tr{\cm{D^\al D^\be U}(\cm{D^\ga U}\cp{D_\be H^{\mu\nu}} +
\cp{D_\be H^{\mu\nu}}\cm{D^\ga U})} \epsilon_{\al\ga\mu\nu}
\eeq{4df:4bc}

\beq + A_{12} \Tr{\cm{D^\mu U}\cm{D^\nu U}(\cm{\chi}\cp{G^{\ga\de}} +
\cp{G^{\ga\de}}\cm{\chi})} \epsilon_{\mu\nu\ga\de} \eeq{2dcg:3c}
\beq + A_{13} \Tr{\cm{D^\mu U}\cm{\chi}\cm{D^\nu U}\cp{G^{\ga\de}}}
\epsilon_{\mu\nu\ga\de} \eeq{2dcg:3d}
\beq + A_{14} \Tr{\cm{D^\mu U}\cm{D^\nu U}\cp{G^{\ga\de}}}\Tr{\cm{\chi}}
\epsilon_{\mu\nu\ga\de} \eeq{2dmcg:3b}
\beq + A_{15} \Tr{\cm{D^\mu U}\cm{D^\nu U}(\cp{\chi}\cp{H^{\ga\de}} -
\cp{H^{\ga\de}}\cp{\chi})}\epsilon_{\mu\nu\ga\de} \eeq{2dch:3b}
\beq + A_{16} \Tr{\cm{D^\mu U}\cp{H^{\ga\de}}}\Tr{\cm{D^\nu U}\cp{\chi}}
\epsilon_{\mu\nu\ga\de} \eeq{2dmch:3b}

\beq + A_{17} \Tr{\cm{D^\mu U}(\cm{D^\nu \chi}\cp{H^{\ga\de}}-
\cm{\chi}\cp{D^\nu H^{\ga\de}} + \cp{H^{\ga\de}}\cm{D^\nu \chi}
- \cp{D^\nu H^{\ga\de}}\cm{\chi})}\epsilon_{\mu\nu\ga\de}
\eeq{2dch:2ab}
\beq + A_{18} \lb \Tr{\cm{D^\mu U}\cp{H^{\ga\de}}}\Tr{\cm{D^\nu \chi}} -
\Tr{\cm{D^\mu U}\cp{D^\nu H^{\ga\de}}}\Tr{\cm{\chi}} \rb
\epsilon_{\mu\nu\ga\de} \eeq{2dmch:2ab}
\end{tabular}
\end{table}
\vspace{-3ex}
\begin{table} \squeezetable
\begin{tabular}[t]{ld}
\mbox{Terms in the Lagrangian with 3 or more $\phi$'s, an
$\epsilon_{\al\be\ga\de}$}, and at least 2 $A_\mu$'s & \mbox{Eq.} \\
\hline
\beq + iA_{19} \Tr{\cm{D_\mu U}\cm{D^\mu U}(\cp{G^{\al\be}}\cp{H^{\ga\de}} -
\cp{H^{\ga\de}}\cp{G^{\al\be}})}\epsilon_{\al\be\ga\de}  \eeq{2dgh:3a}
\beq + iA_{20} \Tr{\cm{D^\mu U}\cm{D^\nu U}(\cp{G^{\al\be}}\cp{{H_\al}^\ga} +
\cp{{H_\al}^\ga}\cp{G^{\al\be}})}\epsilon_{\mu\nu\be\ga}  \eeq{2dgh:3b}
\beq + iA_{21} \Tr{\cm{D^\mu U}\cp{G^{\al\be}}\cm{D^\nu U}\cp{{H_\al}^\ga}}
\epsilon_{\mu\nu\be\ga}  \eeq{2dgh:3c}
\beq + iA_{22} \Tr{\cm{D^\mu U}(\cm{D_\al U}\cp{G^{\al\be}}\cp{H^{\ga\de}} -
\cp{H^{\ga\de}}\cp{G^{\al\be}}\cm{D_\al U})} \epsilon_{\mu\be\ga\de}
\eeq{2dgh:3d}

\beq + iA_{23} \Tr{\cm{D^\mu U}(\cp{D^\nu H^{\al\be}}\cp{{H_\al}^\ga} -
\cp{H^{\al\be}}\cp{D^\nu {H_\al}^\ga})}\epsilon_{\mu\nu\be\ga}
\eeq{2dhh:2aa}
\beq + iA_{24} \Tr{\cm{D^\mu U}(\cp{D_\al H^{\al\be}}\cp{H^{\ga\de}} -
\cp{H^{\al\be}}\cp{D_\al H^{\ga\de}} -
\cp{D_\al H^{\ga\de}}\cp{H^{\al\be}} + \cp{H^{\ga\de}}\cp{D_\al H^{\al\be}})}
\epsilon_{\mu\be\ga\de} \eeq{2dhh:2ab}

\beq + iA_{25} \Tr{\cm{\chi}\cp{H^{\al\be}}\cp{H^{\ga\de}}}
\epsilon_{\al\be\ga\de}  \eeq{0dchh:b}
\beq + iA_{26} \Tr{\cm{\chi}}\Tr{\cp{H^{\al\be}}\cp{H^{\ga\de}}}
\epsilon_{\al\be\ga\de} \eeq{0dmchh:b}
\end{tabular}
\end{table}
\narrowtext
\renewcommand{\beq}{\begin{equation}}
\renewcommand{\eeq}[1]{\label{#1} \end{equation}}

\renewcommand{\beq}{$}
\renewcommand{\eeq}[1]{$ & \ref{#1} \\ }
\widetext
\begin{table} \squeezetable
\caption[test]{Terms in the Lagrangian with 5 or more $\phi$'s and an
$\epsilon_{\al\be\ga\de}$, ($\epsilon_{0123}=+1$), sorted
according to the minimum number of electromagnetic fields $A_\mu$. The
formulas are general, but the sorting order and number of $\phi$'s
depends on the specific assumptions described in Sec.\ \ref{ssreorg} and the
caption of Table \ref{2phi}.
The double covariant derivative  $D_\mu D_\nu A$ is assumed to be symmetric
in its indices in accord with Eq.\ (\ref{symder}).
\label{eps5phi}}

\begin{tabular}[t]{ld}
\mbox{Terms in the Lagrangian with 5 or more $\phi$'s and an
$\epsilon_{\al\be\ga\de}$, with an $A_\mu$ possible
but not necessary} & \mbox{Eq.} \\
\hline
\beq + iA_{27} \Tr{\cm{D^\mu D^\nu U}(\cm{D_\nu U} \cm{D^\be U} \cm{D^\ga
U} \cm{D^\de U} + \cm{D^\de U} \cm{D^\ga U} \cm{D^\be
U} \cm{D_\nu U} )} \epsilon_{\mu\be\ga\de} \eeq{6d:5b}

\beq + iA_{28} \Tr{\cm{\chi} \cm{D_\al U}\cm{D_\be U} \cm{D_\ga U}
\cm{D_\de U}} \epsilon^{\al\be\ga\de} \eeq{4dc:7d}
\end{tabular}
\end{table}
\vspace{-3ex}
\begin{table} \squeezetable
\begin{tabular}[t]{ld}
\mbox{Terms in the Lagrangian with 5 or more $\phi$'s, an
$\epsilon_{\al\be\ga\de}$}, and at least 1 $A_\mu$ & \mbox{Eq.} \\
\hline
\beq + A_{29} \Tr{\cp{H^{\mu\nu}}(\cm{D^\al U}\cm{D^\be U}\cm{D_\ga
U}\cm{D^\ga U} +
\cm{D_\ga U}\cm{D^\ga U}\cm{D^\be U}\cm{D^\al U})}\epsilon_{\mu\nu\al\be}
\eeq{4df:7ba}
\beq + A_{30} \Tr{\cp{H^{\mu\nu}}(\cm{D^\al U}\cm{D_\ga U}\cm{D^\be
U}\cm{D^\ga U} +
\cm{D_\ga U}\cm{D^\be U}\cm{D^\ga U}\cm{D^\al U})}\epsilon_{\mu\nu\al\be}
\eeq{4df:7bb}

\beq + A_{31} \Tr{\cp{H^{\mu\nu}}\cm{D^\al U}}\Tr{\cm{D^\be U}
\cm{D_\ga U}\cm{D^\ga U}}\epsilon_{\mu\nu\al\be} \eeq{4dmf:7ba}
\beq + A_{32} \Tr{\cp{H^{\mu\nu}}(\cm{D^\al U}\cm{D^\ga U}+
\cm{D^\ga U}\cm{D^\al U})}
\Tr{\cm{D^\be U}\cm{D_\ga U}}\epsilon_{\mu\nu\al\be} \eeq{4dmf:7bb}
\end{tabular}
\end{table}
\narrowtext
\renewcommand{\beq}{\begin{equation}}
\renewcommand{\eeq}[1]{\label{#1} \end{equation}}


\clearpage

\renewcommand{\beq}{$}
\renewcommand{\eeq}[1]{$ & \ref{#1} \\ }
\widetext
\begin{table} \squeezetable
\caption[test]{Terms in the Lagrangian proportional to the
classical equation of motion.
The lowest order equation of motion operator ${\cal O}_{EOM}^{(2)}$ is given
by Eq.\ (\ref{eomop}).
The double covariant derivative  $D_\mu D_\nu A$ is assumed to be symmetric
in its indices in accord with Eq.\ (\ref{symder}), as are all multiple
covariant derivatives.
\label{eom}}

\begin{tabular}[t]{ld}
\mbox{Terms in the Lagrangian without an $\epsilon_{\al\be\ga\de}$
and proportional to the classical equation of motion} & \mbox{Eq.} \\
\hline
\beq + E_{1} \Tr{{\cal O}_{EOM}^{(2)}\cm{D_\al D^\al D_\be D^\be U}}
\eeq{6d:1a}

\beq + E_{2} \Tr{{\cal O}_{EOM}^{(2)}{\cal O}_{EOM}^{(2)}\cm{D_\be U}\cm{D^\be
U}} \eeq{6d:4aa}

\beq + E_{3} \Tr{{\cal O}_{EOM}^{(2)} \cm{D^\al U} \cm{D_\al D_\be U}
\cm{D^\be U}} \eeq{6d:4bb}

\beq + E_{4} \Tr{{\cal O}_{EOM}^{(2)} {\cal O}_{EOM}^{(2)}} \Tr{\cm{D_\be U}
\cm{D^\be U}} \eeq{6dm:4aa}
\beq + E_{5} \Tr{{\cal O}_{EOM}^{(2)} \cm{D_\al D_\be U}} \Tr{\cm{D^\al U}
\cm{D^\be U}} \eeq{6dm:4ab}

\beq + E_{6} \Tr{{\cal O}_{EOM}^{(2)} \cm{D_\al U}}  \Tr{{\cal O}_{EOM}^{(2)}
\cm{D^\al U}} \eeq{6dm:4ba}
\beq + E_{7} \Tr{{\cal O}_{EOM}^{(2)} \cm{D_\al U}}  \Tr{\cm{D^\al D^\be U}
\cm{D_\be U}} \eeq{6dm:4bb}

\beq + E_{8} \Tr{{\cal O}_{EOM}^{(2)} \cm{D_\be D^\be \chi}} \eeq{4dc:1}

\beq + E_{9} \Tr{{\cal O}_{EOM}^{(2)} {\cal O}_{EOM}^{(2)} \cp{\chi}}
\eeq{4dc:2}

\beq + E_{10} \Tr{{\cal O}_{EOM}^{(2)} \cm{D_\be U} \cm{\chi} \cm{D^\be U}}
\eeq{4dc:3b}

\beq + E_{11} \Tr{{\cal O}_{EOM}^{(2)}(\cm{D_\be U}\cp{D^\be \chi} + \cp{D^\be
\chi}\cm{D_\be U})} \eeq{4dc:4a}

\beq + E_{12} \Tr{{\cal O}_{EOM}^{(2)} {\cal O}_{EOM}^{(2)}} \Tr{\cp{\chi}}
\eeq{4dmc:2}

\beq + E_{13} \Tr{{\cal O}_{EOM}^{(2)}\cm{D_\be U}} \Tr{\cm{D^\be U}
\cm{\chi}} \eeq{4dmc:3a}
\beq + E_{14} \Tr{{\cal O}_{EOM}^{(2)}\cm{\chi}} \Tr{\cm{D_\be U}
\cm{D^\be U} } \eeq{4dmc:3b}
\beq + E_{15} \Tr{{\cal O}_{EOM}^{(2)}\cm{D_\be U}\cm{D^\be U}} \Tr{\cm{\chi}}
\eeq{4dmc:3c}
\beq + E_{16} \Tr{{\cal O}_{EOM}^{(2)}\cm{D_\be U}} \Tr{\cp{D^\be \chi}}
\eeq{4dmc:4a}

\beq + iE_{17} \Tr{{\cal O}_{EOM}^{(2)}(\cm{D_\mu U}\cm{D_\nu U}
\cp{H^{\mu\nu}} +
\cp{H^{\mu\nu}}\cm{D_\nu U}\cm{D_\mu U})} \eeq{4df:3ba}
\beq + iE_{18} \Tr{{\cal O}_{EOM}^{(2)}(\cm{D_\mu U}\cp{D_\nu G^{\mu\nu}} -
\cp{D_\nu G^{\mu\nu}}\cm{D_\mu U})} \eeq{4df:4aa}

\beq + iE_{19} \Tr{{\cal O}_{EOM}^{(2)}\cm{D_\mu U}}\Tr{\cm{D_\nu U}
\cp{H^{\mu\nu}}} \eeq{4dmf:3ba}

\beq + E_{20} \Tr{{\cal O}_{EOM}^{(2)}(\cp{\chi}\cm{\chi} +
\cm{\chi}\cp{\chi})} \eeq{2dcc:2b}
\beq + E_{21} \Tr{{\cal O}_{EOM}^{(2)}\cp{\chi}} \Tr{\cm{\chi}} \eeq{2dmcc:2ba}
\beq + E_{22} \Tr{{\cal O}_{EOM}^{(2)}\cm{\chi}} \Tr{\cp{\chi}} \eeq{2dmcc:2bb}

\beq + E_{23} \Tr{{\cal O}_{EOM}^{(2)}(\cp{G^{\al\be}}\cp{H_{\al\be}} -
\cp{H_{\al\be}}\cp{G^{\al\be}})} \eeq{2dgh:2ba}
\end{tabular}
\end{table}
\vspace{-3ex}
\begin{table} \squeezetable
\begin{tabular}[t]{ld}
\mbox{Terms in the Lagrangian with an $\epsilon_{\al\be\ga\de}$,
($\epsilon_{0123}=+1$), and proportional to the classical equation of
motion} & \mbox{Eq.} \\
\hline
\beq + iE_{24} \Tr{{\cal O}_{EOM}^{(2)} \cm{D^\al U}
\cm{D^\be U} \cm{D^\ga U}  \cm{D^\de U}}
\epsilon_{\al\be\ga\de}  \eeq{6d:5a}

\beq + E_{25} \Tr{{\cal O}_{EOM}^{(2)}(\cm{D^\be U}\cm{D^\ga U}
\cp{G^{\mu\nu}} -
\cp{G^{\mu\nu}}\cm{D^\ga U}\cm{D^\be U})} \epsilon_{\be\ga\mu\nu}
\eeq{4df:3aa}
\beq + E_{26} \Tr{{\cal O}_{EOM}^{(2)}\cm{D^\be U}\cp{G^{\mu\nu}}\cm{D^\ga U}}
\epsilon_{\be\ga\mu\nu} \eeq{4df:3ab}

\beq + E_{27} \Tr{{\cal O}_{EOM}^{(2)}(\cm{D^\be U}\cp{D^\ga H^{\mu\nu}} +
\cp{D^\ga H^{\mu\nu}}\cm{D^\be U})} \epsilon_{\be\ga\mu\nu}
\eeq{4df:4ba}

\beq + iE_{28} \Tr{{\cal O}_{EOM}^{(2)}\cp{H^{\al\be}}\cp{H^{\ga\de}}}
\epsilon_{\al\be\ga\de} \eeq{2dhh:2ba}
\beq + iE_{29} \Tr{{\cal O}_{EOM}^{(2)}\cp{G^{\al\be}}\cp{G^{\ga\de}}}
\epsilon_{\al\be\ga\de} \eeq{2dgg:2ba}
\end{tabular}
\end{table}
\narrowtext
\renewcommand{\beq}{\begin{equation}}
\renewcommand{\eeq}[1]{\label{#1} \end{equation}}

\mediumtext
\begin{table}
\caption[test]{\label{tracerel}Application of trace relations: The
first column
contains the relevant equation numbers of structures which are related
by trace relations. The second column refers to the specific trace relation
which has been applied. The third column denotes the equation number
of the structure which we have chosen to eliminate.}
\begin{tabular}{ccc}
related structures &trace relation&structure eliminated\\
\hline
(\ref{6d:4aa}),(\ref{6d:4ba}),(\ref{6dm:4aa}),(\ref{6dm:4ba})&
(\ref{tra4:6})&(\ref{6d:4ba})\\
(\ref{6d:4ab}),(\ref{6d:4bb}),(\ref{6dm:4ab}),(\ref{6dm:4bb})&
(\ref{tra4:2})&(\ref{6d:4ab})\\
(\ref{6d:4ac}),(\ref{6d:4bc}),(\ref{6dm:4ac}),(\ref{6dm:4bc})&
(\ref{tra4:6})&(\ref{6d:4bc})\\
(\ref{6d:4ad}),(\ref{6d:4ae}),(\ref{6d:4bd}),(\ref{6dm:4ad}),
(\ref{6dm:4bd}),(\ref{6dm:4be})&(\ref{tra4:2})&(\ref{6d:4bd})\\
(\ref{6d:6a})-(\ref{6d:6e}),(\ref{6dm:6a})-(\ref{6dm:6g})&
(\ref{tra6:2})&(\ref{6d:6e})\\
(\ref{6dm:6a}),(\ref{6dm:6b}),(\ref{6dm:6h}),(\ref{6dm:6i})&
(\ref{tra4:6})&(\ref{6dm:6b})\\
(\ref{6dm:6c}),(\ref{6dm:6d}),(\ref{6dm:6i}),(\ref{6dm:6j})&
(\ref{tra4:4})&(\ref{6dm:6d})\\
(\ref{4dc:3a}),(\ref{4dc:3b}),(\ref{4dmc:3a}),(\ref{4dmc:3b}),(\ref{4dmc:3c})&
(\ref{tra4:3})&(\ref{4dc:3a})\\
(\ref{4dc:3c}),(\ref{4dc:3d}),(\ref{4dmc:3d}),(\ref{4dmc:3e}),(\ref{4dmc:3f})&
(\ref{tra4:1})&(\ref{4dc:3c})\\
(\ref{4dc:7a})-(\ref{4dc:7c}),(\ref{4dmc:7c})-(\ref{4dmc:7e})
&(\ref{tra5b})&(\ref{4dc:7b})\\
(\ref{4dmc:7a}),(\ref{4dmc:7b}),(\ref{4dmc:7f}),(\ref{4dmc:7g})&
(\ref{tra4:6})&(\ref{4dmc:7b})\\
(\ref{4df:3bb})-(\ref{4df:3bd}),(\ref{4dmf:3bb})-(\ref{4dmf:3bd})&
(\ref{tra4:2})&(\ref{4df:3bd})\\
(\ref{2dcc:3a}),(\ref{2dcc:3b}),(\ref{2dmcc:3a}),(\ref{2dmcc:3c}),
(\ref{2dmcc:3e}),(\ref{2dmcc:3g})&(\ref{tra4:5})&(\ref{2dcc:3b})\\
(\ref{2dcc:3c}),(\ref{2dcc:3d}),(\ref{2dmcc:3b}),(\ref{2dmcc:3d}),
(\ref{2dmcc:3f}),(\ref{2dmcc:3h})&(\ref{tra4:5})&(\ref{2dcc:3d})\\
(\ref{2dgg:3a}),(\ref{2dgg:3b}),(\ref{2dmgg:3a}),(\ref{2dmgg:3b})&
(\ref{tra4:6})&(\ref{2dgg:3b})\\
(\ref{2dgg:3c})-(\ref{2dgg:3e}),(\ref{2dmgg:3c})-(\ref{2dmgg:3e})&
(\ref{tra4:2})&(\ref{2dgg:3d})\\
(\ref{2dhh:3a}),(\ref{2dhh:3b}),(\ref{2dmhh:3a}),(\ref{2dmhh:3b})&
(\ref{tra4:6})&(\ref{2dhh:3b})\\
(\ref{2dhh:3c})-(\ref{2dhh:3e}),(\ref{2dmhh:3c})-(\ref{2dmhh:3e})&
(\ref{tra4:2})&(\ref{2dhh:3d})\\
\end{tabular}
\end{table}
\narrowtext

\mediumtext
\begin{table}
\caption[test]{\label{epsilonrel}Application of epsilon relations: The
first column
contains the relevant equation numbers of structures which are related
by epsilon relations.  The second column denotes the equation number
of the structure which we have chosen to eliminate.}
\begin{tabular}{cc}
related structures & structure eliminated\\
\hline
(\ref{6d:5a}),(\ref{6d:5b}),(\ref{6d:5c})&
(\ref{6d:5c})\\
(\ref{4df:3ab}),(\ref{4df:3ad}),(\ref{4df:3ag})&
(\ref{4df:3ad})\\
(\ref{4df:3aa}),(\ref{4df:3ac}),(\ref{4df:3ae}),(\ref{4df:3af})&
(\ref{4df:3af})\\
(\ref{4df:4ba}),(\ref{4df:4bb}),(\ref{4df:4bc}),(\ref{4df:4bd})&
(\ref{4df:4bd})\\
(\ref{4df:7ba}),(\ref{4df:7bb}),(\ref{4df:7bc}),(\ref{4df:7bd})&
(\ref{4df:7bc}),(\ref{4df:7bd})\\
(\ref{2dgh:3c}),(\ref{2dgh:3e}),(\ref{2dgh:3h})&
(\ref{2dgh:3e}),(\ref{2dgh:3h})\\
(\ref{2dgh:3a}),(\ref{2dgh:3b}),(\ref{2dgh:3d}),
(\ref{2dgh:3f}),(\ref{2dgh:3g}),(\ref{2dgh:3i})&
(\ref{2dgh:3f}),(\ref{2dgh:3g}),(\ref{2dgh:3i})\\
(\ref{2dhh:2aa}),(\ref{2dhh:2ab}),(\ref{2dhh:2ac})&
(\ref{2dhh:2ac})\\
(\ref{2dgg:2aa}),(\ref{2dgg:2ab}),(\ref{2dgg:2ac})&
(\ref{2dgg:2ac})\\
(\ref{2dhh:2ba}),(\ref{2dhh:2bb})&(\ref{2dhh:2bb})\\
(\ref{2dgg:2ba}),(\ref{2dgg:2bb})&(\ref{2dgg:2bb})\\
(\ref{0dggh:a})&(\ref{0dggh:a})\\
\end{tabular}
\end{table}

\end{document}